\documentclass[reqno,12pt,a4]{article}
\usepackage{amsmath, amssymb, slashed,graphicx,color,multicol}
\usepackage{cellspace} 
\usepackage[page,toc]{appendix}
\usepackage[colorlinks=true,urlcolor=black,linkcolor=black,citecolor=black]{hyperref}
\usepackage[utf8]{inputenc}
\newtheorem{definition}{Definition}
\newtheorem{proposition}[definition]{Proposition}

\newtheorem{corollary}[definition]{Corollary}
\newtheorem{lemma}[definition]{Lemma}
\newtheorem{thm}{Theorem}
\newcommand{\ubar} \underline
\newcommand{\wbar}[1]{\underline{\bar{#1}}}
\newcommand{\Gamme}{\mathsf{\Gamma}}
\newcommand{\bunderline}[1]{{\mkern2mu\underline{\mkern-2mu #1\mkern-2mu}\mkern2mu }}
\newcommand{\Tamma}{\bunderline{\Gamma}}
\newtheorem{hyprc}{Hypothesis}

\newcommand{\bim}{\stackrel{\beta}{\sim}}
\newcommand{\sslash}{{\scriptscriptstyle/\mkern-6mu/}}
\newcommand{\N}{\textsc{n}}
\newcommand{\F}{\mathrm{F}}
\newcommand{\Damma}{\digamma}

\begin{document}

\begin{titlepage}
\begin{center}\Large
Renormalization of SU(2) Yang--Mills theory\\with Flow Equations
\end{center}
\vspace{20pt}
{\small
\begin{center}
Alexander N. Efremov
\\[-0.5mm]
\href{http://www.universite-paris-saclay.fr}{ Universit\'e Paris-Saclay}, \href{http://www.edpif.org}{Doctoral School ED564},
\\[-0.5mm]
CPHT, Ecole Polytechnique, F-91128 Palaiseau, France.
\\[-0.5mm]
\href{mailto:alexander@efremov.fr}{alexander@efremov.fr}
\end{center}
\begin{center}
Riccardo Guida
\\[-0.5mm]
Institut de Physique Th\'eorique, Universit\'e Paris-Saclay, CEA, CNRS,
\\[-0.5mm] F-91191 Gif-sur-Yvette, France
\\[-0.5mm] 
\href{http://orcid.org}{ORCID iD}: \href{http://orcid.org/0000-0002-2854-7214}{\texttt{orcid.org/0000-0002-2854-7214}}
\\[-0.5mm]
\href{www.researcherid.com}{ResearcherID}: \href{http://www.researcherid.com/rid/N-7759-2013}{\texttt{N-7759-2013}}
\\[-0.5mm]
\href{mailto:riccardo.guida@cea.fr}{riccardo.guida@cea.fr}
\end{center}
\begin{center}
Christoph Kopper
\\[-0.5mm]
Centre de Physique Th{\'e}orique de l'Ecole Polytechnique,
\\[-0.5mm]
CNRS, UMR 7644, F-91128 Palaiseau, France.
\\[-0.5mm]
\href{mailto:Christoph.Kopper@polytechnique.edu}{Christoph.Kopper@polytechnique.edu}
\end{center}
}
\vspace{20pt}
\begin{abstract}
We give a proof of perturbative renormalizability of SU(2) Yang--Mills theory  in four-dimensional Euclidean space  which is based on the Flow Equations of the renormalization group. The main motivation is to present a proof which does not make appear mathematically undefined objects (as for example dimensionally regularized generating functionals), which permits to parametrize the theory in terms of {\it physical} renormalization conditions, and which allows to control the singularities of the correlation functions of the theory in the infrared domain. Thus a large part of the proof is dedicated to bounds on massless correlation functions. 
\vfill \flushright \begin{tabular}{lr}Preprint&IPhT t17/055\\&RR 018-04-2017\end{tabular}
\end{abstract}
\end{titlepage}

\tableofcontents

\section{Introduction}
Renormalization theory based on the differential Flow Equations \cite{wilson,wehoug,pol}  has allowed to build a unified approach to the analysis of renormalizability for a wide class of theories without recourse to Feynman graphs. This method was applied to show momentum bounds of massive $\phi^4$ theory \cite{kome}, used to prove renormalization of spontaneously broken SU(2) Yang--Mills \cite{komu} and to establish uniform bounds on Schwinger functions of massless $\phi^4$ fields \cite{kogui}.
Starting with the milestone works, \cite{YM54,FP,st2,taylor,LZJ,tHV72,brs,tyutin,st}
(see \cite{Lai81} for more references),
a variety of results on the renormalizability of nonabelian gauge theories flourished in the literature,
in different contexts
and with different level of mathematical rigor.
Work on this problem
in the context of Flow Equations (FE)
includes~\cite{wet,bec,bam,ma}, and,
more recently, \cite{fhh}.
The present work deals with perturbative renormalizability and shares certain aspects with some of these articles:
from~\cite{bec} we borrow the fruitful idea that the local
operator describing the violation of Slavnov--Taylor identities (STI) for the one-particle irreducible (1PI) functions
\cite{st,KSZ75}
is constrained by
the nilpotency of the underlying
''Slavnov differential operator'';
as in \cite{bam} (and in contrast with \cite{fhh}) we define the marginal correlators
by physical boundary conditions at vanishing IR cutoff
and nonexceptional momenta.
Our main result is a proof of the renormalizability of Yang--Mills theory that complements the previous work
in view of the following features:
i) Rigorous control of the IR and UV behavior of the one-particle irreducible functions is established by means of uniform bounds in momentum space. In particular,
from the bounds follow the existence of the IR limit of 1PI functions at nonexceptional momenta and the existence of the subsequent UV limit.
ii) The vanishing of the STI violation in the UV limit is proven;
iii) The relevant part of 1PI functions and STI violation at nonvanishing renormalization momenta, which gives initial conditions for the renormalization group FE, is thoroughly identified.

Our proof refers explicitly to SU(2) Yang--Mills theory. However, it could be extended without important conceptual changes to other semi-simple compact Lie groups.

We proceed as follows. In section \ref{sec_327a} we fix the notations,
introduce the classical Yang--Mills action, generating functionals and
regulators. We define the BRST transformations. Then we derive
the FE of the renormalization group, for the connected amputated Schwinger functions and for the one-particle
irreducible vertex functions. Finally we study the STI, the Antighost Equation (AGE)  and their violation. Remark that
in our context gauge invariance is broken through the presence of
the cutoffs.

Our proof of renormalizability of Yang-Mills theory is based on
momentum bounds for the vertex bounds which permit to take the limits
$\Lambda \to 0$ (IR-cutoff) and $\Lambda_0 \to \infty$ (UV-cutoff)
for nonexceptional external momenta. These bounds are an extension of those in \cite{kogui} and are established
inductively with the aid of the FE. They are expressed in terms
of tree amplitudes and polynomials of logarithms. For our trees
we only have to consider vertices of coordination numbers 1 and 3.
In section \ref{sec_21bnd} we present the definitions of the tree structures
to be considered, and the aforementioned bounds.
These bounds permit to prove the existence of the vertex functions
for nonexceptional external momenta on removal of the cutoffs.
This statement also holds for vertex functions carrying an operator
insertion related the the BRST transformation of the fields.
These vertex functions are required to formulate the STI.
Then we also have to consider vertex functions with operator
insertions which permit to formulate the violation of the STI.
Our bounds then permit to show that, for suitable renormalization
conditions, these functions describing the violation of the STI
vanish in the UV limit.

Section \ref{sec_327b} is dedicated to the proof of those bounds. The rhs
of the FE is a sum of products of vertex functions in lower
loop order joined by free propagators. Our technique of proof
then is based on the fact that applying our inductive bounds
on these chains, which still have to be closed by a $\Lambda$-derived
propagator and then integrated over the circulating loop
momentum and over $\Lambda$, reproduce our inductive bounds
at loop order $l$. The proof treats irrelevant terms first,
then  marginal and finally strictly relevant ones.
Particular attention has to be paid to the renormalization conditions.
Section~\ref{sec_327b} ends with a proof of UV-convergence for $\Lambda_0 \to \infty$.

In section \ref{sec_sti} we prove that the renormalization conditions
required to prove the bounds of the previous sections can
actually be imposed. These renormalization conditions are
such that they leave us free to fix the physical coupling
of the theory. At the same time they permit us to make
vanish the relevant part of all functions describing
the violation of the STI. This is required in the previous
proof.

In the Appendices \ref{sec_gauss}, \ref{ex_AAcc}, \ref{sec_t}, \ref{sec_b} we present some facts on Gaussian measures,
an example of chains of vertex functions, an analysis of linear independence of euclidean tensor structures, a large number of elementary bounds on integrals we encounter in the proofs.
We also add a list with bounds on the propagators and their
derivatives. In the two subsequent appendices~\ref{sec_gamma},~\ref{sec_gamma2} we analyse
the generating functionals of the (inserted) vertex functions,
as far as they have relevant content. Appendix \ref{sec_rr} is a supplement to section \ref{sec_vsti} and contains the generalization of the STI and AGE to the case of nonvanishing IR cutoff. In the last two appendices~\ref{sec_ren},~\ref{sec_op}
we present the list of renormalization points and operator
insertions to be considered.
\section{The formalism}\label{sec_327a}
\subsection{Notations}
$\mathbb{N}$ is the set of nonnegative integers. $|S|$ is the cardinality of a set $S$. Furthermore, $(a,b,c,...)$, $\{a,b,c,...\}$  denote a sequence and a set, respectively. Unless otherwise stated, sequence stands for finite sequence. For shortness we set
$[a:b]:=\{i\in \mathbb{Z}: a\leqslant i \leqslant b\}$ and $[b]:=[1: b]$.
Repeated indices are implicitly summed over, e.g. $A^a t_a:=\Sigma_a A^a t_a$. We choose the following basis of the Lie algebra
\begin{align}
t_a&:=\frac{1}{2}\sigma_a,&[t_a,t_b]&=i\epsilon_{abc}t_c,&a,b,c &\in \{1,2,3\},
\end{align}
where $\sigma_a$ are the Pauli matrices and $\epsilon_{abc}$ is the Levi--Civita symbol, $\epsilon_{123}=1$. In this article we will deal with tensor fields on $\mathbb{R}^4$ in Cartesian coordinate systems with metric tensor $\delta_{\mu \nu}$. If $A,B$ are two Cartesian tensors of $\mathbb{R}^4$ of rank $r$ with components $A_{\mu_1...\mu_r}$ and $B_{\mu_1...\mu_r}$, respectively, then the scalar product $(A,B)$ is the contraction
\begin{equation}
(A,B) := A^*_{\mu_1...\mu_r}B_{\mu_1...\mu_r}.\label{scalar_product}
\end{equation}
Given a Cartesian tensor $T$, we use the norm
\begin{equation}
|T|:=(T,T)^{\frac{1}{2}}.
\end{equation}
For instance, for $p \in \mathbb{R}^4$, $|p|^2= \Sigma_\mu  p^2_\mu$. Let $T$, $A$, and $B$ be Cartesian tensors such that $T_{\Vec{\mu} \Vec{\nu}}=A_{\Vec{\mu} \Vec{\rho}}\,B_{\Vec{\rho} \Vec{\nu}}$ where $\Vec{\mu}$, $\Vec{\rho}$, $\Vec{\nu}$ are multi-indices, for example $\Vec{\mu}:=(\mu_1,...,\mu_n)$. Then using the Cauchy--Schwarz inequality
\begin{equation}
|T|^2=\sum_{\Vec{\mu},\Vec{\nu}} |\sum_{\Vec{\rho}}A_{\Vec{\mu}\Vec{\rho}}B_{\Vec{\rho}\Vec{\nu}}|^2\leqslant \sum_{\Vec{\mu},\Vec{\nu}}\sum_{\Vec{\rho},\Vec{\sigma}} |A_{\Vec{\mu}\Vec{\rho}}|^2|B_{\Vec{\sigma}\Vec{\nu}}|^2=|A|^2|B|^2\,.\label{eq_holder}
\end{equation}
The integral over $\mathbb{R}^4$ of the product of two functions is denoted by
\begin{equation}
\langle f_1,f_2 \rangle  := \int d^4 x \, f_1(x)f_2(x) \,,
\end{equation}
and the Fourier transform of a function is defined by
\begin{equation}
f(p):=\int d^4x \, e^{-ipx} f(x).
\end{equation}
The convolution of two functions $f$, $g$ is denoted as below
\begin{equation}
(f * g)(x)= \int d^4 y \, f(y) g(x-y).
\end{equation}
For functions $\phi_i(p_i)$ and $F(p_1,...,p_k)$ with $p_i \in \mathbb{R}^4$ the symbol $\langle F| \phi_1...\phi_{k};p\rangle$  denotes the following integral in momentum space
\begin{flalign}
&
\langle F| \phi_1...\phi_{k};p\rangle
\mspace{-4mu}:= \mspace{-8mu}\int \mspace{-4mu}
(2\pi)^4\delta\big(\sum^k_{j=1} p_j+p\big)
F(p_1,...,p_k) \phi_1(p_1)...\phi_k(p_{k})
\prod^k_{i=1}
\mspace{-2mu}
\frac{d^4 p_i}{(2\pi)^4}
\mspace{1mu}.\mspace{-26mu}&
\end{flalign}
We also use the shorthands
\begin{align}
\langle \phi_1...\phi_{k};p\rangle&:=\langle 1| \phi_1...\phi_{k};p\rangle,&\langle \phi_1...\phi_{k}\rangle&:=\langle 1| \phi_1...\phi_{k};0\rangle.
\end{align}
A cumulative notation for the elementary fields and corresponding sources is
\begin{align}
\Phi&:=\begin{pmatrix}A^a_\mu,&B^a,&c^a,&\bar{c}^a\end{pmatrix},&K&:=\begin{pmatrix}j^a_\mu,&b^a,&\bar{\eta}^a,&\eta^a\end{pmatrix},
\end{align}
where $c,\bar{c},\eta,\bar{\eta}$ are generators of an infinite-dimensional anticommuting algebra. Furthermore, we use the following shorthand
\begin{equation}
K \cdot \Phi :=\langle j^a_\mu, A^a_\mu \rangle + \langle b^a, B^a \rangle + \langle \bar{\eta}^a, c^a \rangle  + \langle \bar{c}^a, \eta^a \rangle.\label{eq_jA}
\end{equation}
We will have to consider one-particle irreducible functions, also known as vertex functions, whose generating functional is denoted by $\Gamma$. These functions are translation-invariant in position space. Their \emph{reduced} Fourier transforms $\Gamma^{\Vec{\phi}}$ are defined as follows
\begin{equation}
\Gamma^{\Vec{\phi}}(p_1,\dots,p_{\N-1}):=  \int \Big(\prod^{\N-1}_{i=1} d^4x_i\, e^{-i p_i x_i} \Big) \;  \Gamma^{\Vec{\phi}}(0,x_1,\dots,x_{\N-1})\,,
\end{equation}
where $\Vec{\phi}:=(\phi_0,...,\phi_{\N-1})$ is a sequence of field labels, $\phi_i \in \{A, B, c, \bar{c}\}$, and
\begin{equation}
\Gamma^{\Vec{\phi}}(x_0,x_1,\dots,x_{\N-1}):= \Big( \frac{\delta}{\delta \phi_{0}(x_{0})} \dots \frac{\delta}{\delta \phi_{\N-1}(x_{\N-1})}\Big) \Gamma \Big|_{\Vec{\phi}=0}\,.
\end{equation}
Note that, with the exception of appendix~\ref{sec_gauss}, derivatives wrt Grassmannian variables are always intended to be left derivatives, see e.g.~\cite{ber}. The complete Fourier transformed $\N$-point vertex function then satisfies
\begin{equation}
(2\pi)^{4(\N-1)} \Big(\frac{\delta }{\delta \phi_{0}(-p_{0})} \dots \frac{\delta }{\delta \phi_{\N-1}(-p_{\N-1})}\Big) \Gamma \Big |_{\Vec{\phi}=0}=\delta(\sum \limits^{\N-1}_{i=0} p_i)\,\Gamma^{\Vec{\phi}}\,, \label{eq_f1}
\end{equation}
where $\Gamma^{\Vec{\phi}}$ stands for $\Gamma^{\Vec{\phi}}(p_1,...,p_{\N-1})$. The reduced and complete Fourier transforms with $n_\chi \geqslant 1$ composite operator insertions of sources $\Vec{\chi}=(\chi_0,...,\chi_{n_\chi-1})$ and $\Vec{\phi}:=(\phi_{n_\chi},...,\phi_{n-1})$ are correspondingly related by
\begin{equation}
(2\pi)^{4(n-1)} \Big(\prod \limits^{n-1}_{i=n_\chi} \frac{\delta }{\delta \phi_i(-p_i)}\,\prod \limits^{n_\chi-1}_{i=0}  \frac{\delta }{\delta \chi_i(-p_i)} \Big) \Gamma \Big|_{\substack{\Vec{\phi}=0\\\Vec{\chi}=0}}=\delta(\sum \limits^{n-1}_{i=0} p_i)\,\Gamma^{\Vec{\phi}}_{\Vec{\chi}}\,, \label{eq_f2}
\end{equation}
where the order of the derivatives $\delta/\delta \phi_i$ is the same as before, the derivatives $\delta/\delta \chi_i$ are ordered with left-to-right increasing indices, and $\Gamma^{\Vec{\phi}}_{\Vec{\chi}}$ stands for $\Gamma^{\Vec{\phi}}_{\Vec{\chi}}(p_1,...,p_{n-1})$.
Note that $n-1$ is the total number of arguments, e.g. for $\Gamma^{\Vec{\phi}}_{\Vec{\chi}}(p_1,...,p_{n-1})$ we have $n=n_\chi + \N$.

It will be useful to keep a bijective relation between momenta and field labels (including possible insertion labels), $p_i \leftrightarrow \phi_i$. Hence, we assume that $p_0$ is the negative subsum of all other momenta,
\begin{align}
\mathbb{P}_n&:=\{\Vec{p} \in \mathbb{R}^{4n}: \Vec{p}=(p_0,...,p_{n-1}),\; p_0=-\sum^{n-1}_{i=1} p_i\},& |\Vec{p}|^2&:=\sum \limits^{n-1}_{i=0} p^2_i,\label{eq_pn}
\end{align}
and, referring to \eqref{eq_f1}, \eqref{eq_f2}, we use the notation
\begin{align}
\Gamma^{\Vec{\phi}}(\Vec{p})&=\Gamma^{\Vec{\phi}}(p_1,...,p_{\N-1}),&\Vec{p}&=(p_0,...,p_{\N-1}) \in \mathbb{P}_{\N},\\
\Gamma^{\Vec{\phi}}_{\Vec{\chi}}(\Vec{p})&=\Gamma^{\Vec{\phi}}_{\Vec{\chi}}(p_1,...,p_{n-1}),&\Vec{p}&=(p_0,...,p_{n-1}) \in \mathbb{P}_{n}.
\end{align}

A momentum configuration $\vec{p}\in\mathbb{P}_n$ is said \textbf{exceptional} iff there exists a nonempty $S\subset[1:n-1]$ such that $\sum_{i\in S}p_i=0$. If this is not the case $\vec{p}$ is said \textbf{nonexceptional}.

We rely on the multi-index formalism for derivatives with respect to momenta. Taking in account that there are no derivatives wrt $p_0$, we set:
\begin{align}
\textbf{w} &:= (w_{0,1} , w_{0,2},  ... , w_{n-1,4} ),& w_{i,\mu } &\in \mathbb{N}, \quad w_{0,\mu}:=0,\\
w &:= (w_{0} , ... , w_{n-1} ),& w_i &:= \sum^{4}_{\mu=1 } w_{i,\mu },\\
\mathbb{W}_n&:=\{w\in \{0\} \times \mathbb{N}^{n-1}:\|w\|\leqslant w_{max} \},&\|w\|&:=\sum^{n-1}_{i=0} w_i,\label{eq_17w}\\
\partial ^{\,\textbf{w}} \Gamma& := \prod^{n-1}_{i=0} \prod_{\mu =1}^{4}\Bigl(\frac{\partial }{\partial p_{i,\mu }} \Bigr)^{w_{i,\mu }}\Gamma,&\partial^w\Gamma &:=  \Big(\prod^{n-1}_{i=0} \partial^{w_i}_{p_i}\Big)\Gamma,
\end{align}
where $\partial^{k}_{p}$ is the tensor with components $\partial_{p_{\mu_1}}...\partial_{p_{\mu_k}}$, and $w_{max}$ is an arbitrary integer $\geqslant4$ fixed throughout the paper. The following shorthands will be helpful: 
\begin{align}
\dot{\Gamma}^{\Lambda\Lambda_0}&:=\partial_\Lambda \Gamma^{\Lambda\Lambda_0},&\partial_{\phi_i} \Gamma^{\Vec{\phi}}&:=\frac{\partial \Gamma^{\Vec{\phi}}}{\partial p_i},&\Gamma^{;w}&:=\partial^w \Gamma,\label{eq_11ns}\\
\tilde{\delta}_{\phi(p)}&:=(2\pi)^4\frac{\delta}{\delta \phi(-p)}\label{eq_409t},&\check{\phi}(p)&:=\phi(-p),&\log_+x&:=\log \max(1,x).
\end{align}
\subsection{The Classical Action}
We start with a four-dimensional Euclidean classical field theory invariant under local SU(2) transformations and with a coupling $g \neq 0$. Given $G=e^{i g \alpha}$ with $\alpha=\alpha^a t_a$ an element of the Lie algebra, a gauge transformation reads:
\begin{align}
A_\mu & \mapsto  \frac{i}{g}\, G\partial_\mu G^{-1} + G A_\mu G^{-1},& \phi & \mapsto G \phi G^{-1},
\end{align}
and in infinitesimal form
\begin{align}
A_\mu &\mapsto A_\mu+\partial_\mu \alpha-ig[A_\mu,\alpha],& \phi & \mapsto \phi +ig[\alpha,\phi],
\end{align}
where we also show the transformation for a field $\phi$ in the adjoint representation of the Lie algebra. The quantization of gauge theories using auxiliary ghost fields is presented, for example, by Slavnov and Faddeev in their book \cite{slfd}. Following \cite{fad,thooft}, with an appropriate linear gauge-fixing functional the semiclassical Lagrangian density takes the form
\begin{align}
\mathcal{L}^{tot}_{0}&=\frac{1}{4}F^a_{\mu \nu} F^a_{\mu \nu} + \frac{1}{2\xi}(\partial_\mu A^a_\mu)^2 - \partial_\mu \bar{c}^a (D_\mu c)^a,\\
D_\mu c &=\partial_\mu c -ig[A_\mu,c],\\
F_{\mu \nu} &= \partial_\mu A_\nu-\partial_\nu A_\mu - ig[A_\mu,A_\nu],
\end{align}
where $\xi>0$ is the Feynman parameter and $\{F_{\mu \nu},A_\mu,c\}$ are elements of the su(2) algebra, e.g. $A_\mu=t^a A^a_\mu$. We will study the quantum theory in a framework which breaks local gauge invariance due to the presence of momentum space regulators. However, the Lagrangian respects global SU(2) symmetry, Euclidean isometries ISO(4) and has ghost number zero. We admit all counterterms compatible with these symmetries:
\begin{align}
\mathcal{L}_{ct}=&r^{0,\bar{c}c\bar{c}c} \bar{c}^bc^b \bar{c}^a c^a  + r^{0,\bar{c}cAA}_1 \bar{c}^bc^b A^a_\mu A^a_\mu  + r^{0,\bar{c}cAA}_2 \bar{c}^ac^b A^a_\mu A^b_\mu   \nonumber\\
&+ r^{0,AAAA}_1 A^b_\mu A^b_\nu A^a_\mu A^a_\nu  + r^{0,AAAA}_2 A^b_\nu A^b_\nu A^a_\mu A^a_\mu  + 2\epsilon_{abc} r^{0,AAA} (\partial_\mu A^a_\nu) A^b_\mu A^c_\nu \nonumber\\
&-r^{0,A\bar{c}c}_1 \epsilon_{abd} (\partial_\mu \bar{c}^a) A^b_\mu c^d -r^{0,A\bar{c}c}_2 \epsilon_{abd}  \bar{c}^a A^b_\mu \partial_\mu c^d   + \Sigma^{0,\bar{c}c} \bar{c}^a \partial^2 c^a \nonumber \\
&- \frac{1}{2}\Sigma^{0,AA}_T A^a_{\mu}(\partial^2 \delta_{\mu \nu} - \partial_\mu \partial_\nu)A^a_{\nu} + \frac{1}{2\xi}\Sigma^{0,AA}_L(\partial_\mu A^a_\mu)^2\nonumber\\
&+\frac{1}{2}\delta m^2_{AA} A^a_\mu A^a_\mu - \delta m^2_{\bar{c} c} \bar{c}^a c^a. \label{eq_ct}
\end{align}
There are eleven marginal counterterms and two strictly relevant counterterms. To denote the marginal counterterms we use the symbols~$r^0$ and~$\Sigma^0$. For the strictly relevant counterterms we use~$\delta m^2_{AA}$ and~$\delta m^2_{\bar{c} c}$.
\subsection{Generating functionals and Flow equations}\label{sec_gf}
In the next two sections we will introduce the essential
structural tools required for our proof of renormalizability of
nonabelian Yang--Mills theory. These tools are on
the one hand the differential Flow Equations of
the renormalization group, and on the other hand
the (violated) Slavnov--Taylor identities. They are
both obtained from the functional integral representation
of the theory.

Functional integrals are know to exist,
beyond perturbation theory, if suitably regularized \cite{gljf}.
To restrict their support to sufficiently regular
functions in position space one has to introduce an
ultraviolet (UV) cutoff. To avoid infrared problems one
has to introduce a corresponding infrared cutoff, the
most practical one being a finite volume cutoff, introduced
by putting the system on a torus with periodic boundary
conditions. For an analysis of the support properties of
(Gaussian) functional integrals see~\cite{ree}.

We do not make explicit the finite volume cutoff in
this paper, which would still blow up the manuscript
to some degree.  For some more details in this respect
see \cite{mul}.
On the other hand the UV cutoff is explicit.

Once the finite volume cutoff has been introduced,
the FEs (with UV regulator) can be shown to hold up to $\Lambda = 0$, $\Lambda$ being the flow
parameter, even beyond perturbation theory
(as far as the bosonic degrees of freedom are
concerned). One can then construct perturbative solutions
of the FEs, which are well-defined and regular
in the open domain of nonexceptional external momenta,
and which satisfy the Slavnov--Taylor identities, in the limits where the flow parameter is sent to zero
and the UV cutoff is sent to infinity.
If keeping the finite volume until the end,
one then has to show that these solutions have uniform
limits (in compact subsets of the open domain of
nonexceptional momenta) once the volume is sent to
infinity. Short-circuiting this reasoning, as we do here,
is tantamount to postulating from the beginning that the FEs and the (violated)
Slavnov-Taylor identities hold, for nonexceptional
momenta, also in the limit $\Lambda \to 0$.

By definition, the characteristic function of a finite measure $d \nu$ is its Fourier transform, see e.g. \cite{dlfm,gljf},
\begin{align}
\chi(K)&:=\int d\nu(\Phi)\,e^{ \frac{i}{\hbar}  K \cdot \Phi}\,. \label{def_cf}
\end{align}
We consider a Gaussian measure $d\nu_{\Lambda \Lambda_0}$ with characteristic function
\begin{align}
\chi^{\Lambda\Lambda_0}(j,b,\bar{\eta},\eta)&=e^{\frac{1}{\hbar}\langle \bar{\eta}, S^{\Lambda\Lambda_0} \eta \rangle - \frac{1}{2\hbar}\langle j, C^{\Lambda\Lambda_0} j \rangle -  \frac{1}{2 \hbar \xi} \langle b,b \rangle }, \label{def_gauss}
\end{align}
and we denote as usual
\begin{align}
(C^{\Lambda \Lambda_0}j)_\mu(x)&=\int d^4 y\, C^{\Lambda \Lambda_0}_{\mu \nu}(x,y)j_\nu(y),\\
C^{\Lambda \Lambda_0}_{\mu \nu}(x,y)&=\int \frac{d^4 p}{(2 \pi)^4} \, e^{ip(x-y)}C^{\Lambda \Lambda_0}_{\mu \nu}(p),
\end{align}
where in momentum space the regularized propagators are defined by the expressions
\begin{align}
C_{\mu \nu}(p)&:=\frac{1}{p^2}(\delta_{\mu \nu} + (\xi -1 )\frac{p_\mu p_\nu}{p^2}),&C^{\Lambda\Lambda_0}_{\mu \nu}(p)&:=C_{\mu \nu}(p)\,\sigma_{\Lambda\Lambda_0}(p^2)\,, \label{def_c}\\
S(p)&:=\frac{1}{p^2},&S^{\Lambda\Lambda_0}(p)&:=S(p)\,\sigma_{\Lambda\Lambda_0}(p^2)\,, \label{def_s}\\
\sigma_{\Lambda\Lambda_0}(s)&:=\sigma_{\Lambda_0}(s)-\sigma_{\Lambda}(s),&\sigma_{\lambda}(s)&:=\exp(-\frac{s^2}{\lambda^4})\,.
\end{align}
For shortness we will also write $C^{-1}_{\Lambda \Lambda_0}$ instead of $(C^{\Lambda \Lambda_0})^{-1}$. The parameters $\Lambda$, $\Lambda_0$, such that $0 < \Lambda \leqslant \Lambda_0$, are respectively IR and UV cutoffs.
\begin{definition} Let $d\mu_{\Lambda\Lambda_0}$ be the measure defined by
\begin{align}
d \mu_{\Lambda\Lambda_0}(A,B, c,\bar{c}) := d \nu_{\Lambda\Lambda_0}(A,B-i\frac{1}{\xi}\partial A,c,\bar{c}). \label{eq_dm}
\end{align}
\end{definition}
For $\Phi=(A,B,c,\bar{c})$ and $K=(j,b , \bar{\eta}, \eta)$ and an infinitesimal variation $\delta \Phi=(\delta A,0,\delta c,\delta \bar{c})$, using the properties of Gaussian measures from appendix \ref{sec_gauss}, we have
\begin{align}
d\mu_{\Lambda\Lambda_0}(\Phi+\delta \Phi)=&d\mu_{\Lambda\Lambda_0}(\Phi)\Big(1 + \frac{1}{\hbar}\langle \bar{c}, S^{-1}_{\Lambda\Lambda_0} \delta c \rangle + \frac{1}{\hbar}\langle \delta \bar{c}, S^{-1}_{\Lambda\Lambda_0} c \rangle \nonumber \\
&-\frac{1}{\hbar}\langle A, C^{-1}_{\Lambda\Lambda_0}\delta A\rangle  -\frac{1}{\hbar} \langle i\partial(B -i \frac{1}{\xi}\partial A), \delta A \rangle \Big)\,. \label{eq_dm_delta}
\end{align}
\begin{definition} \label{def_z} The free partition function $Z^{\Lambda\Lambda_0}_0$ is defined by
\begin{align}
Z^{\Lambda\Lambda_0}_0(K)&:=\int d \mu_{\Lambda\Lambda_0}(\Phi) e^{\frac{1}{\hbar} K \cdot \Phi}.
\end{align}
\end{definition}
It follows that
\begin{align}
Z^{\Lambda\Lambda_0}_0(j,b,\bar{\eta},\eta)&= \chi^{\Lambda\Lambda_0}(-i\,j-\frac{1}{\xi}\partial b,-i\,b,-i\,\bar{\eta},-i\,\eta),\label{eq_18z}\\
Z^{\Lambda \Lambda_0}_0(K)&=e^{\frac{1}{2\hbar}\langle K, \mathbf{C}^{\Lambda \Lambda_0} K \rangle},
\end{align}
where $\mathbf{C}^{\Lambda \Lambda_0}$ is a 7x7 matrix,
\begin{equation}
\mathbf{C}^{\Lambda \Lambda_0}:=\begin{pmatrix}C^{\Lambda \Lambda_0}_{\mu \nu},&S^{\Lambda \Lambda_0}p_\mu,&0,&0 \\-S^{\Lambda \Lambda_0}p_\nu,&\dfrac{1}{\xi}(1-\sigma_{\Lambda \Lambda_0}),&0,&0\\ 0,&0,&0,&-S^{\Lambda \Lambda_0}\\0,&0,&S^{\Lambda \Lambda_0},&0\end{pmatrix}.\label{eq_328s}
\end{equation}
For $\Lambda<\Lambda_0$, $\mathbf{C}^{\Lambda \Lambda_0}$ is invertible:
\begin{equation}
\mathbf{C}^{-1}_{\Lambda \Lambda_0}=\begin{pmatrix}C^{-1}_{\Lambda \Lambda_0; \mu \nu} - \dfrac{1}{\xi}p_\mu p_\nu,&-p_\mu,&0,&0\\p_\nu,&\xi,&0,&0\\0,&0,&0,&S^{-1}_{\Lambda \Lambda_0}\\0,&0,&-S^{-1}_{\Lambda \Lambda_0},&0\end{pmatrix} \label{eq_qform}.
\end{equation}
We write the bosonic part of $\mathbf{C}^{-1}_{\Lambda \Lambda_0}$ as $\mathbf{P}^T \mathbf{Q}^{\Lambda \Lambda_0} \mathbf{P}$, where $\mathbf{P}$ is a diagonal matrix with $\mathbf{P}_{\mu \mu}=|p|$ for each index $\mu \in \{1,...4\}$ and $\mathbf{P}_{55}=1$. The eigenvalues $q_\alpha$ of $\mathbf{Q}^{\Lambda \Lambda_0}$ are:
\begin{align*}
q_{1,2,3}&=\sigma^{-1}_{\Lambda \Lambda_0},&q_{4,5} &= \frac{\xi_{\Lambda \Lambda_0} \pm (\xi^2_{\Lambda \Lambda_0} -4\sigma^{-1}_{\Lambda \Lambda_0})^{\frac{1}{2}}}{2},&\xi_{\Lambda \Lambda_0}&:=\xi+\frac{1}{\xi}(\sigma^{-1}_{\Lambda \Lambda_0}-1).
\end{align*}
The fact that the real part of these  eigenvalues is positive is known to be a prerequisite for the definition of a Gaussian measure for the bosonic part of the theory.

A useful relation follows from \eqref{def_gauss}, \eqref{eq_18z}:
\begin{equation}
\int d \mu_{\Lambda \Lambda_0}(\Phi) \, B \, e^{\frac{1}{\hbar}K  \cdot \Phi }= \frac{1}{\xi} \int d \mu_{\Lambda \Lambda_0}(\Phi) \,(b + i\partial A ) \, e^{\frac{1}{\hbar}K  \cdot \Phi }\,. \label{eq_17shift}
\end{equation}
\begin{definition}The partition function $Z^{\Lambda\Lambda_0}$ of $SU(2)$ Yang--Mills field theory is given by
\begin{align}
Z^{\Lambda\Lambda_0}(K)& := \int d \mu_{\Lambda\Lambda_0}(\Phi)\,e^{- \frac{1}{\hbar}L^{\Lambda_0\Lambda_0}}\,e^{\frac{1}{\hbar} K \cdot \Phi},& L^{\Lambda_0\Lambda_0}&:=\int d^4x \, \mathcal{L}^{\Lambda_0\Lambda_0} \label{eq_z}.
\end{align}
The interaction Lagrangian density $\mathcal{L}^{\Lambda_0\Lambda_0}:=\mathcal{L}^{\Lambda_0\Lambda_0}_0 + \mathcal{L}^{\Lambda_0\Lambda_0}_{ct}$ is given by \eqref{eq_ct} and
\begin{equation}
\mathcal{L}^{\Lambda_0\Lambda_0}_0:=g\epsilon_{abc}(\partial_\mu A^a_\nu) A^b_\mu A^c_\nu + \frac{g^2}{4}\epsilon_{cab} \epsilon_{cds}A^a_\mu A^b_\nu A^d_\mu A^s_\nu - g\epsilon_{abc}(\partial_\mu\bar{c}^a) A^b_\mu c^c \,. \label{eq_l0}
\end{equation}
\end{definition}
Since we restrict to perturbation theory, all generating functionals are formal series in terms of $\hbar$ and of their source/field arguments. We also emphasize that $\mathcal{L}^{\Lambda_0\Lambda_0}_0$ does not depend on the $B$ field.\\
From the expression for $\mathbf{C}^{-1}_{0 \infty}$ and equation \eqref{eq_l0} one deduces the full tree level Lagrangian density
\begin{equation}
\mathcal{L}^{tot}_{0}=\frac{1}{4}F_{\mu \nu}F_{\mu \nu} + \frac{\xi}{2}B^2 -iB\partial_\mu A_\mu - \partial_\mu \bar{c} D_\mu c\;.
\end{equation}
The density $\mathcal{L}^{tot}_0$ is invariant under the infinitesimal BRST transformation \cite{brs}, \cite{tyutin}
\begin{align}
\delta^{BRS} A&= \epsilon\,Dc,&\delta^{BRS} c&= \epsilon\,\frac{1}{2} ig \{c , c\} ,& \delta^{BRS} \bar{c}&= \epsilon\, i B ,& \delta^{BRS}B&=0,
\end{align}
where $\epsilon$ is a Grassmann parameter, and $\{c,c\}^d=i\epsilon_{abd} c^a c^b $. Defining the classical operator $s$ by $\delta^{BRS} \Phi=\epsilon s\Phi$ one shows that $s$ is nilpotent.
\begin{definition} The generating functional of the Connected Schwinger (CS) functions is 
\begin{align}
e^{\frac{1}{\hbar} W^{\Lambda\Lambda_0}(K)}:=Z^{\Lambda\Lambda_0}(K).
\end{align}
\end{definition}
The derivation of the FE is usually given considering the generating functional $L^{\Lambda\Lambda_0}$ of the Connected Amputated Schwinger  (CAS) functions.
\begin{definition} The generating functional $L^{\Lambda\Lambda_0}$ of CAS functions is
\begin{equation}
e^{-\frac{1}{\hbar}L^{\Lambda\Lambda_0}(\Phi)} := \int d \mu_{\Lambda\Lambda_0}(\Phi^\prime)\,e^{- \frac{1}{\hbar}L^{\Lambda_0\Lambda_0}(\Phi^\prime + \Phi)}.
\end{equation}
\end{definition}
From definition \ref{def_z} we have for any polynomial $P(\Phi)$
\begin{equation}
\frac{d}{d \Lambda} \int d \mu_{\Lambda \Lambda_0}(\Phi) P(\Phi)= \hbar \int d\mu_{\Lambda \Lambda_0}(\Phi) \langle \frac{\delta}{\delta \Phi},\hat{\mathbf{1}} \dot{\mathbf{C}}^{\Lambda \Lambda_0} \frac{\delta}{\delta \Phi} \rangle P(\Phi), \label{eq_poly}
\end{equation}
where
\begin{equation}
(\hat{\mathbf{1}})_{\phi \phi^\prime} = \left\{\begin{matrix} -\delta_{\phi \phi^\prime} & \mbox{if }\phi, \phi^\prime \in \{c,\bar{c}\},\\ \delta_{\phi \phi^\prime} & otherwise\,.\end{matrix}\right.
\end{equation}
Using equation \eqref{eq_poly} one obtains the FE, see e.g. \cite{gljf}, \cite{kks}, 
\begin{equation}
\dot{L}^{\Lambda \Lambda_0}(\Phi)= \frac{\hbar}{2} \langle \frac{\delta}{\delta \Phi}, \hat{\mathbf{1}} \dot{\mathbf{C}}^{\Lambda\Lambda_0} \frac{\delta}{\delta \Phi} \rangle L^{\Lambda \Lambda_0} -\frac{1}{2} \langle \frac{\delta L^{\Lambda \Lambda_0}}{\delta \Phi}, \hat{\mathbf{1}} \dot{\mathbf{C}}^{\Lambda\Lambda_0} \frac{\delta L^{\Lambda \Lambda_0}}{\delta \Phi} \rangle\;.\label{eq_fll}
\end{equation}
From appendix \ref{sec_gauss} on Gaussian measures it follows that
\begin{equation}
d \mu_{\Lambda \Lambda_0}(\Phi - \mathbf{C}^{\Lambda \Lambda_0}\delta \Phi)=d \mu_{\Lambda \Lambda_0}(\Phi) e^{-\frac{1}{2 \hbar} \langle \delta \Phi, \mathbf{C}^{\Lambda \Lambda_0}\delta \Phi\rangle}e^{\frac{1}{\hbar} \langle \delta \Phi, \Phi\rangle}.
\end{equation}
This gives the relation between the generating functionals $W^{\Lambda \Lambda_0}$ and $L^{\Lambda \Lambda_0}$
\begin{equation}
W^{\Lambda \Lambda_0}(K)=\frac{1}{2} \langle K, \mathbf{C}^{\Lambda \Lambda_0} K \rangle - L^{\Lambda \Lambda_0}(\hat{\mathbf{1}}_{\bar{c}} \mathbf{C}^{\Lambda \Lambda_0}K),
\end{equation}
where
\begin{align}
(\hat{\mathbf{1}}_{\bar{c}})_{\phi \phi^\prime} &= \left\{\begin{matrix} -1 & \mbox{if }\phi=\phi^\prime=\bar{c},\\ \delta_{\phi \phi^\prime} & otherwise,\end{matrix}\right.&(\hat{\mathbf{1}}_c)_{\phi \phi^\prime} &= \left\{\begin{matrix} -1 & \mbox{if }\phi=\phi^\prime=c,\\ \delta_{\phi \phi^\prime} & otherwise.\end{matrix}\right.
\end{align}
\begin{definition}[Legendre transform]\label{def_gamma}For $0<\Lambda<\Lambda_0$, let $K^{\Lambda \Lambda_0}(\ubar{\Phi})$ be a solution of the system of equations
\begin{align}
\ubar{A}-\frac{\delta W^{\Lambda \Lambda_0}}{\delta j}\Big|_{K^{\Lambda \Lambda_0}(\ubar{\Phi})}&=0,&\ubar{B}-\frac{\delta W^{\Lambda \Lambda_0}}{\delta b}\Big|_{K^{\Lambda \Lambda_0}(\ubar{\Phi})}&=0,\label{627d}\\
\ubar{c}-\frac{\delta W^{\Lambda \Lambda_0}}{\delta \bar{\eta}}\Big|_{K^{\Lambda \Lambda_0}(\ubar{\Phi})}&=0,&\wbar{c}+\frac{\delta W^{\Lambda \Lambda_0}}{\delta \eta}\Big|_{K^{\Lambda \Lambda_0}(\ubar{\Phi})}&=0.\label{627f}
\end{align}
The effective action is
\begin{equation}
\Gamma^{\Lambda \Lambda_0}(\ubar{\Phi}):=K^{\Lambda \Lambda_0}(\ubar{\Phi}) \cdot \ubar{\Phi} - W^{\Lambda \Lambda_0}(K^{\Lambda \Lambda_0}(\ubar{\Phi}))\,.\label{eq_10gamma}
\end{equation}
\end{definition}
A solution of the system of equations \eqref{627d}, \eqref{627f} always exists as a formal series in $\hbar$ and in the fields. Same statement for the system \eqref{eq_phi}.
\begin{definition} \label{def_gamme} For $0<\Lambda\leqslant \Lambda_0$, let $\Phi^{\Lambda \Lambda_0}(\ubar{\Phi})$ be a solution of the equation
\begin{equation}
\ubar{\Phi}=\left(\Phi - \mathbf{C}^{\Lambda \Lambda_0} \frac{\delta L^{\Lambda \Lambda_0}}{\delta \Phi}\right)\Big|_{\Phi^{\Lambda\Lambda_0}(\ubar{\Phi})}. \label{eq_phi}
\end{equation}
The reduced effective action is
\begin{equation}
\Gamme^{\Lambda \Lambda_0}(\ubar{\Phi}):= \Big(L^{\Lambda \Lambda_0}(\Phi)  - \frac{1}{2} \langle \frac{\delta L^{\Lambda \Lambda_0}}{\delta \Phi},\hat{\mathbf{1}} \mathbf{C}^{\Lambda \Lambda_0} \frac{\delta L^{\Lambda \Lambda_0}}{\delta \Phi}\rangle \Big) \Big|_{\Phi^{\Lambda\Lambda_0}(\ubar{\Phi})}.\label{eq_17g}
\end{equation}
\end{definition}
From definitions \ref{def_gamma} and \ref{def_gamme} it follows that for $0<\Lambda< \Lambda_0$
\begin{align}
\Gamme^{\Lambda\Lambda_0}(\ubar{\Phi})=&\Gamma^{\Lambda\Lambda_0}(\ubar{\Phi}) -\frac{1}{2}\langle \ubar{\Phi}, \mathbf{C}^{-1}_{\Lambda\Lambda_0} \ubar{\Phi} \rangle,\label{eq_17g2}\\
\Phi^{\Lambda\Lambda_0}(\ubar{\Phi})=&\hat{\mathbf{1}}_{\bar{c}} \mathbf{C}^{\Lambda \Lambda_0}K^{\Lambda\Lambda_0}(\ubar{\Phi}),\\
\Gamme^{\Lambda \Lambda_0}(\ubar{\Phi})=& \Big(L^{\Lambda \Lambda_0}(\Phi)  - \frac{1}{2} \langle (\ubar{\Phi}-\Phi),\mathbf{C}^{-1}_{\Lambda \Lambda_0} (\ubar{\Phi}-\Phi)\rangle \Big) \Big|_{\Phi^{\Lambda\Lambda_0}(\ubar{\Phi})}. \label{eq_gamme}
\end{align}
Using equation \eqref{eq_phi} we see that, before the replacement, the rhs of \eqref{eq_gamme}, as a function of $\Phi$ for fixed $\ubar{\Phi}$, has an extremum at $\Phi=\Phi^{\Lambda \Lambda_0}(\ubar{\Phi})$. Applying $\partial_\Lambda$ to equation \eqref{eq_gamme}, substituting $\dot{L}$ with the rhs of \eqref{eq_fll} and using the property of extremum we obtain
\begin{equation}
\dot{\Gamme}^{\Lambda \Lambda_0}(\ubar{\Phi})=  \frac{\hbar}{2} \langle \frac{\delta}{\delta \Phi}, \hat{\mathbf{1}} \dot{\mathbf{C}}^{\Lambda \Lambda_0}  \frac{\delta}{\delta \Phi} \rangle L^{\Lambda \Lambda_0} \Big|_{\Phi^{\Lambda\Lambda_0}(\ubar{\Phi})}\,.
\end{equation}
Defining
\begin{align}
W^{\Lambda \Lambda_0}_{q,-k}&:=(2\pi)^4 \frac{\delta^2 W^{\Lambda \Lambda_0}}{\delta K(-q) \delta K(k)}\,,&\Gamma^{\Lambda \Lambda_0}_{q,-k}&:=(2\pi)^4 \frac{\delta^2 \Gamma^{\Lambda \Lambda_0}}{\delta \ubar{\Phi}(-q) \delta \ubar{\Phi}(k)}\,,
\end{align}
it is easy to see that
\begin{equation}
\int d^4 k \, W^{\Lambda \Lambda_0}_{q,-k} \, \hat{\mathbf{1}}_{\bar{c}}\, \Gamma^{\Lambda \Lambda_0}_{k,-p} \, \hat{\mathbf{1}}_c=\delta(q-p)\,.
\end{equation}
This implies that, using similar notations,
\begin{equation}
L^{\Lambda \Lambda_0}_{q,-p} = \int d^4 k \, \Gamme^{\Lambda \Lambda_0}_{q,-k}  \left(\delta(k-p)+\hat{\mathbf{1}} \mathbf{C}^{\Lambda \Lambda_0}(p) \Gamme^{\Lambda \Lambda_0}_{-p,k}\right)^{-1}\,.
\end{equation}
Eventually, one obtains the FE for the functional $\Gamme$ and also for the functional $\Gamme_\chi$ with one operator insertion of source $\chi$
\begin{align}
\dot{\Gamme}&=\frac{\hbar}{2} \langle \dot{\mathbf{C}} \, \delta_\phi\delta_{\bar{\phi}}\Gamme ,\left(1+\hat{\mathbf{1}}\mathbf{C}\, \delta_{\bar{\phi}} \delta_\phi \Gamme \right)^{-1} \rangle\,,\label{eq_fl}\\
\delta_\chi\dot{\Gamme}&=\frac{\hbar}{2} \langle \dot{\mathbf{C}}\left(1+ \delta_{\bar{\phi}} \delta_\phi\Gamme\, \mathbf{C}\hat{\mathbf{1}} \right)^{-1}  \delta_\chi \delta_{\bar{\phi}} \delta_\phi \Gamme , \left(1+\hat{\mathbf{1}}\mathbf{C} \, \delta_{\bar{\phi}} \delta_\phi \Gamme \right)^{-1} \rangle\,, \label{eq_fl1}
\end{align}
where  $\phi$, $\bar{\phi} \in \{\ubar{A},\ubar{B}, \ubar{c},\wbar{c}\}$ and we omit appropriate sums over field labels. Generalization to $\Vec{\chi}=(\chi_1,...,\chi_{n_\chi})$ with $n_{\chi} > 1$ is straightforward. The FE for $\Gamme$ in modern form has been introduced in \cite{Wet2,bam2,mor,kks2,komu}. Flow equations with composite operator insertions have been introduced in~\cite{keko}.

The mass dimension of a vertex function $\Gamma^{\Vec{\phi};w}_{\Vec{\chi}}(\Vec{p})$ with $\N$ fields $\phi_i \in \{A,B,c,\bar{c}\}$, $n_{\chi}$ insertions of sources $\chi_i$ and $\|w\|$ momentum derivatives is $d:=4-\sum^{n_\chi}_i [\chi_i]- \sum^{\N}_i [\phi_i] - \|w\|$, where $[F]$ stands for the mass dimension of $F$ in position space. We say that such a term is \textbf{irrelevant} if $d<0$, as for example  $\Gamma^{AAc;w}_\gamma$ for $[\gamma]=2$, and \textbf{relevant} otherwise. Furthermore, we call a relevant term \textbf{marginal} if $d=0$, as for example $\Gamma^{AAAA}$, or \textbf{strictly relevant} if $d>0$.\\
Expanding in formal power series in $\hbar$ we have
\begin{equation}
\Gamme^{\Lambda \Lambda_0}(\ubar{\Phi})=\sum^{\infty}_{l=0} \hbar^l \Gamme^{\Lambda \Lambda_0}_l(\ubar{\Phi})\,.
\end{equation}
We also note that the FE \eqref{eq_fl} and \eqref{eq_fl1} admit an inductive structure in the loop number. This property allows us to prove statements by induction, first establishing them at tree-level, then proving that if they hold up to loop order $l-1\geqslant 0$ they are also valid at order $l$.\\
The proposition that follows proves that vertex functions
$\Gamme$ involving $B$ fields do vanish.
We use the notation $B\Vec{\phi}$ to denote sequences
of field labels with $\phi_i\in\{A,B,c,\bar{c}\}$.
\begin{proposition}
Assume vanishing renormalization conditions
for all relevant terms with at least one $B$ field:
\begin{align}
\Gamme^{B\Vec{\phi};0\Lambda_0;w}_l(\Vec{q})&=0\,, \label{eq_16cdt3}
\end{align}
where $\Vec{q}$ is nonexceptional for
marginal terms and vanishing otherwise.
Then,
for all $B\Vec{\phi}$,
$l$, $w$, $\Vec{p}$, and $0<\Lambda\leqslant \Lambda_0$,
\begin{equation}
\Gamme^{B\Vec{\phi};\Lambda\Lambda_0;w}_l(\Vec{p}) = 0\,.
\end{equation}
\end{proposition}
Rank-2 marginal terms are tensors with two vector indices, say $\mu$ and $\nu$, and can be decomposed in the basis $\{ \delta_{\mu\nu}, q_{i\mu} q_{j\nu} \}$, where $q_i,q_j$ range over all momenta in $\vec{q}$ other than $q_0$. For rank-2 marginal terms only the coefficient of $\delta_{\mu\nu}$ is set to zero in \eqref{eq_16cdt3}.

\paragraph{Proof}
We prove the statement by induction, increasing in the loop order, $l-1\mapsto l$.
Given $l$, we proceed by descending from $w_{max}$ in the number of derivatives, $\|w\| \mapsto \|w\|-1$.
For fixed $l$ and $w$, all possible terms
$\Gamme^{B\Vec{\phi};w}_l$ are considered.
By construction, for fixed $l$ and $B\Vec{\phi}$, the inductive scheme deals first with the irrelevant terms and continues, if they exist, with the marginal terms, followed by more and more relevant terms.
The identity $\Gamme^{B\Vec{\phi};\Lambda\Lambda_0;w}_0(\Vec{p})=0$ follows from the definition of $\Gamme$. Assume that the statement of the theorem holds up to loop order $l-1\geqslant 0$. It follows that at order $l$ the rhs of the FE for vertex functions with $B$ fields vanishes. Using the FE we integrate the irrelevant terms from $\Lambda_0$ downward to arbitrary $\Lambda>0$. The boundary conditions $\Gamme^{B\Vec{\phi};\Lambda_0\Lambda_0;w}_l(\Vec{p})=0$ and the vanishing of the rhs of the FE imply that all irrelevant terms with $B$ fields vanish at loop order $l$.
The boundary conditions $\Gamme^{B\Vec{\phi}_r;0\Lambda_0;w}_l(\Vec{q})=0$ and the vanishing of the rhs of the FE imply that all marginal terms vanish at their renormalization point for arbitrary $\Lambda>0$. The derivatives wrt momenta of marginal terms are irrelevant terms. Consequently, we conclude that the marginal $\Gamme^{B\Vec{\phi};\Lambda \Lambda_0;w}_l(\Vec{p})$ vanish for all $\Vec{p}$. Similar arguments hold for all strictly relevant terms.\hfill$\blacksquare$\\
In the following we will always adopt the renormalization conditions \eqref{eq_16cdt3}. Consequently, counterterms involving $B$ fields are not generated.

Let us denote by $\tilde{W}$, $\tilde{Z}$ the functional $W$, $Z$ with $b$ set to zero.
\begin{align}
\tilde{W}(j,\bar{\eta},\eta)&:=W(j,0,\bar{\eta},\eta),&\tilde{Z}(j,\bar{\eta},\eta)&:=Z(j,0,\bar{\eta},\eta)\,. \label{eq_tilda}
\end{align}
The covariance matrix $\tilde{\mathbf{C}}$ is obtained from $\mathbf{C}$\eqref{eq_328s} by removing the column and row which correspond to $b$.
\begin{proposition}
\begin{equation}
W^{\Lambda\Lambda_0}(j,b,\bar{\eta},\eta)=\frac{1}{2\xi}\langle b,b \rangle +  \tilde{W}^{\Lambda\Lambda_0}(j-i\frac{1}{\xi}\partial b,\bar{\eta},\eta)\,. \label{eq_w}
\end{equation}
\end{proposition}
\paragraph{Proof}Using the definition of the partition function $Z^{\Lambda\Lambda_0}$ one computes
\begin{align}
Z^{\Lambda\Lambda_0}(j+i\frac{1}{\xi}\partial b,b,\bar{\eta},\eta)&=e^{- \frac{1}{\hbar}\mathbb{L}^{\Lambda_0\Lambda_0}}\chi^{\Lambda\Lambda_0}(ij,ib,i\bar{\eta},i\eta),\\
\mathbb{L}^{\Lambda_0\Lambda_0}&:=L^{\Lambda_0\Lambda_0}(\frac{\delta}{\delta j},\frac{\delta}{\delta \bar{\eta}},\frac{\delta}{\delta \eta})\,.
\end{align}
From definition \eqref{def_gauss} it follows
\begin{align}
Z^{\Lambda\Lambda_0}(j+i\frac{1}{\xi}\partial b,b,\bar{\eta},\eta)&=e^{\frac{1}{2 \hbar \xi} \langle b,b \rangle }e^{- \frac{1}{\hbar}\mathbb{L}^{\Lambda_0\Lambda_0}}e^{-\frac{1}{\hbar}\langle \bar{\eta}, S^{\Lambda\Lambda_0} \eta \rangle + \frac{1}{2\hbar}\langle j, C^{\Lambda\Lambda_0} j \rangle}\,.
\end{align}
Observing that the expression multiplying $e^{\frac{1}{2 \hbar \xi} \langle b,b \rangle }$ is $\tilde{Z}^{\Lambda\Lambda_0}(j,\bar{\eta},\eta)$ we obtain
\begin{align}
Z^{\Lambda\Lambda_0}(j,b,\bar{\eta},\eta)&=e^{\frac{1}{2 \hbar \xi} \langle b,b \rangle }\tilde{Z}^{\Lambda\Lambda_0}(j-i\frac{1}{\xi}\partial b,\bar{\eta},\eta)\,.
\end{align}
Taking the logarithm finishes the proof. \hfill$\blacksquare$\\
Substitution of $W^{\Lambda\Lambda_0}$ (\ref{eq_w}) into the definition of $\Gamma^{\Lambda\Lambda_0}$ \eqref{eq_10gamma} and integration by parts give
\begin{align}
\Gamma^{\Lambda\Lambda_0}(\ubar{A},\ubar{B},\wbar{c},\ubar{c})&=\langle b, \ubar{B}-i\frac{1}{\xi}\partial \ubar{A} - \frac{1}{2} b \rangle + \tilde{\Gamma}^{\Lambda\Lambda_0}(\ubar{A},\wbar{c},\ubar{c}),\\
\tilde{\Gamma}^{\Lambda\Lambda_0}(\ubar{A},\wbar{c},\ubar{c})&:=\langle j, \ubar{A} \rangle + \langle \bar{\eta}, \ubar{c} \rangle + \langle \wbar{c}, \eta \rangle - \tilde{W}^{\Lambda\Lambda_0}(j,\bar{\eta},\eta)\Big|_{K^{\Lambda \Lambda_0}(\ubar{\Phi})}\,. \label{eq_legendre_w}
\end{align}
From definition \ref{def_gamma} it follows that $b=\xi \ubar{B} -i \partial \ubar{A}$. Consequently, the above expression becomes 
\begin{align}
\Gamma^{\Lambda\Lambda_0}(\ubar{A},\ubar{B},\wbar{c},\ubar{c})&=\frac{1}{2\xi} \langle \xi\ubar{B}-i\partial \ubar{A}, \xi\ubar{B}-i\partial \ubar{A}\rangle + \tilde{\Gamma}^{\Lambda\Lambda_0}(\ubar{A},\wbar{c},\ubar{c})\,.\label{eq_23gamma}
\end{align}
Differentiation wrt $\ubar{A}$ yields an important identity:
\begin{corollary}
\begin{align}
\frac{\delta \Gamma^{\Lambda\Lambda_0}}{\delta \ubar{A}^a_\mu}&=i\partial_\mu(\ubar{B}^a-i\frac{1}{\xi}\partial \ubar{A}^a)+\frac{\delta \tilde{\Gamma}^{\Lambda\Lambda_0}}{\delta \ubar{A}^a_\mu}\,. \label{eq_partial_A}
\end{align}
\end{corollary}
For $\tilde{\Phi}:=(A^a_\mu,c^a,\bar{c}^a)$ the functional $\tilde{\Gamme}^{\Lambda \Lambda_0}(\tilde{\ubar{\Phi}})$ is defined from $L^{\Lambda \Lambda_0}(\tilde{\Phi})$ in analogy with~\eqref{eq_17g}. For ${0<\Lambda<\Lambda_0}$, it follows that
\begin{equation}
\tilde{\Gamma}^{\Lambda \Lambda_0}(\tilde{\ubar{\Phi}})=\tilde{\Gamme}^{\Lambda \Lambda_0}(\tilde{\ubar{\Phi}}) + \frac{1}{2}\langle \tilde{\ubar{\Phi}}, \tilde{\mathbf{C}}^{-1}_{\Lambda \Lambda_0}\, \tilde{\ubar{\Phi}} \rangle.\label{eq_17g3}
\end{equation}
Substitution \eqref{eq_17g2} and \eqref{eq_17g3} into \eqref{eq_23gamma} yields $\Gamme^{\Lambda \Lambda_0}(\ubar{\Phi})=\tilde{\Gamme}^{\Lambda \Lambda_0}(\tilde{\ubar{\Phi}})$. Note also that at $\Lambda=\Lambda_0$ we have $\ubar{\Phi}=\Phi$ and 
\begin{flalign}
&\Gamme^{\Lambda_0 \Lambda_0}(\ubar{\Phi})=L^{\Lambda_0 \Lambda_0}(\Phi)=L^{\Lambda_0 \Lambda_0}(\tilde{\Phi})=\tilde{\Gamme}^{\Lambda_0 \Lambda_0}(\tilde{\ubar{\Phi}})=\int d^4 x \, (\mathcal{L}^{\Lambda_0 \Lambda_0}_0 + \mathcal{L}^{\Lambda_0 \Lambda_0}_{ct})\,.\mspace{-27mu}&\label{eq_27j}
\end{flalign}
\subsection{Violated Slavnov--Taylor identities}\label{sec_vsti}
We are working in a framework where gauge invariance is broken already in the classical lagrangian due to  the gauge fixing term. It has then been realized that invariance of the lagrangian under the BRST transformations ensures the gauge invariance of  physical quantities to be calculated from the theory \cite{niel,pig}. On the level of correlation functions (Green’s functions in the relativistic theory) this invariance leads to a system of identities between different correlation functions which are called Slavnov Taylor identities (STI)  \cite{st,st2,taylor}. These identities may  be used to argue that physical quantities obtained from these functions, as for example the pole of the propagators for all physical fields of the Standard Model, are gauge invariant \cite{gg}.   

In our framework  gauge invariance is also violated in an even more serious way by the presence of the regulators in \eqref{def_c}, \eqref{def_s}. We want to show that for a suitable class of renormalization conditions, which does not restrict the freedom in fixing the physical coupling constant and the normalization of the fields, gauge invariance can be recovered in the renormalized theory. This means we want to show that the STI hold  once we take the limits $\Lambda \to 0$ and $\Lambda_0 \to \infty$. 

The first step is then to write a system of violated STI  suitable  for our subsequent analysis of their restoration. To do so we thus analyse the behavior of the regularized generating functionals of the correlation functions under BRST transformations. The infinitesimal BRST transformations can be generated by composite operator insertions for which we also have a freedom of normalization, as encoded by the constants $R_i$ introduced below \cite{komu}.

We derive the \textit{violated} STI setting $\Lambda=0$. Remember the comments in the beginning of section~\ref{sec_gf} as regards the implications of this choice for the generating functionals on the IR side. We consider the functional~$Z^{0\Lambda_0}_{vst}$ defined with the modified Lagrangian density
\begin{align}
\mathcal{L}^{\Lambda_0\Lambda_0}_{vst}&:=\mathcal{L}^{\Lambda_0\Lambda_0}  +  \gamma \psi^{\Lambda_0}  +  \omega \Omega^{\Lambda_0},&L^{\Lambda_0\Lambda_0}_{vst}&:=\int d^4 x \, \mathcal{L}^{\Lambda_0\Lambda_0}_{vst},\label{eq_17vst}
\end{align}
where $\gamma$, $\omega$ are external sources, and
\begin{align}
\psi^{\Lambda_0}&:=R^{\Lambda_0}_1 \partial c -igR^{\Lambda_0}_2[A,c],&\Omega^{\Lambda_0}&:=\frac{1}{2i}gR^{\Lambda_0}_3\{c,c\},&R^{\Lambda_0}_i&=1+O(\hbar)\,. \label{eq_21brs}
\end{align}
The requirement that at zero loop order $\psi^{\Lambda_0}$ and $\Omega^{\Lambda_0}$ correspond to the classical BRST variation implies that the constants $R^{\Lambda_0}_i$ are equal to one at tree level.  At higher orders we admit counterterms for the BRST transformation~\cite{komu}. Performing the change of variables $\Phi \mapsto \Phi + \delta_\epsilon \Phi$ we obtain the identity
\begin{align}
-\hbar \int \delta_\epsilon (d\mu_{0\Lambda_0}e^{- \frac{1}{\hbar} L^{\Lambda_0\Lambda_0}_{vst}})\,e^{\frac{1}{\hbar}K \cdot \Phi }=\int d\mu_{0\Lambda_0}e^{- \frac{1}{\hbar} L^{\Lambda_0\Lambda_0}_{vst} + \frac{1}{\hbar}K \cdot \Phi } K\cdot \delta_\epsilon (\Phi),\label{eq_17delta}
\end{align}
where 
\begin{align}
\delta_\epsilon A&:=\epsilon \, \sigma_{0\Lambda_0} * \psi^{\Lambda_0},&\delta_\epsilon c&:=-\epsilon \, \sigma_{0\Lambda_0} * \Omega^{\Lambda_0},&\delta_\epsilon \bar{c}&:=\epsilon \, \sigma_{0\Lambda_0} * iB\,\label{eq_20brs}.
\end{align}
Using formula \eqref{eq_dm_delta} for an infinitesimal change of variables and substituting the variation $\delta_\epsilon \Phi$ with its explicit form \eqref{eq_20brs} we have
\begin{align}
-\hbar\, \delta_\epsilon(d\mu_{0 \Lambda_0}(\Phi))=&d\mu_{0 \Lambda_0}(\Phi)\,\epsilon \, I^{\Lambda_0}_1(\Phi)\,,\\
-\hbar\, \delta_\epsilon(e^{-\frac{1}{\hbar}L^{\Lambda_0 \Lambda_0}_{vst}(\Phi)} )=&e^{-\frac{1}{\hbar}L^{\Lambda_0 \Lambda_0}_{vst}(\Phi)} \epsilon \, I^{\Lambda_0}_2(\Phi)\,,
\end{align}
where
\begin{align}
I^{\Lambda_0}_1(\Phi):=&\langle A,C^{-1} \psi^{\Lambda_0}\rangle-\langle \bar{c}, S^{-1} \Omega^{\Lambda_0}\rangle - i \langle B, S^{-1} c\rangle\nonumber\\
&+i\langle\partial(B-i\frac{1}{\xi}\partial A), \sigma_{0\Lambda_0} * \psi^{\Lambda_0} \rangle\,,\\
I^{\Lambda_0}_2(\Phi):=&\langle \sigma_{0\Lambda_0} *  \frac{\delta L^{\Lambda_0 \Lambda_0}}{\delta A}, \psi^{\Lambda_0} \rangle - \langle \sigma_{0\Lambda_0} *  \frac{\delta L^{\Lambda_0 \Lambda_0}}{\delta c}, \Omega^{\Lambda_0} \rangle\nonumber\\
&+i\langle B,\sigma_{0\Lambda_0} *  \frac{\delta L^{\Lambda_0 \Lambda_0}}{\delta \bar{c}}\rangle + \langle \gamma, Q^{\Lambda_0}_{\rho \gamma} \rangle + \langle \omega, Q^{\Lambda_0}_{\rho\omega} \rangle \,,\\
Q^{\Lambda_0}_{\rho \gamma^a_\mu}:=& g(R^{\Lambda_0}_3-R^{\Lambda_0}_2)\epsilon^{abs} (\sigma_{0\Lambda_0}*\psi^{\Lambda_0})^b_\mu c^s \nonumber\\
&+gR^{\Lambda_0}_1R^{\Lambda_0}_3\epsilon^{abs}( c^s (\sigma_{0\Lambda_0}*\partial_\mu c^b) - \sigma_{0\Lambda_0}* (c^s \partial_\mu c^b)) \nonumber\\
&+g^2R^{\Lambda_0}_2R^{\Lambda_0}_3\epsilon^{ase}\epsilon^{ebt}((\sigma_{0\Lambda_0}* (A^b_\mu  c^t)) c^s - A^b_\mu (\sigma_{0\Lambda_0}* (c^t c^s)))\,,\label{eq_11rg}\\
Q^{\Lambda_0}_{\rho\omega^a}:=& (gR^{\Lambda_0}_3)^2 c^s (\sigma_{0\Lambda_0} * (c^a c^s))\,.\label{eq_11rw}
\end{align}
The terms $Q^{\Lambda_0}_{\rho \gamma}$ and $Q^{\Lambda_0}_{\rho\omega}$ originate from the variations $\delta_\epsilon(\gamma \psi^{\Lambda_0})$, $\delta_\epsilon(\omega \Omega^{\Lambda_0})$, see~\eqref{eq_17vst}. Substituting  $B \mapsto \frac{1}{\xi}(b + i \partial A)$, see \eqref{eq_17shift}, in both terms $I^{\Lambda_0}_i$ we get
\begin{equation}
\int d\mu_{0 \Lambda_0}\, I^{\Lambda_0}_i \, e^{-\frac{1}{\hbar}L^{\Lambda_0 \Lambda_0}_{vst} + \frac{1}{\hbar}K \cdot \Phi}  = \int d\mu_{0 \Lambda_0}\, J^{\Lambda_0}_i\, e^{-\frac{1}{\hbar}L^{\Lambda_0 \Lambda_0}_{vst} + \frac{1}{\hbar}K \cdot \Phi},
\end{equation}
where
\begin{align}
J^{\Lambda_0}_1(\Phi):=&\langle A,C^{-1} \psi^{\Lambda_0}\rangle-\langle \bar{c}, S^{-1} \Omega^{\Lambda_0}\rangle + \frac{1}{\xi} \langle \partial A, S^{-1} c\rangle\nonumber\\
&-i\frac{1}{\xi}\langle b,S^{-1} c +  \sigma_{0\Lambda_0} * \partial \psi^{\Lambda_0} \rangle\,,\\
J^{\Lambda_0}_2(\Phi):=&\langle \sigma_{0\Lambda_0} * \frac{\delta L^{\Lambda_0 \Lambda_0}}{\delta A},\psi^{\Lambda_0}\rangle - \langle \sigma_{0\Lambda_0} * \frac{\delta L^{\Lambda_0 \Lambda_0}}{\delta c},\Omega^{\Lambda_0}\rangle\nonumber\\
&+i\frac{1}{\xi}\langle b + i\partial A, \sigma_{0\Lambda_0} * \frac{\delta L^{\Lambda_0 \Lambda_0}}{\delta \bar{c}}\rangle + \langle \gamma, Q^{\Lambda_0}_{\rho \gamma} \rangle + \langle \omega, Q^{\Lambda_0}_{\rho\omega} \rangle\,.
\end{align}
Introducing the operators $Q^{\Lambda_0}_\rho$ and $Q^{\Lambda_0}_{\beta}$,
\begin{align}
Q^{\Lambda_0}_\rho:=&\frac{\delta L^{\Lambda_0\Lambda_0}}{\delta A_\mu}\sigma_{0\Lambda_0}*\psi^{\Lambda_0}_\mu - \frac{\delta L^{\Lambda_0\Lambda_0}}{\delta c}\sigma_{0\Lambda_0}*\Omega^{\Lambda_0} - \frac{1}{\xi} \frac{\delta L^{\Lambda_0 \Lambda_0}}{\delta \bar{c}} \sigma_{0\Lambda_0}* \partial A \nonumber\\
&+A C^{-1} \psi^{\Lambda_0} -  \bar{c} S^{-1}\Omega^{\Lambda_0} + \frac{1}{\xi} \partial A S^{-1} c\,,\label{eq_328a}\\
Q^{\Lambda_0}_{\beta}:=&\sigma_{0\Lambda_0}* \Big(\frac{\delta L^{\Lambda_0\Lambda_0}}{\delta \bar{c}} - \partial \psi^{\Lambda_0}\Big) - S^{-1}c\,,\label{eq_10a}
\end{align}
we have
\begin{equation}
J^{\Lambda_0}_1 + J^{\Lambda_0}_2= i \frac{1}{\xi}\langle b, Q^{\Lambda_0}_{\beta} \rangle + \langle \gamma, Q^{\Lambda_0}_{\rho \gamma} \rangle + \langle \omega, Q^{\Lambda_0}_{\rho\omega} \rangle + \int d^4x \, Q^{\Lambda_0}_\rho(x) \,.
\end{equation}
We may now express the lhs of~\eqref{eq_17delta} as
\begin{equation}
\epsilon \int d \mu_{0 \Lambda_0} (J^{\Lambda_0}_1 + J^{\Lambda_0}_2) \, e^{-\frac{1}{\hbar}L^{\Lambda_0 \Lambda_0}_{vst} + \frac{1}{\hbar} K \cdot \Phi} =-\epsilon \, \mathcal{D} Z^{0 \Lambda_0}_{aux}\Big|_{\rho,\beta=0},\label{eq_20l}
\end{equation}
where
\begin{align}
\mathcal{D}:=&\int d^4x \, \hbar \frac{\delta}{\delta \rho(x)} \,+ i\frac{1}{\xi}\langle b, \hbar\frac{\delta}{\delta \beta} \rangle,\label{eq_20d}\\
Z^{0\Lambda_0}_{aux}:=&\int d\mu_{0\Lambda_0}(\Phi)\,e^{- \frac{1}{\hbar} L^{\Lambda_0\Lambda_0}_{aux}+\frac{1}{\hbar}K \cdot \Phi },\\
\mathcal{L}^{\Lambda_0\Lambda_0}_{aux}:=&\mathcal{L}^{\Lambda_0\Lambda_0}_{vst} +\rho Q^{\Lambda_0}_\rho + \rho \gamma^a_\mu Q^{\Lambda_0}_{\rho \gamma^a_\mu} + \rho \omega^a Q^{\Lambda_0}_{\rho \omega^a}+ \beta Q^{\Lambda_0}_{\beta}\,.\label{eq_17aux}
\end{align}
Defining 
\begin{align}
\mathcal{S}:=&\langle j,\sigma_{0\Lambda_0} * \hbar \frac{\delta}{\delta \gamma} \rangle + \langle \bar{\eta},\sigma_{0\Lambda_0}* \hbar \frac{\delta}{\delta \omega} \rangle - i\langle  \sigma_{0\Lambda_0} * \hbar \frac{\delta}{\delta b},\eta \rangle,\label{eq_20s}
\end{align}
we write the rhs of \eqref{eq_17delta} in the following form
\begin{equation}
\int d\mu_{0\Lambda_0}\,e^{-\frac{1}{\hbar}L^{\Lambda_0 \Lambda_0}_{vst} + \frac{1}{\hbar} K \cdot \Phi} K \cdot \delta_\epsilon \Phi  =-\epsilon \, \mathcal{S}Z^{0 \Lambda_0}_{vst}=-\epsilon \,\mathcal{S}Z^{0 \Lambda_0}_{aux}\Big|_{\rho,\beta=0}.\label{eq_20r}
\end{equation}
Using equations \eqref{eq_20l} and \eqref{eq_20r} we write identity \eqref{eq_17delta} as follows
\begin{equation}
\mathcal{D} Z^{0\Lambda_0}_{aux}\Big|_{\rho,\beta=0} = \mathcal{S} Z^{0\Lambda_0}_{aux}\Big|_{\rho,\beta=0}\,.\label{eq_21z}
\end{equation}
Then we perform the Legendre transform on the auxiliary functional $W_{aux}:=\hbar \log Z_{aux}$. This defines the functional $\Gamma_{aux}$, from which are obtained $\tilde{\Gamma}_{aux}$, see~\eqref{eq_23gamma}, and the functionals
\begin{align}
\tilde{\Gamme}^{0\Lambda_0}_{\chi}&:=\frac{\delta \tilde{\Gamma}^{0\Lambda_0}_{aux}}{\delta \chi}\Big|_{\chi=0},&\tilde{\Gamme}^{0\Lambda_0}_1&:= \int d^4 x \, \tilde{\Gamma}^{0\Lambda_0}_{\rho(x)}\,,\label{eq_11ga}
\end{align}
where $\chi \in \{\rho,\beta\}$ and $\gamma$, $\omega$ are arbitrary. Equation \eqref{eq_21z} then gives
\begin{equation}
\tilde{\Gamme}^{0\Lambda_0}_1 + \langle (i\ubar{B}+\frac{1}{\xi}\partial \ubar{A}),\tilde{\Gamme}^{0\Lambda_0}_\beta\rangle =\frac{1}{2}\mathcal{S}\Tamma^{0 \Lambda_0}, \label{eq_vst}
\end{equation}
where  
\begin{align}
\Tamma^{0 \Lambda_0}&:=i\langle \ubar{B},\bar{\omega} \rangle + \tilde{\Tamma}^{0 \Lambda_0},&\tilde{\Tamma}^{0 \Lambda_0}&:= \tilde{\Gamma}^{0 \Lambda_0} + \frac{1}{2\xi}\langle \ubar{A}, \partial \partial \, \ubar{A} \rangle\,,
\end{align}
and $\mathcal{S}:=\mathcal{S}_{\tilde{c}}  + \mathcal{S}_A - \mathcal{S}_c$, with
\begin{align}
\mathcal{S}_A&:=\langle \frac{\delta \Tamma^{0 \Lambda_0}}{\delta \ubar{A}},\sigma_{0\Lambda_0}*\frac{\delta}{\delta \gamma}\rangle + \langle \frac{\delta \Tamma^{0\Lambda_0}}{\delta \gamma},\sigma_{0\Lambda_0}*\frac{\delta}{\delta \ubar{A}}\rangle,\\
\mathcal{S}_{\tilde{c}}&:=\langle \frac{\delta \Tamma^{0 \Lambda_0}}{\delta \tilde{c}},\sigma_{0\Lambda_0}*\frac{\delta}{\delta \bar{\omega}} \rangle + \langle \frac{\delta \Tamma^{0 \Lambda_0}}{\delta \bar{\omega}},\sigma_{0\Lambda_0}* \frac{\delta}{\delta \tilde{c}} \rangle,\\
\mathcal{S}_c&:=\langle \frac{\delta \Tamma^{0\Lambda_0}}{\delta \ubar{c}},\sigma_{0\Lambda_0}*\frac{\delta}{\delta \omega}\rangle + \langle \frac{\delta \Tamma^{0\Lambda_0}}{\delta \omega},\sigma_{0\Lambda_0}*\frac{\delta}{\delta \ubar{c}}\rangle\,,\\
\frac{\delta}{\delta \tilde{c}}&:=\frac{\delta }{\delta \wbar{c}} - \partial \frac{\delta }{\delta \gamma}\,.\label{eq_409s}
\end{align}
We rewrite equation~\eqref{eq_vst} in the following form
\begin{align}
\langle i\ubar{B},\tilde{\Gamme}^{0\Lambda_0}_\beta \rangle &=\frac{1}{2}\mathcal{S}_{\tilde{c}} \Tamma^{0\Lambda_0}=\langle i\ubar{B},\sigma_{0\Lambda_0}*\frac{\delta}{\delta \tilde{c}} \tilde{\Tamma}^{0\Lambda_0}\rangle,\label{eq_407c}\\
\tilde{\F}^{0 \Lambda_0}_1&=\frac{1}{2}\tilde{S}\tilde{\Tamma}^{0\Lambda_0},\label{eq_407b}
\end{align}
where
\begin{align}
\tilde{S}&:=\mathcal{S}_A - \mathcal{S}_c,&\tilde{\F}^{0 \Lambda_0}_1:=\tilde{\Gamme}^{0 \Lambda_0}_1 + \frac{1}{\xi}\langle \partial A, \tilde{\Gamme}^{0\Lambda_0}_\beta \rangle\,.\label{eq_407a}
\end{align}
The introduction of the functional $\tilde{\Tamma}$ leads to relation \eqref{eq_407b} and to the consistency condition given in \eqref{eq_cohom} below. They are important in the analysis of the renormalization conditions for $\tilde{\Gamme}^{0\Lambda_0;\Vec{\phi}}_1$, see section \ref{sec_sti}.\smallskip\\An algebraic computation shows that
\begin{align}
(\mathcal{S}_i\mathcal{S}_j + \mathcal{S}_j\mathcal{S}_i)\Tamma^{0\Lambda_0}&=0,& \forall i,j &\in \{A,c,\tilde{c}\}\,.\label{eq_88gr} 
\end{align}
Consequently $\tilde{S}^2 \tilde{\Tamma}^{0\Lambda_0}=0$. Thus application of the operator $\tilde{S}$ to equation~\eqref{eq_407b} yields
\begin{equation}
\tilde{S}\tilde{\F}^{0\Lambda_0}_1=0.\label{eq_cohom}
\end{equation}
Using again \eqref{eq_88gr}, we also have
\begin{equation}
\frac{1}{2}(\tilde{S}\mathcal{S}_{\tilde{c}}+\mathcal{S}_{\tilde{c}}\tilde{S})\Tamma^{0 \Lambda_0}=0,\quad\mbox{and thus}\quad \tilde{S} \tilde{\Gamme}_\beta + \sigma_{0\Lambda_0}*\frac{\delta}{\delta \tilde{c}} \tilde{\F}^{0\Lambda_0}_1 =0\,.\label{eq_404a}
\end{equation}
Finally we set  $\gamma,\omega=0$ in \eqref{eq_407c}, \eqref{eq_407b} to get the AGE and the STI:
\begin{align}
\tilde{\Gamme}^{0\Lambda_0}_\beta=&\sigma_{0\Lambda_0}* \Big(\frac{\delta \tilde{\Tamma}^{0 \Lambda_0}}{\delta \wbar{c}} - \partial \tilde{\Tamma}^{0\Lambda_0}_\gamma \Big)&&(AGE)\,, \label{eq_21gh}\\
\tilde{\F}^{0 \Lambda_0}_1=&\langle \frac{\delta \tilde{\Tamma}^{0\Lambda_0}}{\delta \ubar{A}},\sigma_{0\Lambda_0}*\tilde{\Tamma}^{0\Lambda_0}_\gamma\rangle -\langle \frac{\delta \tilde{\Tamma}^{0\Lambda_0} }{\delta \ubar{c}},\sigma_{0\Lambda_0}*\tilde{\Tamma}^{0\Lambda_0}_\omega \rangle&&(STI). \label{eq_21st}
\end{align}
The goal is to show that $\tilde{\Gamme}^{0\infty}_{\beta}=0$ and $\tilde{\Gamme}^{0\infty}_{1}=0$, in the sense of theorems~\ref{thm_3},~\ref{thm_4}. In appendix \ref{sec_rr} we give a version of these two equations valid for $0 < \Lambda < \Lambda_0$.
\subsubsection{The Antighost Equation}
In this section we extend the AGE for arbitrary $\Lambda \in [0,\Lambda_0]$. In notations~\eqref{eq_jA}, \eqref{eq_tilda}, \eqref{eq_11ga} and with $\mathcal{L}^{\Lambda_0 \Lambda_0}_{aux}$ given in \eqref{eq_17aux} we have
\begin{equation}
-\tilde{W}^{\Lambda \Lambda_0}_{\beta}\tilde{Z}^{\Lambda \Lambda_0}_{vst}=-\hbar \frac{\delta}{\delta \beta} \tilde{Z}^{\Lambda \Lambda_0}_{aux}\Big|_{\rho,\beta=0}=\int d \tilde{\mu} \, Q^{\Lambda_0}_\beta e^{-\frac{1}{\hbar} L^{\Lambda_0 \Lambda_0}_{vst} + \frac{1}{\hbar} \tilde{K} \cdot \tilde{\Phi}}\,,\label{eq_325a}
\end{equation}
where $d \tilde{\mu}:=d \mu_{\Lambda \Lambda_0}(\tilde{\Phi})$, $\tilde{\Phi}:=(A,c,\bar{c})$, $\tilde{K}:=(j,\bar{\eta},\eta)$. Integration by parts, see \eqref{eq_parts2b}, yields
\begin{equation}
\int d \tilde{\mu} \, \left(\frac{\delta L^{\Lambda_0 \Lambda_0}_{vst}}{\delta \bar{c}} - S^{-1}_{\Lambda \Lambda_0} c - \eta \right)e^{-\frac{1}{\hbar} L^{\Lambda_0 \Lambda_0}_{vst} + \frac{1}{\hbar} \tilde{K} \cdot \tilde{\Phi}}=0.\label{eq_325b}
\end{equation}
From definition \eqref{eq_17vst} we also have
\begin{equation}
\frac{\delta L^{\Lambda_0 \Lambda_0}_{vst}}{\delta \bar{c}}=\frac{\delta L^{\Lambda_0 \Lambda_0}}{\delta \bar{c}}\,.\label{eq_325c}
\end{equation}
Substituting $Q_{\beta}$ from \eqref{eq_10a} into \eqref{eq_325a}, then using \eqref{eq_325b}, \eqref{eq_325c} and putting $\gamma,\omega=0$, we obtain
\begin{equation}
-\tilde{W}^{\Lambda \Lambda_0}_{\beta}=\sigma_{0 \Lambda_0}* \Big(S^{-1}_{\Lambda \Lambda_0} \frac{\delta \tilde{W}^{\Lambda \Lambda_0}}{\delta \bar{\eta}} + \eta + \partial \tilde{W}^{\Lambda \Lambda_0}_\gamma \Big) - S^{-1}\frac{\delta \tilde{W}^{\Lambda \Lambda_0}}{\delta \bar{\eta}}\,.
\end{equation}
Finally we perform the Legendre transform 
\begin{align}
\tilde{\Gamme}^{\Lambda \Lambda_0}_{\beta}&=\sigma_{0 \Lambda_0}* \Big(S^{-1}_{\Lambda \Lambda_0}\ubar{c} +  \frac{\delta \tilde{\Gamma}^{\Lambda \Lambda_0}}{\delta \wbar{c}} - \partial \tilde{\Gamme}^{\Lambda \Lambda_0}_\gamma \Big) - S^{-1}\ubar{c}\nonumber\\
&=\sigma_{0 \Lambda_0}* \Big(\frac{\delta \tilde{\Gamme}^{\Lambda \Lambda_0}}{\delta \wbar{c}} - \partial \tilde{\Gamme}^{\Lambda \Lambda_0}_\gamma \Big) - S^{-1}\ubar{c} \,.\label{eq_20ge}
\end{align}
One can also prove identity \eqref{eq_20ge} by noting that its lhs and the $\Gamme$-functionals on the rhs obey FEs of form~\eqref{eq_fl1}. Moreover, identity \eqref{eq_21gh} implies that both sides have the same boundary condition at $\Lambda=0$.
\section{Momentum bounds} \label{sec_21bnd}
From now on we use the following conventions:
\begin{itemize}
\item $0<\Lambda\leqslant\Lambda_0$, unless otherwise stated.
\item $M$ is a fixed mass parameter such that $0<M\leqslant \Lambda_0$.
\item We omit the tilda for all vertex functions and insertions, for example  $\tilde{\Gamma}^{\Vec{\phi};w} \mapsto \Gamma^{\Vec{\phi};w}$, $\tilde{\Gamme}^{\Vec{\phi};w} \mapsto \Gamme^{\Vec{\phi};w}$, $\tilde{\Gamma}^{\Vec{\phi};w}_\chi \mapsto \Gamma^{\Vec{\phi};w}_\chi$.
\item We use $A$, $c$, $\bar{c}$ instead of $\ubar{A}$, $\ubar{c}$, $\wbar{c}$, respectively.
\item A tensor monomial is a tensor product of Kronecker $\delta$’s and momentum variables in $p:=(p_1,...,p_{n-1})$, for example $\delta_{\mu \nu}p_{1\rho}p_{2\sigma}$. Let $\{\delta^s p^k\}$ be the set of all monomials being a product of $s$ Kronecker $\delta$'s and $k$ momenta $p_i$ and let $\{\delta^s p^k\}_r$ be the union of the sets $\{\delta^s p^k\}$ such that $r$ equals the rank of the monomials: $r=2s+k$. For example $\{\delta^2\}_4=\{\delta_{\mu \nu} \delta_{\rho \sigma},\delta_{\mu \sigma} \delta_{\rho \nu},\delta_{\mu \rho} \delta_{\nu \sigma}\}$.
\item For $\Vec{p} \in \mathbb{P}_n$ we define the $\eta$-function by
\begin{equation}
\eta(\Vec{p}):= \min_{S \in \wp_{n-1} \backslash \{\emptyset\}}(|\sum_{i \in S} p_i|,M)\,.
\end{equation}
where $\wp_{n-1}$ denotes the power set of $[n-1]$ (the sum does not include $p_0=-\sum_1^{n-1}p_i$). A momentum configuration $\Vec{p}$ is nonexceptional iff $ \eta(\Vec{p}) \neq 0$ and exceptional otherwise.
\item For a fixed constant $c$ such that $0<c<1$  we define
\begin{equation}
\mathbb{M}_n:=\{\Vec{p} \in \mathbb{P}_n: \eta(\Vec{p})>c M \mbox{ and } p^2_i\leqslant M^2\; \forall i \in [n-1]\}\label{eq_2ren}\,.
\end{equation}
Every $\Vec{p} \in \mathbb{M}_n$ is nonexceptional.
\item $\forall n\geqslant 2$, a momentum configuration $\Vec{p}\in \mathbb{M}_n$ is \textbf{symmetric} iff $\Vec{p} \in \mathbb{M}^s_n$,
\begin{equation}
\mathbb{M}^s_n:=\{\Vec{p} \in \mathbb{M}_n: p_i p_j=\frac{M^2}{n-1}(n\,\delta_{ij}-1)\; \forall i,j \in [n-1]\}\,.\label{eq_9sym}
\end{equation} 
\item $\forall n \geqslant 3$, a momentum configuration $\Vec{p} \in \mathbb{M}_n$ is \textbf{coplanar} iff $\Vec{p}\in \mathbb{M}^{cp}_n$,
\begin{equation}
\mathbb{M}^{cp}_n:=\{\Vec{p} \in \mathbb{M}_n: \mathrm{dim}(\mathrm{span}(p_0,...,p_{n-1}))=2 \}\,.\label{eq_9cp}
\end{equation}
\item In the following, a renormalization point is denoted by $\Vec{q}\in\mathbb{P}_n$. See appendix \ref{sec_ren} for the list of all relevant terms and their renormalization points.
\end{itemize}
\subsection{Weighted trees}
The bounds on the vertex functions presented in section \ref{sec_thms} are expressed in terms of sets of weighted trees that are introduced by definitions~\ref{def_tree0},~\ref{def_tree} below. As seen from \eqref{eq_16bnd}, to each edge $e$ of a weighted tree is associated a factor $(|p_e|+\Lambda)^{-\theta(e)}$, $p_e$ being the momentum traversing the edge and $\theta(e)$ being the $\theta$-weight of the edge, expressed as a sum of the $\rho$ and $\sigma$-weights of the edge, see \eqref{eq_16theta}. The relation \eqref{eq_409a} expresses the fact that the total $\theta$-weight of a tree is in agreement with power counting. Nonvanishing $\sigma$-weights are introduced in order to define viable tree bounds for momentum derived vertex functions. The definition of the $\sigma$-weight is inspired by how momentum derivatives are distributed along a tree, taking care of momentum conservation. Before giving the definition of the weighted trees we set up some necessary notations.
\begin{itemize}
\item A tree $\tau$ is a connected graph with no cycles. The sets of vertices and edges of a tree $\tau$ are denoted respectively by $V(\tau)$ and $E(\tau)$, or shortly $V$, $E$. In the following, the terms ``edge`` and ``line`` are equivalent.
\item Let $V_m$ be the set of vertices of valence $m$. Then, $V=\bigcup_{m \geqslant 1} V_m$.
\item Let $E_1$ be the set of edges incident to vertices of valence 1. In other words, $E_1$ is the set of external edges.
\item Each tree has a bijection $\psi:\{0,...,n-1\} \to V_1$, $i \mapsto v_i$ and a sequence of $n$ field labels $\Vec{\varphi}=(\varphi_0,...,\varphi_{n-1})$, $\varphi_i \in \{A,c,\bar{c},\gamma,\omega,\beta\}$ and $n=|V_1|$. The field label $\varphi_i$ defines the type of the vertex $v_i \in V_1$. Let $V_{\varphi} \subseteq V_1$ be the set of all vertices of type $\varphi$, for example $V_A$. Furthermore, let $E_\varphi \subseteq E_1$ be the set of all edges incident to vertices in $V_\varphi$, for example~$E_{\varkappa}$ with $\varkappa \in \{\gamma,\omega\}$.
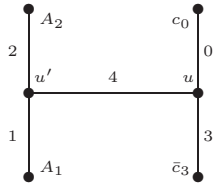
\begin{center} \tiny \begin{picture}(128,64)
%
\put(32,0){\line(0,1){64}}
\put(32,32){\line(1,0){64}}
\put(96,0){\line(0,1){64}}
%
\put(32,64){\circle*{4}}
\put(32,0){\circle*{4}}
\put(32,32){\circle*{4}}
\put(96,64){\circle*{4}}
\put(96,0){\circle*{4}}
\put(96,32){\circle*{4}}
%
\put(24,14){1}
\put(24,46){2}
\put(98,14){3}
\put(98,46){0}
\put(62,36){4}
%
\put(36,2){$A_1$}
\put(36,58){$A_2$}
\put(86,2){$\bar{c}_3$}
\put(86,58){$c_0$}
\put(34,36){$u^\prime$}
\put(90,36){$u$}
\end{picture} \end{center}
When needed, the edges are labeled by integers and the vertices by symbols. The edge incident to a vertex $v_i \in V_1$ has the same index $i$. As an example, for the tree above we have: \smallskip \\
\begin{tabular}{llll}
$V=\{c_0,A_1,A_2, \bar{c}_3,u, u^\prime\}$,&$V_3=\{u^\prime,u\}$,&\\
$V_1=\{c_0,A_1,A_2, \bar{c}_3\}$,&$V_A=\{A_1,A_2\}$,&$V_{\bar{c}}=\{\bar{c}_3\}$,&$V_c=\{c_0\}$,\\
$E_1=\{0,1,2,3\}$,&$E_A=\{1,2\}$,&$E_{\bar{c}}=\{3\}$,&$E_{c}=\{0\}$.
\end{tabular}
\item Recalling the definition of $\mathbb{P}_n$ in \eqref{eq_pn}, for every edge $e \in E$ and vertex $v \in V_1$, the momentum assignments $p_e$, $p_v$ are functions from the set $\mathbb{P}_n$ to $\mathbb{R}^4$, with $n=|V_1|$, defined by the following construction:
\begin{itemize}
\item[a)]label the vertices in $V_1$ by means of $\psi: i \mapsto v_i$ and set $p_{v_i}(\Vec{p}):=p_i$,
\item[b)]apply momentum conservation to all vertices to get $p_e(\Vec{p})$.
\end{itemize}
We use similar notations for multi-indices: $w_v:=w_{\psi^{-1}(v)}$ for $w \in \mathbb{W}_n$ and $v \in V_1$. Given the momentum assignments, a set-valued function $K$ on $E$ is defined by
\begin{equation}
K_e:=\{v \in V_1 \backslash \{v_0\}: \frac{\partial  p_e}{\partial p_v} \neq 0 \}.
\end{equation}
For the tree given above we have $p_{c_0}=-(p_{A_1}+ p_{A_2} + p_{\bar{c}_3})$ and
\begin{align}
K_0&=\{A_1,A_2,\bar{c}_3\},&K_1&=\{A_1\},&K_2&=\{A_2\},\\
K_3&=\{\bar{c}_3\},&K_4&=\{A_1,A_2\}\,.
\end{align}
\end{itemize}
Some additional structure is needed, always in view of the bounds.
\begin{itemize}
\item The vertices in $V_3$ are additionally labeled either as "\textbf{regular}" ($\bullet$) or as "\textbf{hollow}" ($\circ$). The sets of regular and hollow vertices are respectively denoted by $V_\bullet$ and $V_\circ$, hence $V_3=V_\bullet \cup V_\circ$. In terms of our bounds, regular vertices do change the $\rho$-weight of incident edges, while hollow vertices do not, see definition \ref{def_rho}. We use hollow vertices for the bounds on 3-point functions and in the proof of the theorems, see for example section \ref{sec_junc} on the junction of weighted trees and definition \ref{def_fgm}.
\item Edges carry zero or more labels "*". Edges are referred to as "\textbf{*-edges}" if they have one or more labels "*", and as "\textbf{regular edges}" otherwise. The set of all *-edges is denoted by $E_*$. The *-edges play a special role in our bounds, because to each $e\in E_*$ is associated a supplementary factor $|p_e|+\Lambda$, see \eqref{eq_16bnd} and theorem \ref{thm_4}.
\end{itemize}
\begin{definition}\label{def_tree0}  Let be given a sequence of $n\geqslant 3$ field labels, $\Vec{\varphi}=(\varphi_0,...,\varphi_{n-1})$, with $\varphi_i \in \{A,c,\bar{c},\gamma,\omega,\beta\}$. Let $\mathbb{T}_{\Vec{\varphi}}$ denote the set of all trees that satisfy the following rules:
\begin{itemize}
\item There is a bijection $\psi:\{0,...,n-1\} \to V_1$. Each $v_i \in V_1$ has type $\varphi_i$.
\item $ V=  V_1 \cup V_3$.
\item If $n=3$ then $V_3=V_\circ$.
\item $|E_*| \in \mathbb{N}$.
\end{itemize}
\end{definition}
\begin{definition} \label{def_tree}
In the notations of definition \ref{def_tree0}, 
let $\mathcal{T}^{(s)}_{\Vec{\varphi}}$ denote the set of all trees in $\mathbb{T}_{\Vec{\varphi}}$
with total number of labels "*" equal to $s$
and such that $V_3=V_\bullet$ whenever $n>3$.
For shortness we set
$\mathcal{T}_{\Vec{\varphi}}
:=\mathcal{T}_{\Vec{\varphi}}^{(0)}$
and
 $\mathcal{T}_{1\Vec{\varphi}}:=\mathcal{T}_{\Vec{\varphi}}^{(1)}$.
\end{definition}
As an example, below we show two trees, $\tau_3 \in \mathcal{T}_{1c A A}$ and $\tau_7 \in \mathcal{T}_{\beta c  c \bar{c} A A A}$.\\
\null\hfill\begin{minipage}[b][76pt]{32pt}\tiny\begin{picture}(32,64)
%
\put(0,0){\line(0,1){64}}
\put(0,32){\line(1,0){32}}
\put(0,32){\circle*{8}}
\put(0,32){\color{white}\circle*{6}}
\put(0,0){\circle*{4}}
\put(0,64){\circle*{4}}
\put(32,32){\circle*{4}}
\put(2,50){*}
\put(2,42){0}
\put(14,34){1}
\put(2,14){2}
\put(2,60){$c$}
\put(2,4){$A_2$}
\put(24,26){$A_1$}
\end{picture}
\end{minipage}\hfill\begin{minipage}[b][76pt]{128pt}\tiny\begin{picture}(128,64)
%
\put(0,0){\line(0,1){64}}
\put(0,32){\line(1,0){128}}
\put(32,32){\line(0,1){32}}
\put(64,32){\line(0,1){32}}
\put(96,32){\line(0,1){32}}
\put(128,0){\line(0,1){64}}
\put(0,0){\circle*{4}}
\put(0,64){\circle*{4}}
\put(0,32){\circle*{4}}
\put(32,32){\circle*{4}}
\put(32,64){\circle*{4}}
\put(64,64){\circle*{4}}
\put(96,32){\circle*{4}}
\put(96,64){\circle*{4}}
\put(64,32){\circle*{4}}
\put(128,64){\circle*{4}}
\put(128,32){\circle*{4}}
\put(128,0){\circle*{4}}
\put(120,46){3}
\put(2,46){5}
\put(2,14){2}
\put(34,46){0}
\put(66,46){4}
\put(90,46){1}
\put(120,14){6}
\put(14,34){9}
\put(46,34){8}
\put(78,34){7}
\put(110,34){10}
\put(2,60){$A_5$}
\put(34,60){$\beta$}
\put(66,60){$A_4$}
\put(86,60){$c_1$}
\put(118,60){$\bar{c}_3$}
\put(2,4){$c_2$}
\put(116,4){$A_6$}
\end{picture}\end{minipage}\hfill\null
\begin{definition} \label{def_rho} Fix a tree from $\mathbb{T}_{\Vec{\varphi}}$. A $\rho$-weight is a function $\rho:E \to \{0,1,2\}$ with the following properties:
\begin{enumerate}
\item $\forall e \in E_1$, $\rho(e)=0$.
\item There exists a map $\chi: V_\bullet \to E \backslash E_1$ such that
\begin{itemize}
\item[a)] if $\chi(v)=e$, then $e$ is incident to $v$,
\item[b)] $ \forall e \in E \backslash E_1$, $\rho(e)=2- \left\vert{\chi^{-1}(\{e\})}\right\vert.$
\end{itemize}
\end{enumerate}
\end{definition}
\begin{definition}\label{def_sigma} Let be given a tree from $\mathbb{T}_{\Vec{\varphi}}$ and $w \in \mathbb{W}_n$, with $n=|V_1|$. A $\sigma$-weight is a function $\sigma: E \to \mathbb{N}$ defined by
\begin{equation}
\sigma(e):=\sum_{v \in V_1} \sigma_v(e),
\end{equation}
where $(\sigma_v: E \to \mathbb{N})_{v \in  V_1}$ is a family of functions such that
\begin{align}
\sum \limits_{e \in E} \sigma_v(e)&=w_v\;,&\sigma_v(e)&=0\mbox{ if } v \not \in K_e\;.
\end{align}
\end{definition}
By definition \eqref{eq_17w}, $w_{0}=0$ for every $w \in \mathbb{W}_n$. Hence, $\sigma_{v_0}(e)=0$ $\forall e \in E$.
\begin{definition} Let be given a tree $\tau \in \mathbb{T}_{\Vec{\varphi}}$ and $w \in \mathbb{W}_n$, with $n=|V_1|$. A $\theta$-weight is a function $\theta:E \to \mathbb{N}$ defined by
\begin{equation}
\theta(e):=\rho(e)+\sigma(e),\label{eq_16theta}
\end{equation}
where $\rho$ and $\sigma$ are a $\rho$-weight and a $\sigma$-weight corresponding to $w$, respectively. The pair $(\tau, \theta)$ is a weighted tree.  The total $\theta$-weight of $(\tau,\theta)$ is
\begin{equation}
\theta(\tau):=\sum_{e \in E} \theta(e).
\end{equation}
The set of all $\theta$-weights corresponding to given $\tau$ and $w$ is denoted by $\Theta^w_\tau$.
\end{definition}
For every tree $\tau \in \mathcal{T}^{(s)}_{\Vec{\varphi}}$ with $n\geqslant 4$ the total $\theta$-weight is given by the formula
\begin{equation}
\theta(\tau)=n + \|w\| - 4\,.\label{eq_409a}
\end{equation}
This relation follows from definitions \ref{def_tree}, \ref{def_rho}, \ref{def_sigma}, which give the sum rule $\sum_{e \in E} \theta(e)=\|w\|+2|E\setminus E_1|-|V_3|$, and from the relations $|E\setminus E_1|-|V_3|+1=0$ and $|V_3|=n-2$.

As an example we consider three trees $\tau_1,\tau_2,\tau_3 \in \mathcal{T}_{AAAA}$. We give three different weights $\theta_a$, $\theta_b$, $\theta_c$, which all correspond to the derivative wrt the momentum $p_1$, literally $w_1=1$ and $w=(0,1,0,0)$. We find a family of weighted trees $\{(\tau_i,\theta): \theta \in \Theta^w_{\tau_i}\}_{i \in \{1,2,3\}}$, where
\begin{align}
\Theta^w_{\tau_3}=\Theta^w_{\tau_1}&=\{\theta_a,\theta_b,\theta_c\},&\Theta^w_{\tau_2}&=\{\theta_a,\theta_c\}.
\end{align}
\\
\null\hfill{\tiny\begin{tabular}{ccccc}
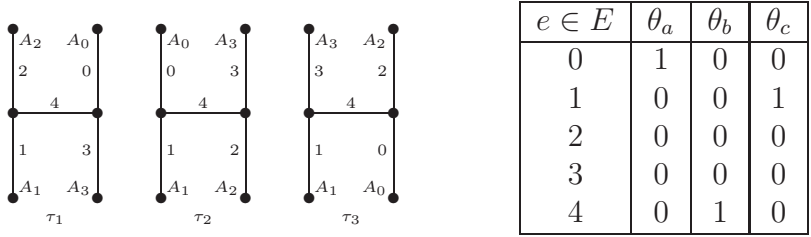
\begin{minipage}[c][72pt]{32pt}\begin{picture}(32,64)
%
\put(0,64){\line(0,-1){64}}
\put(0,32){\line(1,0){32}}
\put(32,64){\line(0,-1){64}}
%
\put(0,32){\circle*{4}}
\put(32,32){\circle*{4}}
\put(0,0){\circle*{4}}
\put(0,64){\circle*{4}}
\put(32,0){\circle*{4}}
\put(32,64){\circle*{4}}
%
\put(2,16){1}
\put(2,46){2}
\put(26,16){3}
\put(26,46){0}
\put(14,34){4}
%
\put(2,2){$A_1$}
\put(20,2){$A_3$}
\put(20,58){$A_0$}
\put(2,58){$A_2$}
\end{picture}\end{minipage}&&\begin{minipage}[c][72pt]{32pt}\centering\begin{picture}(32,64)
%
\put(0,64){\line(0,-1){64}}
\put(0,32){\line(1,0){32}}
\put(32,64){\line(0,-1){64}}
%
\put(0,32){\circle*{4}}
\put(32,32){\circle*{4}}
\put(0,0){\circle*{4}}
\put(0,64){\circle*{4}}
\put(32,0){\circle*{4}}
\put(32,64){\circle*{4}}
%
\put(2,16){1}
\put(2,46){0}
\put(26,16){2}
\put(26,46){3}
\put(14,34){4}
%
\put(2,2){$A_1$}
\put(20,2){$A_2$}
\put(20,58){$A_3$}
\put(2,58){$A_0$}
\end{picture}\end{minipage}&&\begin{minipage}[c][72pt]{32pt}\centering\begin{picture}(32,64)
%
\put(0,64){\line(0,-1){64}}
\put(0,32){\line(1,0){32}}
\put(32,64){\line(0,-1){64}}
%
\put(0,32){\circle*{4}}
\put(32,32){\circle*{4}}
\put(0,0){\circle*{4}}
\put(0,64){\circle*{4}}
\put(32,0){\circle*{4}}
\put(32,64){\circle*{4}}
%
\put(2,16){1}
\put(2,46){3}
\put(26,16){0}
\put(26,46){2}
\put(14,34){4}
%
\put(2,2){$A_1$}
\put(20,2){$A_0$}
\put(20,58){$A_2$}
\put(2,58){$A_3$}
\end{picture}\end{minipage}\\
$\tau_1$&&$\tau_2$&&$\tau_3$
\end{tabular}}\hfill\begin{tabular}{|c|c|c|c|}\hline
$e \in E$&$\theta_a$&$\theta_b$&$\theta_c$\\
\hline
0&1&0&0\\
1&0&0&1\\
2&0&0&0\\
3&0&0&0\\
4&0&1&0\\
\hline
\end{tabular}\hfill\null
\subsection{Theorems}\label{sec_thms}
We always assume that the renormalization constants are independent of~$\Lambda_0$ (though weakly $\Lambda_0$-dependent "renormalization constants" can also be accommodated for, see \cite{kks}). From now on we denote $\Vec{\varkappa}:=(\varkappa_1,....,\varkappa_{n_\varkappa})$ with $\varkappa_i \in \{\gamma, \omega\}$, $n_{\varkappa} \geqslant 0$.
\begin{hyprc}\label{rc1}We impose on all strictly relevant terms vanishing renormalization conditions at zero momentum and $\Lambda=0$:
\begin{align}
\Gamme^{0\Lambda_0;\Vec{\phi};w}_{\Vec{\varkappa}}(0)&=0,&\mbox{if}\quad 2n_\varkappa +\N+\|w\|<4\,. \label{eq_16cdt2}
\end{align}
\end{hyprc}
\begin{hyprc}\label{rc2}
On the following marginal terms we impose renormalization conditions at zero momentum and $\Lambda=M$:
\begin{align}
\Gamme^{M\Lambda_0;c\bar{c} c \bar{c}}(0)&=0,&\Gamme^{M\Lambda_0;c\bar{c} A A}(0)&=0,&\partial_A\Gamme^{M\Lambda_0;c \bar{c} A}(0)&=0\,, \label{eq_16cdt1}
\end{align}
for the notation see \eqref{eq_11ns}.
\end{hyprc}
Remark that Bose--Fermi symmetry and translation invariance imply that $\partial_A\Gamme^{M\Lambda_0;c \bar{c} A}(0)=0$ iff  $\partial_c\Gamme^{M\Lambda_0;A \bar{c} c}(0)=0$. To prove proposition \ref{prop_barc} and theorem~\ref{thm_1} all remaining marginal renormalization constants are chosen at $\Lambda=0$ arbitrarily but in agreement with the global symmetries of the regularized theory:
SU(2), Euclidean isometries ISO(4), ghost number conservation.
For instance, all renormalization conditions
must comply with the vanishing of the ghost number violating functions,
like
$\Gamme^{\Lambda,\Lambda_0;cccc}$
or
$\Gamme^{\Lambda,\Lambda_0}_{(\varkappa_0,\varkappa_1)}$
for $\varkappa_i \in \{\gamma,\omega\}$.
The list of the remaining marginal renormalization constants follows literally from \eqref{eq_11r} and  appendix \ref{sec_gamma2}.
\begin{proposition}\label{prop_barc}Assume the validity of hypotheses \ref{rc1} and \ref{rc2}. For all sequences of $\N \geqslant 3$ field labels in $\{A,c,\bar{c}\}$ with $\phi_{\N -1}=\bar{c}$, denoted by $\Vec{\phi}\bar{c}$, all $w=(w',0)\in\mathbb{W}_{\N}$, all $(\Vec{p},0)\in\mathbb{P}_{\N}$, and all positive $\Lambda$, $\Lambda_0$ s.t. ${\max(\Lambda, M) \leqslant \Lambda_0}$,
\begin{equation}
\Gamme^{\Lambda\Lambda_0;\Vec{\phi} \bar{c};w}(\Vec{p},0)=0\,.\label{eq_anti}
\end{equation}
\end{proposition}
Note that in \eqref{eq_anti} the momentum of the indicated antighost $\bar{c}$ vanishes, and there is no derivative wrt this momentum.
\paragraph{Proof}
We prove the statement by induction, increasing in the loop order, $l-1\mapsto l$.
Given $l$, we proceed by descending from $w_{max}$ in the number of derivatives, $\|w\| \mapsto \|w\|-1$.
For fixed $l$ and $w$, all possible terms
$\Gamme^{\Vec{\phi}\bar{c};w}_l$ are considered.
By construction, for fixed $l$ and $\Vec{\phi}\bar{c}$, the inductive scheme deals first with the irrelevant terms and continues, if they exist, with the marginal terms, followed by more and more relevant terms.
Since the momentum of the antighost has been assumed to vanish, the statement holds at loop order $l=0$. The validity of the statement for all loop orders smaller than $l$ implies that $\dot{\Gamme}^{\Lambda\Lambda_0;\Vec{\phi} \bar{c};w}_l(\Vec{p},0)=0$.
The irrelevant terms have vanishing boundary conditions, hence $\Gamme^{\Lambda_0\Lambda_0;\Vec{\phi} \bar{c};w}_l(\Vec{p},0)=0$. Integrating the FE from $\Lambda_0$ downwards to arbitrary $\Lambda>0$, we get $\Gamme^{\Lambda \Lambda_0;\Vec{\phi} \bar{c};w}_l(\Vec{p},0)=0$ for the irrelevant terms. Next we consider the marginal terms. Since the corresponding irrelevant terms have already been shown to vanish at vanishing antighost momentum, we use the Taylor formula to extend \eqref{eq_16cdt1} to arbitrary momenta $(\Vec{p},0)$, still preserving the vanishing antighost momentum. Then, we integrate the FE from $M$ to arbitrary $\Lambda>0$, which completes the proof that
$\Gamme^{\Lambda \Lambda_0;\Vec{\phi} \bar{c};w}_l(\Vec{p},0)=0$
for marginal terms. Similar arguments hold for all the strictly relevant terms~$\Gamme^{\Vec{\phi} \bar{c};w}_l$.\hfill$\blacksquare$
\begin{corollary} The following counterterms vanish, see \eqref{eq_ct},
\begin{align}
r^{0,\bar{c} c A A}_{1}&=0,&r^{0,\bar{c} c \bar{c} c}&=0,&r^{0,\bar{c} c A A}_{2}&=0,&r^{0,A \bar{c} c}_2&=0\label{eq_21r1}.
\end{align}
\end{corollary}
\paragraph{Proof}
Using \eqref{eq_anti} we have, for all $p_2,p_3\in\mathbb{R}^4$ and $\Lambda = \Lambda_0$ (omitted),
\begin{align}
\Gamme^{c \bar{c} c \bar{c} }(0,p_2,p_3)&=0,&\Gamme^{c \bar{c} AA }(0,p_2,p_3)&=0,&\partial_{p_2} \Gamme^{A \bar{c} c}(0,p_2)&=0\,.\label{vangas}
\end{align}
Recall \eqref{eq_27j} and \eqref{eq_ct}.\hfill$\blacksquare$
\begin{corollary}
For all $X \in \{\beta,1\}$,
$\Vec{\varkappa}$,
$\Vec{\phi}\bar{c}$,
$w=(w',0)$,
$\Vec{p}$,
and all positive~$\Lambda$,~$\Lambda_0$ s.t. $\max(\Lambda, M) \leqslant \Lambda_0$:
\begin{equation}
\Gamme^{\Lambda\Lambda_0;\Vec{\phi} \bar{c};w}_{X\Vec{\varkappa}}(\Vec{p},0)=0 \label{eq_21anti}.
\end{equation}
\end{corollary}
\paragraph{Proof}It follows from the definitions of the inserted functions given in \eqref{eq_17vst}, \eqref{eq_21brs}, \eqref{eq_17aux} that at tree level $\Gamme^{\Lambda \Lambda_0;\Vec{\phi} \bar{c};w}_{X\Vec{\varkappa};l=0}(\Vec{p},0)=0$. Then using equation \eqref{eq_21r1} one shows that for all these terms we have vanishing boundary conditions, $\Gamme^{\Lambda_0\Lambda_0;\Vec{\phi} \bar{c};w}_{X\Vec{\varkappa};l}(\Vec{p},0)=0$ $\forall l$. Assuming that the statement is true at the loop order $l-1\geqslant 0$, by induction in $l$, using \eqref{eq_anti} and integrating the FE from $\Lambda_0$ to arbitrary $\Lambda$ one shows that it holds at the loop order $l$.\hfill$\blacksquare$

From now on, for simplicity of notation we write $\mathcal{P}^{(k)}_{s}$ to denote polynomials with nonnegative coefficients and degree $s$ where the superscript is a label to make one polynomial different from another. We define:
\begin{align}
P^{\lambda_1 \lambda_2}_{s}(\Vec{p})&:=\mathcal{P}^{(0)}_s\Big(\log_+ \frac{\max(|\Vec{p}|,M)}{\lambda_1 + \eta(\Vec{p})}\Big) + \mathcal{P}^{(1)}_s\Big(\log_{+} \frac{\lambda_2}{M}\Big)\,,\label{eq_1p}\\
P^{\lambda}_{s}(\Vec{p})&:=P^{\lambda \lambda}_{s}(\Vec{p})\,,\\
\Pi^{\lambda}_{\tau,\theta}(\Vec{p})&:= \prod \limits_{e \in E} \left(\lambda + |p_e|\right)^{-\theta(e)}\,,\label{eq_pi}\\
Q^{\Lambda;w}_{\tau}(\Vec{p})&:=\frac{\prod \limits_{e \in E_*}(\Lambda + |p_e|)}{\prod \limits_{e \in E_\varkappa}(\Lambda + |p_e|)}\left\{\begin{matrix}\inf \limits_{i \in \mathbb{I}} \sum \limits_{\theta \in \Theta^{w^\prime(i)}_{\tau}} \Pi^{\Lambda}_{\tau,\theta}(\Vec{p}),&|V_1|=3\,,\label{eq_16bnd}\\
\sum \limits_{\theta \in \Theta^w_\tau} \Pi^{\Lambda}_{\tau,\theta}(\Vec{p}),&otherwise\,,\end{matrix}\right.
\end{align}
where $\tau \in \mathbb{T}_{\Vec{\varphi}}$, $w^\prime(i)$ is obtained from $w$ by diminishing $w_i$ by one unit, and, for nonvanishing $w$, $\mathbb{I}:=\{i: w_i>0\}$.
The following sets are also used in theorems~\ref{thm_1}--\ref{thm_4}:
\begin{align}
\mathbb{Y}_n^{+}&:=\{(\Lambda,\Lambda_0):0<\Lambda\leqslant\Lambda_0\text{~and~}\Lambda_0\geqslant M\}\times \mathbb{P}_n\,,
\\
\mathbb{Y}_n&:= \mathbb{Y}_n^{+}\;\bigcup\;\{(0,\Lambda_0):\Lambda_0\geqslant M\}\times \{\vec{p}\in\mathbb{P}_n:\eta(\vec{p})\neq0\}.
\end{align}
\begin{thm} \label{thm_1}
There exists a collection of regular vertex functions $\Gamme^{\Vec{\phi}}_{\Vec{\varkappa};l}$ on $\mathbb{Y}_{\N+n_\varkappa}^+$, complying with the global symmetries of the theory, satisfying the FE and the renormalization conditions given by hypotheses \ref{rc1} and \ref{rc2}, and with irrelevant terms vanishing at $\Lambda=\Lambda_0$.
Furthermore, for all $\Vec{\phi}$, $\Vec{\varkappa}$, all $l\in\mathbb{N}$, $w \in \mathbb{W}_{\N+n_\varkappa}$, the following bounds hold on $\mathbb{Y}_{\N+n_\varkappa}^+$:
\begin{description}
\item[a)]$d\geqslant 0$ or $\N+n_\varkappa=2$
\begin{equation}
|\Gamme^{\Lambda\Lambda_0;\Vec{\phi};w}_{\Vec{\varkappa};l}(\Vec{p})| \leqslant (\Lambda + |\Vec{p}|)^d P^{\Lambda}_r(\Vec{p})\,,\\
\end{equation}
\item[b)]$d<0$
\begin{equation}
|\Gamme^{\Lambda\Lambda_0;\Vec{\phi};w}_{\Vec{\varkappa};l}(\Vec{p})| \leqslant \sum \limits_{\tau \in \mathcal{T}_{\Vec{\varkappa}\Vec{\phi}}} Q^{\Lambda;w}_{\tau}(\Vec{p}) \; P^{\Lambda}_r(\Vec{p})\,.
\end{equation}
\end{description}
Here $d:=4-2n_\varkappa-\N-\|w\|$. If $l=0$ then $r:=0$, otherwise $r$ stands for $r(d,l)$.
\begin{equation}
r(d,l):=\left\{\begin{matrix}2l,&d \geqslant 0\,,\\2l-1,&d<0\,.\end{matrix}\right.
\end{equation}
\end{thm}
Theorem \ref{thm_1} shows in particular that the functions $\Gamme^{\Lambda \Lambda_0;\Vec{\phi};w}_{\Vec{\varkappa}}$ are bounded uniformly in $\Lambda_0$. To prove convergence in the limit $\Lambda_0 \to \infty$ we establish the following bounds for their derivatives wrt $\Lambda_0$.
\begin{thm} \label{thm_2}Let be given a collection of vertex functions $\Gamme^{\Vec{\phi}}_{\Vec{\varkappa};l}$ as in theorem~\ref{thm_1}.
Then, for all $\Vec{\phi}$, $\Vec{\varkappa}$, all $l\in\mathbb{N}$, $w \in \mathbb{W}_{\N+n_\varkappa}$, the following bounds hold on $\mathbb{Y}_{\N+n_\varkappa}^+$:
\begin{description}
\item[a)]$d \geqslant 0$ or $\N+n_\varkappa=2$
\begin{equation}
|\partial_{\Lambda_0} \Gamme^{\Lambda\Lambda_0;\Vec{\phi};w}_{\Vec{\varkappa};l}(\Vec{p})| \leqslant \frac{\Lambda + M+|\Vec{p}|}{\Lambda^2_0} (\Lambda + |p|)^d P^{\Lambda\Lambda_0}_{r}(\Vec{p})\,,
\end{equation}
\item[b)]$d<0$
\begin{equation}
|\partial_{\Lambda_0}\Gamme^{\Lambda\Lambda_0;\Vec{\phi};w}_{\Vec{\varkappa};l}(\Vec{p})| \leqslant  \frac{\Lambda + M+|\Vec{p}|}{\Lambda^2_0} \sum \limits_{\tau \in \mathcal{T}_{\Vec{\varkappa}\Vec{\phi}}} Q^{\Lambda;w}_{\tau}(\Vec{p}) \, P^{\Lambda\Lambda_0}_{r}(\Vec{p})\,.
\end{equation}
\end{description}
See theorem \ref{thm_1} for the definition of $d$ and $r$.
\end{thm}
Convergence of the limit $\Lambda \to 0^+$ of the terms
$\Gamme^{\Lambda\Lambda_0;\Vec{\phi};w}_{\Vec{\varkappa}}(\Vec{p})$,
$\partial_{\Lambda_0}\Gamme^{\Lambda\Lambda_0;\Vec{\phi};w}_{\Vec{\varkappa}}(\Vec{p})$
when $\Vec{p}$ is nonexceptional (or $d>0$)
follows from the Cauchy criterion
\begin{equation}
|\partial^k_{\Lambda_0} \Gamme^{\Lambda\Lambda_0;\Vec{\phi};w}_{\Vec{\varkappa}}(\Vec{p})-\partial^k_{\Lambda_0}  \Gamme^{\Lambda^\prime\Lambda_0;\Vec{\phi};w}_{\Vec{\varkappa}}(\Vec{p})|\leqslant \int \limits^{\Lambda^\prime}_{\Lambda} d \lambda \, |\partial_{\lambda}\partial^k_{\Lambda_0}  \Gamme^{\lambda\Lambda_0;\Vec{\phi};w}_{\Vec{\varkappa}}(\Vec{p})|\,,
\end{equation}
and the bounds from theorems \ref{thm_1}, \ref{thm_2}.
Convergence of the limit $\Lambda_0 \to \infty$ of the terms
$\Gamme^{0\Lambda_0;\Vec{\phi};w}_{\Vec{\varkappa}}(\Vec{p})$
when $\Vec{p}$ is nonexceptional (or $d>0$)
follows from Cauchy criterion and the bounds from theorem \ref{thm_2},
\begin{equation}
|\Gamme^{0\Lambda_0;\Vec{\phi};w}_{\Vec{\varkappa}}(\Vec{p})-\Gamme^{0\Lambda^\prime_0;\Vec{\phi};w}_{\Vec{\varkappa}}(\Vec{p})|\leqslant \int \limits^{\Lambda^\prime_0}_{\Lambda_0} d \lambda_0 \, |\partial_{\lambda_0}\Gamme^{0\lambda_0;\Vec{\phi};w}_{\Vec{\varkappa}}(\Vec{p})|\,.
\end{equation}

In the following we will consider the functions $\Gamme^{0\Lambda_0; \Vec{\phi};w}_{1\Vec{\varkappa}}$, $\Gamme^{0\Lambda_0;\Vec{\phi};w}_{\beta\Vec{\varkappa}}$ which appear on the lhs respectively of the ST identities~\eqref{eq_21st} and of the AGE \eqref{eq_21gh}.  The goal of theorems \ref{thm_3}, \ref{thm_4} is to show that $\Gamme^{0\Lambda_0;\Vec{\phi};w}_{1\Vec{\varkappa}}$ and $\Gamme^{0\Lambda_0;\Vec{\phi};w}_{\beta\Vec{\varkappa}}$ vanish in the limit $\Lambda_0 \to \infty$, which restores the STI and AGE. The renormalization conditions for these functions at $\Lambda=0$ are obtained from the rhs of the STI and AGE. In section \ref{sec_sti} we show that the required boundary marginal terms~$\Gamme^{0\Lambda_0;\Vec{\phi};w}_{1 \Vec{\varkappa}}$,~$\Gamme^{0\Lambda_0;\Vec{\phi};w}_{\beta \Vec{\varkappa}}$ satisfy the bounds of the theorems under the conditions specified  in hypothesis \ref{rc3}.
\begin{hyprc}\label{rc3}
We allow $R^{AAA}$, $r^{AA}_1$, $r^{\bar{c}c}$  to be chosen arbitrarily but the remaining marginal renormalization constants must satisfy a set of equations: $R_1$\eqref{eq_gh_c}, $R_2$\eqref{eq_st_cAA}, $R_3$\eqref{eq_st_cccA}, $r^{AA}_2$\eqref{eq_st_cA}, $R^{A \bar{c} c}_1$\eqref{eq_gh_cA}, $R^{AAAA}_{1,2}$\eqref{eq_st_cAAA}, see appendix~\ref{sec_gamma2} and \eqref{eq_11r} in appendix \ref{sec_gamma}  for notations.
\end{hyprc}
For shortness we also introduce the following definition
\begin{align}
F^{\Lambda\Lambda_0}_s(\Vec{p})&:=\frac{M+|\Vec{p}| + \Lambda}{\Lambda_0}\Big(1+\Big(\frac{|\Vec{p}|}{\Lambda_0}\Big)^{w_{max}}\Big)\mathcal{P}^{(2)}_{s}\Big(\frac{|\Vec{p}|}{\Lambda + M} \Big)\,.
\end{align}
\begin{thm}\label{thm_3}Let be given a collection of vertex functions $\Gamme^{\Vec{\phi}}_{\Vec{\varkappa};l}$, regular on $\mathbb{Y}_{\N+n_\varkappa}$, complying with the global symmetries of the theory, satisfying the hypotheses \ref{rc1}, \ref{rc2}, \ref{rc3} and the bounds of theorems~\ref{thm_1},\ref{thm_2}.
Let $\Gamme^{\Vec{\phi}}_{\beta,\Vec{\varkappa};l}$ be a collection of vertex functions with one insertion of the operator $Q_\beta^{\Lambda_0}$ \eqref{eq_10a}, regular on $\mathbb{Y}_{1+\N+n_\varkappa}$, complying with the global symmetries of the theory, satisfying the FE, and s.t.~the AGE \eqref{eq_21gh} holds. Then, for all $\Vec{\phi}$, $\Vec{\varkappa}$, all $l\in\mathbb{N}$, $w \in \mathbb{W}_{1+\N+n_\varkappa}$, the following bounds hold on $\mathbb{Y}_{1+\N+n_\varkappa}$:
\begin{description}
\item[a)]$d \geqslant 0$ or $\N+n_\varkappa=1$
\begin{equation}
|\Gamme^{\Lambda\Lambda_0;\Vec{\phi};w}_{\beta\Vec{\varkappa};l}(\Vec{p})| \leqslant (\Lambda + |\Vec{p}|)^{d} F^{\Lambda\Lambda_0}_{s_\beta}(\Vec{p})P^{\Lambda\Lambda_0}_{r_\beta}(\Vec{p})\,,
\end{equation}
\item[b)]$d<0$
\begin{equation}
|\Gamme^{\Lambda\Lambda_0;\Vec{\phi};w}_{\beta \Vec{\varkappa};l}(\Vec{p})| \leqslant \sum \limits_{\tau \in \mathcal{T}_{\beta \Vec{\varkappa} \Vec{\phi}}} Q^{\Lambda;w}_{\tau}(\Vec{p}) \, F^{\Lambda\Lambda_0}_{s_\beta}(\Vec{p})P^{\Lambda\Lambda_0}_{r_\beta}(\Vec{p})\,.
\end{equation}
\end{description}
Here $d:=3-2n_\varkappa-\N-\|w\|$ and $s_\beta:=0$. If $l=0$ then $r_\beta:=0$ otherwise $r_\beta$ stands for $r_\beta(d,l)$.
\begin{equation}
r_\beta(d,l):=\left\{\begin{matrix}2l,&d\geqslant 0\,,\\2l-1,&d<0\,.\end{matrix}\right.
\end{equation}
\end{thm}
We note that the 1-point vertex function with integrated insertion~\eqref{eq_11ga} vanishes, $\Gamme^{\Lambda \Lambda_0;c;w}_1=0$, e.g. due to SU(2) symmetry.
\begin{thm}\label{thm_4}Let be given a collection of vertex functions $\Gamme^{\Vec{\phi}}_{\Vec{\varkappa};l}$, $\Gamme^{\Vec{\phi}}_{\beta\Vec{\varkappa};l}$
as in theorem~\ref{thm_3}.
Let $\Gamme^{\Vec{\phi}}_{1,\Vec{\varkappa};l}$ be a collection of vertex functions with one integrated insertion of the appropriate operator among $Q_\rho^{\Lambda_0}$ \eqref{eq_328a}, $Q_{\rho\gamma}^{\Lambda_0}$ \eqref{eq_11rg}, $Q_{\rho\omega}^{\Lambda_0}$ \eqref{eq_11rw}, regular on $\mathbb{Y}_{\N+n_\varkappa}$, complying with the global symmetries of the theory, and satisfying the FE.
Assume that the STI \eqref{eq_21gh} and consistency conditions \eqref{eq_cohom} and \eqref{eq_404a} do hold.
Then, for all $\Vec{\phi}$, $\Vec{\varkappa}$, all $l\in\mathbb{N}$, $w \in \mathbb{W}_{\N+n_\varkappa}$, the following bounds hold on $\mathbb{Y}_{\N+n_\varkappa}$:
\begin{description}
\item[a)]$d>0$ or $\N+n_\varkappa=2$
\begin{equation}
|\Gamme^{\Lambda\Lambda_0;\Vec{\phi};w}_{1\Vec{\varkappa};l}(\Vec{p})| \leqslant (\Lambda + |\Vec{p}|)^d F^{\Lambda\Lambda_0}_{s_1}(\Vec{p})P^{\Lambda\Lambda_0}_{r_1}(\Vec{p})\,,
\end{equation}
\item[b)]$d \leqslant 0$
\begin{equation}
|\Gamme^{\Lambda\Lambda_0;\Vec{\phi};w}_{1\Vec{\varkappa};l}(\Vec{p})| \leqslant  \sum \limits_{\tau \in \mathcal{T}_{1\Vec{\varkappa}\Vec{\phi}}} Q^{\Lambda;w}_{\tau}(\Vec{p}) \, F^{\Lambda\Lambda_0}_{s_1}(\Vec{p})P^{\Lambda\Lambda_0}_{r_1}(\Vec{p})\,.
\end{equation}
\end{description}
Here $d:=5-2n_\varkappa-\N-\|w\|$. If $l=0$ then $r_1:=0$, $s_1:=0$ otherwise $r_1$, $s_1$ stand respectively for $r_1(d,l)$, $s_1(d,l)$. 
\begin{align}
r_1(d,l)&:=\left\{\begin{matrix}3l,&d> 0\,,\\3l-1,&d= 0\,,\\3l-2,&d< 0\,,\end{matrix}\right.&s_1(d,l)&:=\left\{\begin{matrix}l,&d \geqslant 0\,,\\l-1,&d < 0\,.\end{matrix}\right.
\end{align}
\end{thm}
\section{Proof of Theorems \ref{thm_1}--\ref{thm_4}} \label{sec_327b}
In this section we will prove theorems \ref{thm_1}, \ref{thm_2}, \ref{thm_3} and \ref{thm_4} in this order. We proceed by induction in the loop order $l$. We first verify that they hold at tree level $l=0$. Afterwards we assume that they hold true up to loop order $l-1\geqslant 0$, and we will verify the induction step from $l-1$ to $l$.

Put $D_X:=4$ for all vertex functions $\Gamme^{\Vec{\phi}}_{\Vec{\varkappa}}$. For all inserted functions $\Gamme^{\Vec{\phi};w}_{X \Vec{\varkappa}}$ with $X \in \{\beta,1\}$ and $\|w\| \leqslant w_{max}$ let
\begin{align}
D_X&:=\left\{\begin{matrix}3,&X=\beta\,,\\5,&X=1\,,\end{matrix}\right.&d_X&:=D_X - 2n_\varkappa -\N -\|w\|\,.
\end{align}
Note that at zero loop order $\Gamme^{\Lambda \Lambda_0}_{l=0}=\Gamme^{\Lambda_0 \Lambda_0}_{l=0}$. Using \eqref{eq_27j}, \eqref{eq_11ga} and the definition of $\mathcal{L}_{aux}$ in~\eqref{eq_17aux} one finds that in momentum space
\begin{align}
\Gamme^{\Lambda_0 \Lambda_0;\Vec{\phi}}_1&=Q^{\Lambda_0;\Vec{\phi}}_{\rho(0)},&\Gamme^{\Lambda_0 \Lambda_0;\Vec{\phi}}_{1\varkappa}&=Q^{\Lambda_0;\Vec{\phi}}_{\rho(0)\varkappa},&\Gamme^{\Lambda_0 \Lambda_0;\Vec{\phi}}_\beta&=Q^{\Lambda_0;\Vec{\phi}}_\beta\,,
\end{align}
where the momentum variable corresponding to the source $\rho$ is set to zero. Everywhere in the following $Q^{\Lambda_0}_{\rho}$ will stand for $Q^{\Lambda_0}_{\rho(0)}$. From definition of $Q^{\Lambda_0}_\rho$ in \eqref{eq_328a} it follows that the vertex functions $Q^{\Lambda_0;\Vec{\phi}}_{\rho;l=0}$ with $\N=2$ vanish. The nonvanishing functions $Q^{\Lambda_0;\Vec{\phi}}_{\rho;l=0}$, $Q^{\Lambda_0;\Vec{\phi}}_{\beta;l=0}$ have the form $h_s(p,q)(1-\sigma_{0\Lambda_0}(p^2))$ where $h_s$ is a homogeneous tensor polynomial of degree $s\leqslant 2$ in the momentum variables~$p,q \in \mathbb{R}^4$ which depends at most linearly on the momentum~$q$. From the definitions of $Q^{\Lambda_0}_{\rho\omega}$ \eqref{eq_11rw} and $Q^{\Lambda_0}_{\rho\gamma}$ \eqref{eq_11rg} we obtain that $Q^{\Lambda_0;\Vec{\phi}}_{\rho\varkappa;l=0}$ has the form $h_s(p)(\sigma_{0\Lambda_0}((p+q)^2)-\sigma_{0\Lambda_0}(p^2))$ with $s\leqslant 1$. For $\|w\|\leqslant s$ (relevant terms)  using inequalities~\eqref{eq_402t},~\eqref{eq_331b},~\eqref{eq_402s} we have 
\begin{equation}
|\Gamme^{\Lambda \Lambda_0;\Vec{\phi};w}_{X \Vec{\varkappa};l=0}(\Vec{p})|\leqslant c\,\Lambda^{s-\|w\|}_0 \Big(\frac{|\Vec{p}|}{\Lambda_0}\Big)^{s+1-\|w\|} \leqslant c\, \frac{|\Vec{p}|}{\Lambda_0} (|\Vec{p}| + \Lambda)^{s-\|w\|}\,.
\end{equation}
For $\|w\|>s$ (irrelevant terms) the same inequalities yield
\begin{align}
|\Gamme^{\Lambda \Lambda_0;\Vec{\phi};w}_{X \Vec{\varkappa};l=0}(\Vec{p})|&\leqslant c \,\Lambda^{s-\|w\|}_0 \Big(\frac{|\Vec{p}|}{\Lambda_0} + 1\Big)\nonumber\\
&\leqslant c\,(|\Vec{p}|+\Lambda)^{s-\|w\|}\,\frac{|\Vec{p}|+\Lambda}{\Lambda_0} \Big(\frac{|\Vec{p}|}{\Lambda_0}+1\Big)^{\|w\|-s}\,.
\end{align}
Since $s-\|w\|$ is the dimension $d_X$ we have the following bounds
\begin{align}
|\Gamme^{\Lambda \Lambda_0;\Vec{\phi};w}_{X \Vec{\varkappa};l=0}(\Vec{p})|&\leqslant  (\Lambda + |\Vec{p}|)^{d_X}  F^{\Lambda \Lambda_0}_0(\Vec{p}),&X & \in \{\beta,1\}.
\end{align}
Thus the statements of theorems \ref{thm_1}--\ref{thm_4} hold at loop number $l=0$. The proof proceeds by induction on $l$ and on the number of derivatives $\|w\|$, ascending in $l$ and, for fixed $l$, descending in $\|w\|$ from $w_{max}$ to 0.
\subsection{Chains of vertex functions}
\begin{definition}\label{def_split}A division in $m$ parts of a finite set $\mathbb{I}$  is a sequence $S:=(s_j)_{j \in [m]}$
of $m$ disjoint sets $s_j \subseteq \mathbb{I}$, possibly empty, such that $\bigcup_{j \in [m]} s_{j}=\mathbb{I}$.
An ordered partition is a division with all $s_j$ nonempty.
Given a division $S$ as above stated,
a division of a sequence
$\Vec{\Psi}=(\Psi_i)_{i \in \mathbb{I}}$
is the sequence of elements
$\Vec{\Psi}_j := (\Psi_i)_{i \in s_j}$, with $j\in[m]$.
\end{definition}
\begin{definition} \label{def_chain} Let $S$ be a division in $m$ parts of a finite set $\mathbb{I}$, and $\Vec{\Psi}=(\Psi_i)_{i\in \mathbb{I}}$ be a sequence of labels $\Psi_i\in\{A,c,\bar{c},\gamma,\omega,\beta,1\}$. Denote by $(\Vec{\Psi}_j)$ the division of $\Vec{\Psi}$ induced by $S$.
A \textbf{chain} of vertex functions is then defined by the expression
\begin{equation} 
\mathcal{F}^{\zeta_1 \Vec{\Psi} \bar{\zeta}_{m}}_{S}:= \Gamme^{\zeta_1 \Vec{\Psi}_1 \bar{\zeta}_1} \prod \limits^{m}_{j=2} \mathbf{C}_{\zeta_j \bar{\zeta}_{j-1}} \Gamme^{\zeta_j \Vec{\Psi}_j \bar{\zeta}_j}\,,
\end{equation}
where the repeated field labels $\zeta_j,\bar{\zeta}_j$ belong to $\{A, c, \bar{c}\}$ and are summed over (as usual).
\end{definition}
Using this definition the FE \eqref{eq_fl} has the form
\begin{equation}
\dot{\Gamme}^{\Vec{\Psi}}= \frac{\hbar}{2} \sum \limits_{S} (-)^{\pi_a}\langle \dot{C} \mathcal{F}^{A \Vec{\Psi} A}_S + \dot{S}(\mathcal{F}^{\bar{c} \Vec{\Psi} c}_S - \mathcal{F}^{c \Vec{\Psi} \bar{c}}_S) \rangle\,.
\end{equation}
The sum above runs over all possible divisions of $[0: n-1]$, $n$ being the number of components of $\Vec{\Psi}$. The symbol $\pi_a$ denotes the number of transpositions $mod\, 2$ of the anticommuting variables $\{c,\bar{c},\beta,\gamma, 1\}$ in the permutation $i \mapsto \pi(i)$ such that $(\Psi_{\pi(0)},...,\Psi_{\pi(n-1)})=\Vec{\Psi}_{1}\oplus...\oplus\Vec{\Psi}_{m}$, where $(a_1,...,a_p) \oplus (a_{p+1},...,a_q)=(a_1,...,a_q)$.

A preliminary step toward the proof of theorem \ref{thm_1} is to bound $\partial^w (\mathbf{C}_{\zeta \bar{\zeta}}\Gamme^{\zeta \bar{\zeta}})$ with $\|w\| \leqslant w_{max}$.
\begin{proposition}For all $0<k<l$, $0 \leqslant w \leqslant w_{max}$, $p \in \mathbb{R}^4$
\begin{equation}
\Big|\Big(\prod \limits^w_{i=0} \frac{\partial}{\partial p_{\mu_i}} \Big)\left( \Gamme^{\zeta \bar{\zeta};\Lambda\Lambda_0}_{k}(p) \mathbf{C}^{\Lambda \Lambda_0}_{\zeta \bar{\zeta}}(p) \right)\Big| \leqslant \frac{P^{\Lambda}_{2k}}{(|p|+\Lambda)^w}\,.\label{eq_16gc}
\end{equation}
\end{proposition}
\paragraph{Proof} Using inequality \eqref{eq_holder} we see that
\begin{equation}
|\partial^w (\Gamme^{\zeta \bar{\zeta};\Lambda\Lambda_0}_{k} \mathbf{C}^{\Lambda \Lambda_0}_{\zeta \bar{\zeta}})| \leqslant \sum \limits^{w}_{w_1=0}\frac{w!}{w_1! (w-w_1)!}|\partial^{w_1} \Gamme^{\zeta \bar{\zeta};\Lambda\Lambda_0}_{k}| |\partial^{w-w_1} \mathbf{C}^{\Lambda \Lambda_0}_{\zeta \bar{\zeta}}|\,.
\end{equation}
Setting $w_2=w-w_1$ it follows from \eqref{eq_16C} and the bounds of theorem \ref{thm_1} already proved inductively for $k<l$ that
\begin{equation}
\frac{1}{w_2!}|\partial^{w_1} \Gamme^{\zeta \bar{\zeta};\Lambda\Lambda_0}_{k}||\partial^{w_2} \mathbf{C}^{\Lambda \Lambda_0}_{\zeta \bar{\zeta}}|\leqslant \frac{c_\xi d^{w_2}  P^{\Lambda}_{2k}}{(|p|+\Lambda)^{w_1+w_2}}\,.
\end{equation}
Because $w\leqslant w_{max}$ the constants $c_\xi$, $d^{w_2}$ may be absorbed in $P^{\Lambda}_{2k}$.
\hfill$\blacksquare$
\begin{definition} \label{def_bchain}Let be given
\begin{enumerate}
\item a sequence $\Vec{\Psi}=(\Psi_i)_{i \in \mathbb{I}}$ as in Definition~\ref{def_chain};
\item an ordered partition $S=(s_j)_{j \in [m]}$ of  $\mathbb{I}$;
\item the sequences of field labels $(\zeta_j)_{j \in [m]}$, $(\bar{\zeta}_j)_{j \in [m]}$;
\item a multi-index $w \in \mathbb{W}_n$ and a sequence $\textsf{w}:=(\textsf{w}_j)_{j \in [m]}$ such that $\textsf{w}_j \in \mathbb{W}_n$ and $\sum_{j \in [m]} \textsf{w}_{j}=w$.
\end{enumerate}
Then, we define a \textbf{reduced chain} of vertex functions as
\begin{equation} 
\Gamme^{\zeta_1 \Vec{\Psi}_1 \bar{\zeta}_1;\textsf{w}_1} \prod \limits^{m}_{j=2} \mathbf{C}_{\zeta_j \bar{\zeta}_{j-1}} \Gamme^{\zeta_j \Vec{\Psi}_j \bar{\zeta}_j;\textsf{w}_j},\label{eq_bchain}
\end{equation}
where $\Gamme^{\zeta \Vec{\Psi}_{j} \bar{\zeta};\textsf{w}_j}=\partial^{\textsf{w}_j}(\Gamme^{\zeta \Vec{\Psi}_{j} \bar{\zeta}})$ are derivatives wrt the external momenta appearing in ${\dot{\Gamme}}^{\Vec{\Psi}}$, and the sequences $(\zeta_j)_{j \in [m]}$, $(\bar{\zeta}_j)_{j \in [m]}$ are \textbf{fixed}. Introducing the auxiliary quantities $\mathring{\Upsilon}:=(\Vec{\Upsilon}_j)_{j \in [m]}$ and $\Vec{\Upsilon}_j:=\zeta_j \Vec{\Psi}_j \bar{\zeta}_{j}$, we denote \eqref{eq_bchain} by $\mathcal{S}^{\mathring{\Upsilon};\textsf{w}}$ or,  with some abuse of notations, by $\mathcal{S}^{\zeta_1 \Vec{\Psi} \bar{\zeta}_m;\textsf{w}}_S$.
\end{definition}
The adjective "reduced" indicates that the chains contain neither $\Gamme^{AA}$, $\Gamme^{c \bar{c}}$ nor derivatives applied to the propagators $\mathbf{C}$. 

It follows from inequalities \eqref{eq_16C}, \eqref{eq_16gc} and theorems \ref{thm_1}--\ref{thm_4} proved in loop order $l-1$ that there exists a common bound for the terms $\Gamme^{\zeta \Vec{\Psi}\bar{\zeta};w_1 + w_2+ w_3}_{l_1 + l_2}\mathbf{C}$ and $\Gamme^{\zeta \Vec{\Psi}\bar{\zeta};w_1}_{l_1}\partial^{w_2}(\mathbf{C}\Gamme^{\zeta \bar{\zeta}}_{l_2})\partial^{w_3} \mathbf{C}$. This property basically follows from
an explicit check of the degree of the polynomials in the bounds
and from
the fact that, for $w_2,w_3$ supported on $p$, a factor $(|p|+\Lambda)^{-\|w_2+w_3\|}$
times a weighted tree $(\tau,\theta)$ in the bound of~$\Gamme^{\zeta \Vec{\Psi}\bar{\zeta};w_1}_{l_1}$
may be bounded by the weighted tree $(\tau,\theta')$ in the bound of~$\Gamme^{\zeta \Vec{\Psi}\bar{\zeta};w_1 + w_2+ w_3}_{l_1 + l_2}$,
with $\theta'$ equal to $\theta$ everywhere but on the edge $e\in E_1$ carrying
momentum~$p$, for which $\theta'(e)=\theta(e)+\|w_2+w_3\|$.
Hence to bound $\dot{\Gamme}^{\Vec{\Psi}}_{l}$ it is enough to consider a loop integral with a reduced chain
\begin{equation}
\int \frac{d^4k}{(2\pi)^4} \; \dot{\mathbf{C}}_{\zeta \bar{\zeta}}(k) \, \mathcal{S}^{\zeta \Vec{\Psi} \bar{\zeta};\textsf{w}}_{S;l-1}(k,\Vec{p}_S,-k), \label{eq_iloop}
\end{equation}
where $\Vec{p}_{s_j}=(p_i)_{ i\in s_j}$. As an example we give in Appendix \ref{ex_AAcc} the complete list of chains for $\Gamme^{AA c \bar{c}}$. The appellative "reduced" may be omitted in the following,
since it is always clear from the context whether a chain is reduced or not.
\subsection{Junction of weighted trees}\label{sec_junc}
Given a reduced chain $\mathcal{S}^{\mathring{\Upsilon};\textsf{w}}$ we define its amplitude $\hat{\mathcal{S}}^{\mathring{\Upsilon};\textsf{w}}$ 
by substituting the vertex functions  and propagators with their corresponding bounds taken from theorems~\ref{thm_1}--\ref{thm_4} and inequalities~\eqref{eq_16C}.
Recalling that $\Vec{\Upsilon}_j:=\zeta_j \Vec{\Psi}_j \bar{\zeta}_{j}$, we then set
\begin{align} 
\hat{\mathcal{S}}^{\Lambda;\mathring{\Upsilon};\textsf{w}}&:=\hat{S}^{\Lambda;\Vec{\Upsilon}_1;\textsf{w}_1} \prod \limits^{m}_{j=2} \frac{1}{
(\Lambda + |p_{\zeta_j}|)^2} \hat{S}^{\Lambda;\Vec{\Upsilon}_j;\textsf{w}_j}\label{eq_tchain}
\\
\hat{S}^{\Lambda;\Vec{\Upsilon}_j;w}&:=\left\{\begin{matrix}(\Lambda + |\Vec{p}|)^{d_X},&case\ a,\\\sum \limits_{\tau \in \mathcal{T}_{\Vec{\Upsilon}_j}} Q^{\Lambda;w}_{\tau},&case\ b,\end{matrix}\right.
\,.
\label{eq_hatp}
\end{align}
Here the cases $a$ and $b$ refer to the respective parts in theorems~\ref{thm_1}--\ref{thm_4}.

The tree structure of the bound
is spoiled if there exists an interval $\mathbb{J}^a:=[j^a:j^a+m^a-1] \subset [m]$ 
such that
all $\hat{S}^{\Lambda;\Vec{\Upsilon}_j;w}$ for $j\in\mathbb{J}^a$ correspond to a strictly relevant contribution, associated to the cases $a$ in the theorems. A workaround for this difficulty will start with the following definition. For every tree $\tau \in \mathbb{T}_{\Vec{\varphi}}$, set $E_{1;v}:=\{e \in E_1:e \mbox{ incident to } v\}$.
\begin{definition}\label{def_fgm}
Let $\Vec{\varphi}$ be an arbitrary sequence with $\varphi_i \in \{A,c,\bar{c},\beta,\gamma,\omega\}$. A tree
 $f \in \mathbb{T}_{\Vec{\varphi}}$
is a \textbf{fragment} if
\begin{itemize}
\item[a)]  there exists $s \in \{0,1\}$ s.t.~
$|V_\circ|+s$ equals the total number of ''*'' labels,
\item[b)] $\forall v \in V_\circ$: $|E_{1;v}|\geqslant 2$ and $E_{1;v}\cap E_* \not= \emptyset$.
\end{itemize}
The set of all such fragments is denoted by $\mathcal{F}^{(s)}_{\Vec{\varphi}}$. Moreover we set
$\mathcal{F}_{\Vec{\varphi}}:=\mathcal{F}^{(0)}_{\Vec{\varphi}}$
and
$\mathcal{F}_{1\Vec{\varphi}}:=\mathcal{F}^{(1)}_{\Vec{\varphi}}$.
\end{definition}
For each $s\in\{0,1\}$, $\mathcal{T}^{(s)}_{\Vec{\varphi}} \subset \mathcal{F}^{(s)}_{\Vec{\varphi}} \subset \mathbb{T}_{\Vec{\varphi}}$.

\smallskip Let us now state a useful property which relies on the above definition.
Let be given a subsequence $\mathbb{J}^a:=[j^a:j^a+m^a-1] \subset [m]$ with any number of elements $m^a$.
Let $S$ and $\Vec{\Psi}$ be as in definition~\ref{def_bchain}.
Define the following restrictions:
$S^a=S|_{\mathbb{J}^a}$, $\textsf{w}^a=\textsf{w}|_{\mathbb{J}^a}$.
Set
$w^a:=\sum_{j \in \mathbb{J}^a}\textsf{w}_j$
and
$\Vec{\Psi}^a:=(\Psi_i)$
with $i\in  \cup_{j \in \mathbb{J}^a} \;s_j$.
Then, there exists a set of fragments 
$\mathcal{F}_{\zeta \Vec{\Psi}^a \bar{\zeta}}$ such that
\begin{align}
\hat{\mathcal{S}}^{\Lambda;\zeta \Vec{\Psi}^a \bar{\zeta};\textsf{w}^a}_{S^a} &\leqslant 2^{\frac{|S_a|}{2}} \sum \limits_{f \in \mathcal{F}_{\zeta \Vec{\Psi}^a \bar{\zeta}}} Q^{\Lambda;w^a}_{f}\,. \label{eq_26f}
\end{align}
Here we give a proof of \eqref{eq_26f} for an example, generalisation is clear. Consider an amplitude $\hat{\mathcal{S}}^{\zeta_1 A^4 \bar{\zeta}_4;w}$ composed of four elements $\hat{S}^{\zeta_j A \bar{\zeta}_j;w}$ where $j \in \{1,...,4\}$ and $w=0$. First let us define the following set of fragments $\mathcal{F}$ (here $s=0$):\\
\null\hfill\begin{tabular}{cc}
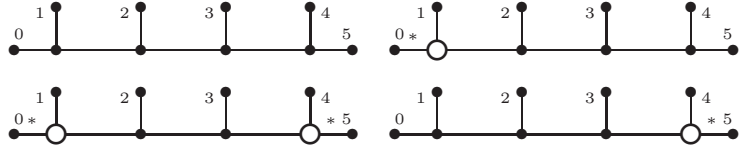
\begin{minipage}[c][32pt]{132pt}\centering\tiny\begin{picture}(128,16)
\put(0,0){\line(1,0){128}}
\put(0,0){\circle*{4}}
\put(128,0){\circle*{4}}
\multiput(16,0)(32,0){4}{
  \put(0,0){\line(0,1){16}}
  \put(0,16){\circle*{4}}
  \put(0,0){\circle*{4}}
}
\put(0,4){0}
\put(8,12){1}
\put(40,12){2}
\put(72,12){3}
\put(116,12){4}
\put(124,4){5}
\end{picture}\end{minipage}&\begin{minipage}[c][32pt]{132pt}\centering\tiny\begin{picture}(128,16)
\put(0,0){\line(1,0){128}}
\put(0,0){\circle*{4}}
\put(128,0){\circle*{4}}
\multiput(16,0)(32,0){4}{
  \put(0,0){\line(0,1){16}}
  \put(0,16){\circle*{4}}
  \put(0,0){\circle*{4}}
}
\put(0,4){0}
\put(8,12){1}
\put(40,12){2}
\put(72,12){3}
\put(116,12){4}
\put(124,4){5}
\put(16,0){\circle*{8}}
\put(16,0){\color{white}\circle*{6}}
\put(5,2){*}
\end{picture}\end{minipage}\\
\begin{minipage}[c][32pt]{132pt}\centering\tiny\begin{picture}(128,16)
\put(0,0){\line(1,0){128}}
\put(0,0){\circle*{4}}
\put(128,0){\circle*{4}}
\multiput(16,0)(32,0){4}{
  \put(0,0){\line(0,1){16}}
  \put(0,16){\circle*{4}}
  \put(0,0){\circle*{4}}
}
\put(0,4){0}
\put(8,12){1}
\put(40,12){2}
\put(72,12){3}
\put(116,12){4}
\put(124,4){5}
\put(16,0){\circle*{8}}
\put(16,0){\color{white}\circle*{6}}
\put(112,0){\circle*{8}}
\put(112,0){\color{white}\circle*{6}}
\put(118,2){*}
\put(5,2){*}
\end{picture}\end{minipage}&\begin{minipage}[c][32pt]{132pt}\centering\tiny\begin{picture}(128,16)
\put(0,0){\line(1,0){128}}
\put(0,0){\circle*{4}}
\put(128,0){\circle*{4}}
\multiput(16,0)(32,0){4}{
  \put(0,0){\line(0,1){16}}
  \put(0,16){\circle*{4}}
  \put(0,0){\circle*{4}}
}
\put(0,4){0}
\put(8,12){1}
\put(40,12){2}
\put(72,12){3}
\put(116,12){4}
\put(124,4){5}
\put(112,0){\circle*{8}}
\put(112,0){\color{white}\circle*{6}}
\put(118,2){*}
\end{picture}\end{minipage}
\end{tabular}\hfill\null\\
In each fragment the *-edge corresponds to a factor $\Lambda + |p_e|$ in the corresponding amplitude $Q^{\Lambda;w=0}_{f \in \mathcal{F}}$. One shows the following bound:
\begin{equation}
\mbox{\begin{minipage}[c][48pt]{164pt}\tiny\centering\begin{picture}(144,24)
\put(0,8){\line(1,0){144}}
\put(0,8){\circle*{4}}
\put(144,8){\circle*{4}}
\multiput(18,8)(36,0){4}{
  \put(0,0){\line(0,1){16}}
  \put(0,16){\circle*{4}}
  \put(0,0){\circle*{8}}
  \put(0,0){\color{white}\circle*{6}}
  \put(2,0){\line(-1,0){4}}
  \put(0,2){\line(0,-1){4}}
}
\multiput(14,8)(36,0){3}{
  \put(16,4){$|C|$}
}
\put(8,-5){$\hat{S}^{\zeta_1 A \bar{\zeta}_1}$}
\put(46,-5){$\hat{S}^{\zeta_2 A \bar{\zeta}_2}$}
\put(80,-5){$\hat{S}^{\zeta_3 A \bar{\zeta}_3}$}
\put(116,-5){$\hat{S}^{\zeta_4 A \bar{\zeta}_4}$}
\put(0,12){0}
\put(10,20){1}
\put(46,20){2}
\put(82,20){3}
\put(118,20){4}
\put(138,12){5}
\end{picture}\end{minipage}}\leqslant 2^{\frac{4}{2}} \sum \limits_{f \in \mathcal{F}} Q^{\Lambda;w=0}_{f}\,,
\end{equation}
or more explicitly:
\begin{equation}
\frac{\prod \limits_{v \in V_3} (\Lambda + |\Vec{p}_v|)}{ \prod \limits_{e \in E \backslash E_1} (\Lambda + |p_e|)^2} \leqslant 2^{\frac{|V_3|}{2}} \sum \limits_{\{\chi\}} \frac{\prod \limits_{v \in V_3} (\Lambda + |p_{\chi(v)}|)}{\prod \limits_{e \in E \backslash E_1} (\Lambda + |p_e|)^2}\,.\label{704a}
\end{equation}
Here the $\oplus$-vertices stand each for a corresponding $\hat{S}^{\zeta_j A \bar{\zeta}_j;w=0}$, see \eqref{eq_hatp}; the set of $\oplus$-vertices is identified with $V_3$ in~\eqref{704a}; $\Vec{p}_v$ indicates the set of incoming momenta of the vertex~$v$; $p_e$ denotes the momentum corresponding to an edge $e$; the sum runs over the set of functions $\chi:V_3 \to E\backslash E_1 \cup \{e_0,e_5\}$ which map every vertex $v \in V_3$ to an edge incident to $v$; the $|C|$'s stand for the usual bounds on the corresponding propagators.

Let be given a sequence of fragments $\Vec{f}:=(f_1,...,f_{m'})$ with $f_j\in\mathcal{F}_{\zeta \Vec{\Psi}_j \bar{\zeta}}$ and a sequence $\textsf{w}^\prime=(\textsf{w}^\prime_j)_{j \in [m^\prime]}$ with $\textsf{w}^\prime_j \in \mathbb{W}_n$. We define the amplitude $\hat{\mathcal{Q}}^{\Lambda;\textsf{w}^\prime}_{\Vec{f}}$ by
\begin{equation} 
\hat{\mathcal{Q}}^{\Lambda;\textsf{w}^\prime}_{\Vec{f}}:=\,Q^{\Lambda;\textsf{w}^\prime_1}_{f_1} \prod \limits^{m'}_{j=2} \frac{1}{(\Lambda + |p_{\zeta_j}|)^2} Q^{\Lambda;\textsf{w}^\prime_j}_{f_j}\,.
\end{equation}
\begin{lemma}Given an amplitude $\hat{\mathcal{Q}}^{\Lambda;\textsf{w}^\prime}_{\Vec{f}}$ as above there exists a fragment $f \in \mathcal{F}_{\zeta \Vec{\Psi}\bar{\zeta}}$ such that 
\begin{equation}
\hat{\mathcal{Q}}^{\Lambda;\textsf{w}^\prime}_{\Vec{f}} \leqslant Q^{\Lambda;\textsf{w}^\prime}_{f}\,. \label{eq_fgm}
\end{equation}
\end{lemma}
\paragraph{Proof}We proceed by induction in $m'$. If there are no joints, $m'=1$, the statement is evident. Assume it is true for some $m'-1 \geqslant 0$ and consider a sequence of $m'$ fragments. Let $v_l,v_r \in V_3$ be the left and right vertices of a joint. Recall definition in~\eqref{eq_pi}.
\begin{itemize}
\item $v_l,v_r \in V_3\backslash V_\circ$.
\begin{align}
\mbox{\begin{minipage}[c][32pt]{112pt}\centering\tiny\begin{picture}(96,32)
%
\put(0,0){\line(0,1){32}}
\put(0,16){\line(1,0){32}}
\put(96,0){\line(0,1){32}}
\put(96,16){\line(-1,0){32}}
\multiput(32,16)(8,0){4}{\line(1,0){4}}
%
\put(32,16){\circle*{4}}
\put(64,16){\circle*{4}}
\put(0,16){\circle*{4}}
\put(96,16){\circle*{4}}
%
\put(2,18){$v_l$}
\put(86,18){$v_r$}
\put(46,18){$C$}
\put(14,10){$\ubar{0}$}
\put(78,10){$\ubar{0}$}
\end{picture}\end{minipage}}&\leqslant\mbox{\begin{minipage}[c][32pt]{48pt}\centering\tiny\begin{picture}(32,32)
%
\put(0,0){\line(0,1){32}}
\put(0,16){\line(1,0){32}}
\put(32,0){\line(0,1){32}}
%
\put(32,16){\circle*{4}}
\put(0,16){\circle*{4}}
%
\put(2,18){$v_l$}
\put(22,18){$v_r$}
\put(14,10){$\ubar{2}$}
\end{picture}\end{minipage}}\\
\Pi^{\Lambda}_{f_l,\theta_l} \;\frac{1}{(\Lambda + |p|)^2} \;\Pi^{\Lambda}_{f_r,\theta_r}&= \Pi^{\Lambda}_{f,\theta}
\end{align}
Here the corresponding external edges have been merged together to form a new internal edge with $\theta$-weight equal to 2. The $\theta$-weight of all other edges is unchanged.
\item $v_l \in V_3\backslash V_\circ,v_r \in V_\circ$ or vice versa.
\begin{align}
\mbox{\begin{minipage}[c][38pt]{112pt}\centering\tiny\begin{picture}(96,32)
%
\put(0,0){\line(0,1){32}}
\put(0,16){\line(1,0){32}}
\put(96,0){\line(0,1){32}}
\put(96,16){\line(-1,0){32}}
\multiput(32,16)(8,0){4}{\line(1,0){4}}
%
\put(32,16){\circle*{4}}
\put(64,16){\circle*{4}}
\put(0,16){\circle*{4}}
\put(96,16){\circle*{8}}
\put(96,16){\color{white}\circle*{6}}
\put(96,32){\circle*{4}}
%
\put(2,18){$v_l$}
\put(84,18){$v_r$}
\put(46,18){$C$}
\put(14,10){$\ubar{0}$}
\put(76,18){*}
\put(76,10){$\ubar{0}$}
\end{picture}\end{minipage}}&\leqslant\mbox{\begin{minipage}[c][38pt]{48pt}\centering\tiny\begin{picture}(32,32)
%
\put(0,0){\line(0,1){32}}
\put(0,16){\line(1,0){32}}
\put(32,0){\line(0,1){32}}
%
\put(32,16){\circle*{4}}
\put(0,16){\circle*{4}}
\put(32,32){\circle*{4}}
%
\put(2,18){$v_l$}
\put(22,18){$v_r$}
\put(14,10){$\ubar{1}$}
\end{picture}\end{minipage}}\\
\Pi^{\Lambda}_{f_l,\theta_l} \;\frac{1}{(\Lambda +|p|)^2} \;(\Lambda + |p|) \Pi^{\Lambda}_{f_r,\theta_r}&=\Pi^{\Lambda}_{f,\theta}\\
\mbox{\begin{minipage}[c][38pt]{112pt}\centering\tiny\begin{picture}(96,32)
%
\put(0,0){\line(0,1){32}}
\put(0,16){\line(1,0){32}}
\put(96,0){\line(0,1){32}}
\put(96,16){\line(-1,0){32}}
\multiput(32,16)(8,0){4}{\line(1,0){4}}
%
\put(32,16){\circle*{4}}
\put(64,16){\circle*{4}}
\put(0,16){\circle*{4}}
\put(96,16){\circle*{8}}
\put(96,16){\color{white}\circle*{6}}
\put(96,32){\circle*{4}}
\put(96,0){\circle*{4}}
%
\put(2,18){$v_l$}
\put(84,12){$v_r$}
\put(46,18){$C$}
\put(14,10){$\ubar{0}$}
\put(78,10){$\ubar{0}$}
\put(90,24){*}
\end{picture}\end{minipage}}&\leqslant\mbox{\begin{minipage}[c][38pt]{48pt}\centering\tiny\begin{picture}(32,36)
%
\put(0,0){\line(0,1){32}}
\put(0,16){\line(1,0){32}}
\put(32,0){\line(0,1){32}}
%
\put(32,16){\circle*{8}}
\put(32,16){\color{white}\circle*{6}}
\put(0,16){\circle*{4}}
\put(32,32){\circle*{4}}
\put(32,0){\circle*{4}}
%
\put(2,18){$v_l$}
\put(20,12){$v_r$}
\put(26,24){*}
\put(14,10){$\ubar{2}$}
\end{picture}\end{minipage}}\\
\Pi^{\Lambda}_{f_l,\theta_l}\; \frac{1}{(\Lambda +|p|)^2}\; (\Lambda + |p_*|) \Pi^{\Lambda}_{f_r,\theta_r}  &=(\Lambda + |p_*|) \Pi^{\Lambda}_{f,\theta}
\end{align}
\item $v_l,v_r \in V_\circ$.
\begin{align}
\mbox{\begin{minipage}[c][40pt]{112pt}\centering\tiny\begin{picture}(96,32)
%
\put(0,0){\line(0,1){32}}
\put(0,16){\line(1,0){32}}
\put(96,0){\line(0,1){32}}
\put(96,16){\line(-1,0){32}}
\multiput(32,16)(8,0){4}{\line(1,0){4}}
%
\put(32,16){\circle*{4}}
\put(64,16){\circle*{4}}

\put(0,16){\circle*{8}}
\put(0,16){\color{white}\circle*{6}}

\put(96,16){\circle*{8}}
\put(96,16){\color{white}\circle*{6}}
\put(96,32){\circle*{4}}
\put(96,0){\circle*{4}}
\put(0,32){\circle*{4}}
%
\put(4,18){$v_l$}
\put(84,12){$v_r$}
\put(46,18){$C$}
\put(14,18){*}
\put(14,10){$\ubar{0}$}
\put(78,10){$\ubar{0}$}
\put(90,24){*}
\end{picture}\end{minipage}}&\leqslant\mbox{\begin{minipage}[c][40pt]{48pt}\centering\tiny\begin{picture}(32,36)
%
\put(0,0){\line(0,1){32}}
\put(0,16){\line(1,0){32}}
\put(32,0){\line(0,1){32}}
%
\put(32,16){\circle*{8}}
\put(32,16){\color{white}\circle*{6}}
\put(0,16){\circle*{4}}
\put(32,32){\circle*{4}}
\put(0,32){\circle*{4}}
\put(32,0){\circle*{4}}
%
\put(2,18){$v_l$}
\put(20,12){$v_r$}
\put(26,24){*}
\put(14,10){$\ubar{1}$}
\end{picture}\end{minipage}}\\
\Pi^{\Lambda}_{f_l,\theta_l} (\Lambda + |p|)\;\frac{1}{(\Lambda +|p|)^2}\; (\Lambda + |p_*|) \Pi^{\Lambda}_{f_r,\theta_r}  &=(\Lambda + |p_*|) \Pi^{\Lambda}_{f,\theta}\\
\mbox{\begin{minipage}[c][40pt]{112pt}\tiny\centering\begin{picture}(96,32)
%
\put(0,0){\line(0,1){32}}
\put(0,16){\line(1,0){32}}
\put(96,0){\line(0,1){32}}
\put(96,16){\line(-1,0){32}}
\multiput(32,16)(8,0){4}{\line(1,0){4}}
%
\put(32,16){\circle*{4}}
\put(64,16){\circle*{4}}

\put(0,16){\circle*{8}}
\put(0,16){\color{white}\circle*{6}}

\put(96,16){\circle*{8}}
\put(96,16){\color{white}\circle*{6}}
\put(96,32){\circle*{4}}
\put(0,32){\circle*{4}}
%
\put(4,18){$v_l$}
\put(84,18){$v_r$}
\put(46,18){$C$}
\put(14,18){*}
\put(14,10){$\ubar{0}$}
\put(76,18){*}
\put(76,10){$\ubar{0}$}
\end{picture}\end{minipage}}&\leqslant\mbox{\begin{minipage}[c][40pt]{48pt}\centering\tiny\begin{picture}(32,32)
%
\put(0,0){\line(0,1){32}}
\put(0,16){\line(1,0){32}}
\put(32,0){\line(0,1){32}}
%
\put(32,16){\circle*{4}}
\put(0,16){\circle*{4}}
\put(32,32){\circle*{4}}
\put(0,32){\circle*{4}}
%
\put(2,18){$v_l$}
\put(22,18){$v_r$}
\put(14,10){$\ubar{0}$}
\end{picture}\end{minipage}}\\
\Pi^{\Lambda}_{f_l,\theta_l} (\Lambda + |p|)\;\frac{1}{(\Lambda +|p|)^2} \;(\Lambda + |p|) \Pi^{\Lambda}_{f_r,\theta_r}  &=\Pi^{\Lambda}_{f,\theta}
\end{align}
\end{itemize}
Hence, by merging two fragments we can decrease the number of joints by one and then apply the induction hypothesis.
\hfill$\blacksquare$\\
In the simpler context of $\phi^4$ theory, a completely explicit description of the junction of trees can be found in~\cite{ric}.

According to equations \eqref{eq_frac}, \eqref{eq_plog} the loop integral in \eqref{eq_iloop} is bounded by the following expression
\begin{equation}
\int \dot{\mathbf{C}}^{\Lambda\Lambda_0}_{\zeta \bar{\zeta}} \hat{\mathcal{S}}^{\Lambda;\mathring{\Upsilon};\textsf{w}} \, \mathfrak{P}^{\Lambda}_{l-1} \leqslant \Lambda \,\hat{\mathcal{S}}^{\Lambda;\mathring{\Upsilon};\textsf{w}} \, \mathfrak{P}^{\Lambda}_{l-1}\Big|_{p_{\zeta}, p_{\bar{\zeta}}=0}\,,\label{eq_8l}
\end{equation}
\begin{equation}
\mathfrak{P}^{\Lambda}_{l}:=\left\{\begin{matrix}F^{\Lambda \Lambda_0}_{l} P^{\Lambda \Lambda_0}_{3l},&X=1,\\F^{\Lambda \Lambda_0}_0 P^{\Lambda \Lambda_0}_{2l},&X=\beta,\\P^{\Lambda\Lambda}_{2l},&otherwise,\end{matrix}\right.\label{ric_11}
\end{equation}
which follows directly from the definition of $r$, $r_X$, $s_X$ given in theorems \ref{thm_1}--\ref{thm_4}.

For any function $f$ depending on the variable $p_\zeta$, and other variables which we need not specify, we define $\mathcal{R}_{\zeta}: f(...,p_\zeta,...) \mapsto f(...,0,...)$ where $\mathcal{R}$~stands for restriction. Then we set $\mathcal{R}_{\zeta \bar{\zeta}}:=\mathcal{R}_{\zeta} \circ \mathcal{R}_{\bar{\zeta}}$.
\begin{proposition}For an amplitude $\hat{\mathcal{S}}^{\Lambda;\zeta \Vec{\Psi}\bar{\zeta};\textsf{w}}_S$ there exists a set of trees such that
\begin{equation}
\mathcal{R}_{\zeta \bar{\zeta}}(\hat{\mathcal{S}}^{\Lambda;\zeta \Vec{\Psi}\bar{\zeta};\textsf{w}}_S) \leqslant 2^{\frac{|S|}{2}} \sum_{\tau \in \mathcal{T}_{\zeta\Vec{\Psi}\bar{\zeta}}}  \mathcal{R}_{\zeta \bar{\zeta}}(Q^{\Lambda;w}_\tau)\,.\label{eq_8r}
\end{equation}
\end{proposition}
\paragraph{Proof}Using definition \eqref{eq_tchain} and inequalities \eqref{eq_26f}, \eqref{eq_fgm} we obtain that $\hat{\mathcal{S}}^{\Lambda;\zeta \Vec{\Psi}\bar{\zeta};\textsf{w}}_S$ is bounded by a sum of fragment amplitudes. It remains to show that for a fragment $f\in \mathcal{F}_{\zeta \Vec{\Psi}\bar{\zeta}}$ there exists a tree $\tau \in \mathcal{T}_{\zeta \Vec{\Psi}\bar{\zeta}}$ such that $\mathcal{R}_{\zeta \bar{\zeta}}(Q^{\Lambda;w}_{f}) \leqslant \mathcal{R}_{\zeta \bar{\zeta}}(Q^{\Lambda;w}_\tau)$. We denote by a double bar line the edges $\zeta,\bar{\zeta}$ and consider the case when $f \not \in \mathcal{T}_{\zeta \Vec{\Psi}\bar{\zeta}}$. Based on the inequality $\frac{\Lambda}{\Lambda + |p_\phi|}\leqslant 1$ we find
\begin{equation}
\mbox{\begin{minipage}[c][48pt]{96pt}\centering\tiny\begin{picture}(64,34)
%
\put(0,2){\line(1,0){64}}
\put(32,2){\line(0,1){32}}
\put(32,2){\line(0,1){32}}
\multiput(14,0)(2,0){2}{\thicklines\line(0,1){4}}
%
\put(32,34){\circle*{4}}
\put(0,2){\circle*{4}}
\put(32,2){\circle*{8}}
\put(32,2){\color{white}\circle*{6}}
%
\put(14,4){*}
\put(40,6){$(\ubar{1},\ubar{2})$}
\put(34,30){$\phi$}
\end{picture}\end{minipage}}\leqslant\mbox{\begin{minipage}[c][48pt]{96pt}\centering\tiny\begin{picture}(64,34)
%
\put(0,2){\line(1,0){64}}
\put(32,2){\line(0,1){32}}
\put(32,2){\line(0,1){32}}
\multiput(14,0)(2,0){2}{\thicklines\line(0,1){4}}
%
\put(32,34){\circle*{4}}
\put(0,2){\circle*{4}}
\put(32,2){\circle*{4}}
%
\put(34,30){$\phi$}
\put(40,6){$(\ubar{0},\ubar{1})$}
\end{picture}\end{minipage}}
\end{equation}
\hfill$\blacksquare$\\
At any loop order $l^\prime < l$ using the bound of theorem \ref{thm_1} and the inequality
\begin{equation}
1\leqslant \frac{1}{\Lambda + |p_\varkappa|}\left((\Lambda + |p_1|) + (\Lambda + |p_2|)\right),
\end{equation}
where $p_{\varkappa}+p_1 + p_2=0$, one realizes that for the marginal vertex function $\Gamme^{\Lambda\Lambda_0;\phi_1\phi_2}_{\varkappa;l^\prime}$ the following inequality holds
\begin{equation}
|\Gamme^{\Lambda\Lambda_0;\phi_1\phi_2}_{\varkappa;l^\prime}(\Vec{p})|\leqslant \frac{1}{\Lambda + |p_\varkappa|} \Big(\mbox{\begin{minipage}[c][26pt]{64pt}\tiny\centering\begin{picture}(48,26)
%
\put(0,2){\line(1,0){48}}
\put(24,2){\line(0,1){24}}
%
\put(24,24){\circle*{4}}
\put(0,2){\circle*{4}}
\put(24,2){\circle*{8}}
\put(24,2){\color{white}\circle*{6}}
\put(48,2){\circle*{4}}
%
\put(26,22){$\varkappa$}
\put(34,4){*}
\end{picture}\end{minipage}}+\mbox{\begin{minipage}[c][26pt]{64pt}\tiny\centering\begin{picture}(48,26)
%
\put(0,2){\line(1,0){48}}
\put(24,2){\line(0,1){24}}
%
\put(24,24){\circle*{4}}
\put(0,2){\circle*{4}}
\put(24,2){\circle*{8}}
\put(24,2){\color{white}\circle*{6}}
\put(48,2){\circle*{4}}
%
\put(26,22){$\varkappa$}
\put(8,4){*}
\end{picture}\end{minipage}}\Big) P^\Lambda_{r_\varkappa(l^\prime)}\,.
\end{equation}
Similarly, substituting the relevant terms $\Gamme^{\Lambda\Lambda_0;\Vec{\phi};w}_{\beta;l^\prime}$, $\N+\|w\|\leqslant 3$,  at loop number $l^\prime<l$ with the bounds of theorem \ref{thm_3} we have
\begin{align}
|\Gamme^{\Lambda\Lambda_0;\phi_1 \phi_2 \phi_3}_{\beta;l^\prime}|&\leqslant\mbox{\begin{minipage}[c][32pt]{64pt}\centering\tiny\begin{picture}(48,26)
%
\put(0,2){\line(1,0){48}}
\put(16,2){\line(0,1){24}}
\put(32,2){\line(0,1){24}}
%
\put(16,2){\circle*{4}}
\put(32,2){\circle*{4}}
\put(16,24){\circle*{4}}
\put(32,24){\circle*{4}}
\put(0,2){\circle*{4}}
\put(48,2){\circle*{4}}
%
\put(8,22){$\beta$}
\put(36,22){$\phi_1$}
\put(0,8){$\phi_2$}
\put(44,8){$\phi_3$}
\end{picture}
\end{minipage}}F^{\Lambda\Lambda_0}_{0}P^{\Lambda\Lambda_0}_{r_\beta(l^\prime)}\,,\\
|\Gamme^{\Lambda\Lambda_0;\phi_1 \phi_2}_{\beta;l^\prime}|&\leqslant\Big(\mbox{\begin{minipage}[c][32pt]{64pt}\tiny\centering\begin{picture}(48,26)
%
\put(0,2){\line(1,0){48}}
\put(24,2){\line(0,1){24}}
%
\put(24,24){\circle*{4}}
\put(0,2){\circle*{4}}
\put(24,2){\circle*{8}}
\put(24,2){\color{white}\circle*{6}}
\put(48,2){\circle*{4}}
%
\put(27,22){$\beta$}
\put(-2,8){$\phi_1$}
\put(44,8){$\phi_2$}
\put(8,4){*}
\end{picture}
\end{minipage}}+\mbox{\begin{minipage}[c][32pt]{64pt}\tiny\centering\begin{picture}(48,26)
%
\put(0,2){\line(1,0){48}}
\put(24,2){\line(0,1){24}}
%
\put(24,24){\circle*{4}}
\put(0,2){\circle*{4}}
\put(24,2){\circle*{8}}
\put(24,2){\color{white}\circle*{6}}
\put(48,2){\circle*{4}}
%
\put(27,22){$\beta$}
\put(-2,8){$\phi_1$}
\put(44,8){$\phi_2$}
\put(34,4){*}
\end{picture}
\end{minipage}}\Big)F^{\Lambda\Lambda_0}_{0}P^{\Lambda\Lambda_0}_{r_\beta(l^\prime)}\,,\\
|\partial \Gamme^{\Lambda\Lambda_0;\phi_1 \phi_2}_{\beta;l^\prime}|&\leqslant\mbox{\begin{minipage}[c][32pt]{64pt}\tiny\centering\begin{picture}(48,26)
%
\put(0,2){\line(1,0){48}}
\put(24,2){\line(0,1){24}}
%
\put(24,24){\circle*{4}}
\put(0,2){\circle*{4}}
\put(24,2){\circle*{8}}
\put(24,2){\color{white}\circle*{6}}
\put(48,2){\circle*{4}}
%
\put(27,22){$\beta$}
\put(-2,8){$\phi_1$}
\put(44,8){$\phi_2$}
\put(9,4){*}
\put(9,-4){$\ubar{1}$}
\end{picture}
\end{minipage}}F^{\Lambda\Lambda_0}_{0}P^{\Lambda\Lambda_0}_{r_\beta(l^\prime)}\,.
\end{align}
Furthermore using the bounds of theorem \ref{thm_4} which are assumed to be true for any loop order $l^\prime<l$, we obtain the following inequalities for strictly relevant terms $\Gamme^{\Lambda\Lambda_0 \Vec{\phi};w}_{1;l^\prime}$, $\N+\|w\|<5$,
\begin{align}|\Gamme^{\Lambda\Lambda_0;\phi_0 \phi_1 \phi_2 \phi_3}_{1;l^\prime}|&\leqslant\Big(\mbox{\begin{minipage}[c][32pt]{64pt}\tiny\centering\begin{picture}(48,26)
%
\put(0,2){\line(1,0){48}}
\put(16,2){\line(0,1){24}}
\put(32,2){\line(0,1){24}}
%
\put(16,2){\circle*{4}}
\put(32,2){\circle*{4}}
\put(16,24){\circle*{4}}
\put(32,24){\circle*{4}}
\put(0,2){\circle*{4}}
\put(48,2){\circle*{4}}
%
\put(4,22){$\phi_0$}
\put(18,10){*}
\put(36,22){$\phi_1$}
\put(0,8){$\phi_2$}
\put(44,8){$\phi_3$}
\end{picture}\end{minipage}}+\mbox{\begin{minipage}[c][32pt]{64pt}\tiny\centering\begin{picture}(48,26)
%
\put(0,2){\line(1,0){48}}
\put(16,2){\line(0,1){24}}
\put(32,2){\line(0,1){24}}
%
\put(16,2){\circle*{4}}
\put(32,2){\circle*{4}}
\put(16,24){\circle*{4}}
\put(32,24){\circle*{4}}
\put(0,2){\circle*{4}}
\put(48,2){\circle*{4}}
%
\put(4,22){$\phi_0$}
\put(34,10){*}
\put(36,22){$\phi_1$}
\put(0,8){$\phi_2$}
\put(44,8){$\phi_3$}
\end{picture}\end{minipage}}+...\Big)F^{\Lambda\Lambda_0}_{s_1(l^\prime)}P^{\Lambda\Lambda_0}_{r_1(l^\prime)},\\
|\Gamme^{\Lambda\Lambda_0;\phi_0 \phi_1 \phi_2}_{1;l^\prime}|&\leqslant\Big(\mbox{\begin{minipage}[c][32pt]{64pt}\tiny\centering\begin{picture}(48,26)
%
\put(0,2){\line(1,0){48}}
\put(24,2){\line(0,1){24}}
%
\put(24,24){\circle*{4}}
\put(0,2){\circle*{4}}
\put(24,2){\circle*{8}}
\put(24,2){\color{white}\circle*{6}}
\put(48,2){\circle*{4}}
%
\put(27,22){$\phi_0$}
\put(-2,8){$\phi_1$}
\put(44,8){$\phi_2$}
\put(8,4){**}
\end{picture}\end{minipage}}+\mbox{\begin{minipage}[c][32pt]{64pt}\tiny\centering\begin{picture}(48,26)
%
\put(0,2){\line(1,0){48}}
\put(24,2){\line(0,1){24}}
%
\put(24,24){\circle*{4}}
\put(0,2){\circle*{4}}
\put(24,2){\circle*{8}}
\put(24,2){\color{white}\circle*{6}}
\put(48,2){\circle*{4}}
%
\put(27,22){$\phi_0$}
\put(-2,8){$\phi_1$}
\put(44,8){$\phi_2$}
\put(9,4){*}
\put(34,4){*}
\end{picture}\end{minipage}}+...\Big)F^{\Lambda\Lambda_0}_{s_1(l^\prime)}P^{\Lambda\Lambda_0}_{r_1(l^\prime)},\\
|\partial\Gamme^{\Lambda\Lambda_0;\phi_0 \phi_1 \phi_2}_{1;l^\prime}|&\leqslant\Big(\mbox{\begin{minipage}[c][32pt]{64pt}\tiny\centering\begin{picture}(48,26)
%
\put(0,2){\line(1,0){48}}
\put(24,2){\line(0,1){24}}
%
\put(24,24){\circle*{4}}
\put(0,2){\circle*{4}}
\put(24,2){\circle*{8}}
\put(24,2){\color{white}\circle*{6}}
\put(48,2){\circle*{4}}
%
\put(27,22){$\phi_0$}
\put(-2,8){$\phi_1$}
\put(44,8){$\phi_2$}
\put(8,4){**}
\put(9,-4){$\ubar{1}$}
\end{picture}\end{minipage}}+\mbox{\begin{minipage}[c][32pt]{64pt}\tiny\centering\begin{picture}(48,26)
%
\put(0,2){\line(1,0){48}}
\put(24,2){\line(0,1){24}}
%
\put(24,24){\circle*{4}}
\put(0,2){\circle*{4}}
\put(24,2){\circle*{8}}
\put(24,2){\color{white}\circle*{6}}
\put(48,2){\circle*{4}}
%
\put(27,22){$\phi_0$}
\put(-2,8){$\phi_1$}
\put(44,8){$\phi_2$}
\put(33,4){**}
\put(34,-4){$\ubar{1}$}
\end{picture}\end{minipage}}\Big)F^{\Lambda\Lambda_0}_{s_1(l^\prime)}P^{\Lambda\Lambda_0}_{r_1(l^\prime)},\\
|\Gamme^{\Lambda\Lambda_0;\phi_0 \phi_1;w}_{1;l^\prime}\mathbf{C}^{\Lambda \Lambda_0}|&\leqslant \frac{\Lambda + |p|}{(\Lambda + |p|)^{\|w\|}}F^{\Lambda\Lambda_0}_{s_1(l^\prime)}P^{\Lambda\Lambda_0}_{r_1(l^\prime)}\,,
\end{align}
where the dots stand for omitted fragments obtained by permuting the "*" over the external edges. Substituting the bounds for $\Gamma^{\Vec{\phi};w}_{1\varkappa}$ with those for $\partial_{\bar{c}} \Gamma^{\bar{c}\Vec{\phi};w}_{1}$ yields a similar result for the relevant terms $\Gamma^{\Vec{\phi};w}_{1,\Vec{\varkappa}}$ with $n_\varkappa>0$. Consequently the amplitude $\hat{S}^{\Lambda;\Vec{\phi};w}_{1\Vec{\varkappa}}$ is bounded by a sum of amplitudes~$Q^{\Lambda;w}_{f}$ with $f \in \mathcal{F}_{1 \Vec{\varphi}}$.
\subsection{Irrelevant terms}
The irrelevant terms at arbitrary $0<\Lambda\leqslant \Lambda_0$ are reconstructed by using the FE:
\begin{equation}
\Gamme^{\Lambda \Lambda_0;\Vec{\phi};w}_{X\Vec{\varkappa};l}(\Vec{p})=\Gamme^{\Lambda_0 \Lambda_0;\Vec{\phi};w}_{X\Vec{\varkappa};l}(\Vec{p}) + \int \limits^{\Lambda}_{\Lambda_0} d \lambda  \, \dot{\Gamme}^{\lambda \Lambda_0;\Vec{\phi};w}_{X\Vec{\varkappa};l}(\Vec{p})\,.
\end{equation}
At $\Lambda=\Lambda_0$ for all loop orders $l > 0$, for all $n_\varkappa \geqslant 0$ and $d_X < 0$ we have
\begin{equation}
|\Gamme^{\Lambda_0 \Lambda_0;\Vec{\phi};w}_{X\Vec{\varkappa};l}(\Vec{p})|\leqslant \left\{\begin{matrix}(|\Vec{p}|+\Lambda_0)^{d_X} F^{\Lambda_0 \Lambda_0}_{s_X}P^{\Lambda_0 \Lambda_0}_{r_X}&X \in \{\beta,1\},\\0&otherwise,\end{matrix}\right.
\end{equation}
where the upper inequality is obtained using the bounds on relevant terms from theorem \ref{thm_1}.

To integrate the FE from the boundary $\Lambda_0$ to $\Lambda$ we substitute the chains $\mathcal{S}^{\zeta \phi_0 \phi_1 \bar{\zeta}}$, $\mathcal{S}^{\zeta \varkappa \phi \bar{\zeta}}$ with the bounds given in theorem \ref{thm_1} and use inequalities~\eqref{eq_twin}, \eqref{eq_twin2}. Eventually we get
\begin{align}
|\Gamme^{\Lambda\Lambda_0;\phi_0 \phi_1;w}_l(p)|&\leqslant (\Lambda+|p|)^{2-\|w\|} P^{\Lambda}_{2l-1},&\|w\|&> 2\,,\\
|\Gamme^{\Lambda\Lambda_0;\phi;w}_{\varkappa;l}(p)|&\leqslant (\Lambda+|p|)^{1-\|w\|} P^{\Lambda}_{2l-1},&\|w\|&> 1\,.
\end{align}
In a similar way substitution of $\mathcal{S}^{\zeta \beta \phi \bar{\zeta}}$, $\mathcal{S}^{\zeta \phi_0 \phi_1 \bar{\zeta}}_1$ with the bounds of theorems~\ref{thm_3},~\ref{thm_4} and then integration from $\Lambda_0$ to $\Lambda$, using \eqref{eq_int4a}, give
\begin{align}
|\Gamme^{\Lambda \Lambda_0;\phi;w}_{\beta;l}(p)-\Gamme^{\Lambda_0 \Lambda_0;\phi;w}_{\beta;l}(p)|&<\frac{F^{\Lambda\Lambda_0}_0 P^{\Lambda \Lambda_0}_{2l-1}}{(\Lambda + |p|)^{\|w\|-2}},&\|w\|&>2\,,\\
|\Gamme^{\Lambda \Lambda_0;\phi_0 \phi_1;w}_{1;l}(p)-\Gamme^{\Lambda_0 \Lambda_0;\phi_0 \phi_1;w}_{1;l}(p)|&<\frac{F^{\Lambda\Lambda_0}_{l-1} P^{\Lambda \Lambda_0}_{3l-2}}{(\Lambda + |p|)^{\|w\|-3}},&\|w\|&>3\,.
\end{align}
\begin{proposition}\label{prop_r}Let $\tau_i \in \mathcal{T}_{\zeta \bar{\zeta} \Vec{\phi}}$, $\phi_j \in \{A,c,\bar{c},\gamma,\omega\}$, $|V_1|>4$, $k \in \mathbb{N}$ and $\theta(\tau_i)>2$. Then $\exists \tau_f \in \mathcal{T}_{\Vec{\phi}}$ such that
\begin{equation}
\int \limits^{\Lambda_0}_\Lambda d\lambda \, \lambda \; \mathcal{R}_{\zeta \bar{\zeta}}(Q^{\lambda;w}_{\tau_i}) \; P^{\lambda}_{k}|_{\eta=0} \leqslant  Q^{\Lambda;w}_{\tau_f} P^{\Lambda}_{k+1}. \label{eq_16irr}
\end{equation}
\end{proposition}
\paragraph{Proof} Denote by $v,\bar{v} \in V \backslash V_1$ the vertices incident to $e_\zeta, e_{\bar{\zeta}} \in E_1$. First we assume that $v \neq \bar{v}$. The result of the restriction $\mathcal{R}_\zeta(Q^{\lambda;w}_{\tau_i})$ is the amplitude of the tree $\tau_i$ with two edges $e_1=\{u_1,v\}$, $e_2=\{u_2,v\}$ carrying opposite momenta. Here $\{u,v\}$ denotes the edge which links the vertices $u$, $v$. Without restriction we assume that $e_1=\chi(v)$. Furthermore we define a subtree $\tau^\prime$ of the initial tree $\tau_i$ by the substitution $e_2 \mapsto \{u_1,u_2\}$ such that it does not include the vertices $v$, $\zeta$ nor the edges $e_1$, $e_\zeta$.
\begin{description}
\item  $\rho(e_1)=1$.\\The identity $p_{e_1}=-p_{e_2}$ implies
\begin{equation}
\lambda \frac{1}{(\lambda + |p_{e_1}|)^{\theta(e_1)}} \leqslant \frac{1}{(\lambda + |p_{e_1}|)^{\sigma(e_1)}}=\frac{1}{(\lambda + |p_{e_2}|)^{\sigma(e_1)}}.
\end{equation}
Consequently, $\lambda \mathcal{R}_\zeta (Q^{\lambda;w}_{\tau_i})$ is bounded by the amplitude of the subtree $\tau^\prime $ with $\sigma^\prime(e_2)=\sigma(e_2)+\sigma(e_1)$.
\item $\rho(e_1)=0$ and $\rho(e_2) > 0$.
\begin{equation}
\lambda \frac{1}{(\lambda + |p_{e_2}|)^{\theta(e_2)}} \leqslant \frac{1}{(\lambda + |p_{e_2}|)^{\theta(e_2)-1}}.
\end{equation}
Thus, $\lambda \mathcal{R}_\zeta(Q^{\lambda;w}_{\tau_i})$ is bounded by the amplitude of the subtree $\tau^\prime$ with $\sigma^\prime(e_2)=\sigma(e_2)+\sigma(e_1)$, $\rho^\prime(e_2)=\rho(e_2)-1$ and $\chi^\prime(u_1)=e_2$.
\item $\rho(e_1)=0$ and $\rho(e_2)=0$.\\This implies $e_2 \in E_1$. Because $|V_1|>4$ the vertex $u_1$ is incident to an edge $e_u \in E \backslash E_1$ such that $\rho(e_u)>0$.
\begin{equation}
\lambda \frac{1}{(\lambda + |p_{e_u}|)^{\theta(e_u)}} \leqslant \frac{1}{(\lambda + |p_{e_u}|)^{\theta(e_u) - 1}}.
\end{equation}
It follows that $\lambda \mathcal{R}_\zeta Q^{\lambda;w}_{\tau_i}$ is bounded by the amplitude of the subtree $\tau^\prime$ with $\sigma^\prime(e_2)=\sigma(e_2)+\sigma(e_1)$, $\rho^\prime(e_u)=\rho(e_u)-1$ and $\chi^\prime(u_1)=e_u$.
\end{description}
For the subtree $\tau^\prime$ the total weight satisfies $\theta(\tau^\prime) > 1$. Consequently, we repeat the above reduction by substituting $\lambda$-multiplication with integration. Then using inequalities \eqref{eq_twin}, \eqref{eq_twin2} we have
\begin{equation}
\int \limits^{\Lambda_0}_\Lambda \frac{d \lambda \, P^{\lambda}_{k}|_{\eta=0}}{(\lambda + |p_a|)(\lambda + |p_b|)}  \leqslant \frac{P^{\Lambda}_{k+1}}{\Lambda + |p_a| + |p_b|}\leqslant \frac{P^{\Lambda}_{k+1}}{\Lambda + |\hat{p}|}\,, \label{eq_irr}
\end{equation}
where $\hat{p} \in \{p_a,p_b,p_a + p_b\}$. Here $p_a$ denotes the momentum of the edge used for the reduction, and $\hat{p}=p_b$ is the momentum of an arbitrary edge with nonvanishing $\theta$-weight of the final tree $\tau_f$. Because of the previous edge reduction we can reach the threshold $|V^\prime_1|=4$. In this case we can choose freely $\hat{p}$ to equal $p_a$, $p_b$ or $p_a + p_b$ thus keeping the $\sigma$-weight on the corresponding edge, see \eqref{eq_twin} and \eqref{eq_twin2}.

The case $v=\bar{v}$ follows directly from \eqref{eq_twin}, \eqref{eq_twin2}. To see this we denote by $e \in E \backslash E_1$ the edge incident to $v$ and introduce a vertex $u\in V_3$ adjacent to~$v$. Hence, $\chi(v)=e=\{v,u\}$. If $\tilde{\theta}$, $\theta$ are two $\theta$-weights where the only difference is that $\chi(u)=e$ for the first and $\chi(u) \not = e$ for the second, then
\begin{equation}
\Pi^{\lambda}_{\tau_i,\tilde{\theta}}\leqslant \Pi^\lambda_{\tau_i,\theta}=\frac{1}{\lambda} \frac{1}{\lambda + p_{e \in E(\tau_f)}}\Pi^\lambda_{\tau_f,\theta}\,.
\end{equation}
The final tree $\tau_f$ is a subtree of $\tau_i$ which does not include vertices $\zeta$,$\bar{\zeta}$,$v$ nor the corresponding edges. Moreover, $\theta(\tau_i)>2 \implies \theta(\tau_f)>0$.\hfill$\blacksquare$

Inequality \eqref{eq_16irr} can be applied to bound $\mathcal{S}^{\Lambda \Lambda_0;\Vec{\varkappa}\Vec{\phi};w}$. In this case $\tau_i \in \mathcal{T}_{\zeta \bar{\zeta}\Vec{\varkappa} \Vec{\phi}}$ and the total weight satisfies $\theta(\tau_i)>2-n_\varkappa$. But in theorem~\ref{thm_1} for each edge $e_\varkappa \in E_\varkappa$ we have $(\Lambda + |p_\varkappa|)$ as a denominator which is equivalent to an additional $\sigma$-weight of the edge $e_\varkappa$. An effective tree $\tilde{\tau}$ with $\tilde{\sigma}(e_\varkappa)=\sigma(e_\varkappa)+1$ has $\theta(\tilde{\tau})>2$ and satisfies the conditions of proposition \ref{prop_r}.

Using \eqref{eq_int4a} a similar inequality follows for the irrelevant functions $\Gamme^{\Lambda \Lambda_0;\Vec{\phi};w}_{\beta\Vec{\varkappa};l}$. For any $\tau_i  \in \mathcal{T}_{\zeta\bar{\zeta} \beta\Vec{\varkappa}\Vec{\phi}}$ there exists $\tau_f  \in \mathcal{T}_{\beta\Vec{\varkappa}\Vec{\phi}}$ such that
\begin{equation}
\int \limits^{\Lambda_0}_\Lambda d\lambda \, \lambda \; \mathcal{R}_{\zeta \bar{\zeta}}(Q^{\lambda;w}_{\tau_i}) \; F^{\lambda \Lambda_0}_0 P^{\lambda \Lambda_0}_{2(l-1)}|_{\eta=0} \leqslant  Q^{\Lambda;w}_{\tau_f} F^{\Lambda \Lambda_0}_0 P^{\Lambda\Lambda_0}_{2l-1}\,.
\end{equation}
Before application of \eqref{eq_int4a} to the irrelevant terms $\Gamme^{\Lambda \Lambda_0;\Vec{\phi};w}_{1\Vec{\varkappa};l}$  we need a minor change in \eqref{eq_irr}
\begin{equation}
\int \limits^{\Lambda_0}_\Lambda d \lambda \left(\frac{\lambda + |p_*|}{\lambda + |p_e|}\right)\frac{F^{\lambda\Lambda_0}_{s} P^{\lambda \Lambda_0}_{k}|_{\eta=0}}{(\lambda + |p_a|)(\lambda + |p_b|)} \leqslant \left(\frac{\Lambda + |p^\prime_*|}{\Lambda + |p_e|}\right) \frac{F^{\Lambda\Lambda_0}_{s} P^{\Lambda\Lambda_0}_{k+1}}{\Lambda + |p_b|}\,,
\end{equation}
where $p^\prime_*$ is one  out of $p_*$ or $p_e$ in such way that $|p^\prime_*|:=\max(|p_*|,|p_e|)$. If $p_*=p_e$ the label "*" is moved to edge $e$. Finally, $\forall \tau_i  \in \mathcal{T}_{1\zeta\bar{\zeta} \Vec{\varkappa}\Vec{\phi}}$
\begin{equation}
\int \limits^{\Lambda_0}_\Lambda d\lambda \, \lambda \; \mathcal{R}_{\zeta \bar{\zeta}}(Q^{\lambda;w}_{\tau_i}) \; F^{\lambda \Lambda_0}_{l-1} P^{\lambda \Lambda_0}_{3(l-1)}|_{\eta=0} \leqslant  \sum \limits_{\tau \in \mathcal{T}_{1\Vec{\varkappa}\Vec{\phi}}} Q^{\Lambda;w}_{\tau} F^{\Lambda \Lambda_0}_{l-1}P^{\Lambda\Lambda_0}_{3l-2}\,.
\end{equation}
\subsection{Marginal terms}
\begin{lemma}Let $\Gamme^{\Lambda\Lambda_0;\Vec{\phi};w}_{\Vec{\varkappa};l}$ denote a marginal term, $n_\varkappa \leqslant 1$. Then $\forall \Lambda<\eta(\Vec{p})$
\begin{equation}
|\Gamme^{\Lambda\Lambda_0;\Vec{\phi};w}_{\Vec{\varkappa};l}(\Vec{p})-\Gamme^{\eta \Lambda_0;\Vec{\phi};w}_{\Vec{\varkappa};l}(\Vec{p})|\leqslant  P^{\Lambda}_{2l-2}(\Vec{p}).\label{eq_flw0}
\end{equation}
\end{lemma}
\paragraph{Proof} Note that in theorem \ref{thm_1} for all $l^\prime < l$ the bounds for $\Gamme^{\Lambda\Lambda_0;\Vec{\phi};w}_{\varkappa;l^\prime}$ are more restrictive than the ones for $\partial_{\bar{c}}\Gamme^{\Lambda\Lambda_0;\bar{c}\Vec{\phi};w}_{l^\prime}$. So we will only treat  the case $n_\varkappa=0$ explicitly. In this case we integrate the FE from $\eta$ to~$\Lambda$

\begin{align}
|\Gamme^{\Lambda\Lambda_0;\Vec{\phi};w}_{l}(\Vec{p})-\Gamme^{\eta \Lambda_0;\Vec{\phi};w}_{l}(\Vec{p})|\leqslant& \int \limits^{\eta}_\Lambda d\lambda | \dot{\Gamme}^{\lambda\Lambda_0;\Vec{\phi}}_{l}(\Vec{p})|\nonumber \\
\leqslant& \sum \limits_{\tau \in \mathcal{T}_{\zeta \bar{\zeta} \Vec{\phi}}}\int \limits^{\eta}_\Lambda d\lambda \, \lambda \, \mathcal{R}_{\zeta \bar{\zeta}}(Q^{\lambda;w}_{\tau}) \, P^{\lambda}_{2l-2}|_{\eta_\tau=0},
\end{align}
where $\theta(\tau)=2$. Denoting by $q_a$, $q_b$ the momenta of the edges with nonvanishing $\theta$-weight and using equations \eqref{eq_int3},\eqref{eq_int4} for each term of the sum we obtain
\begin{equation}
\int \limits^{\eta}_\Lambda d\lambda \,  \frac{\lambda \, P^{\lambda}_{k}|_{\eta_\tau=0}}{(\lambda + |q_a|)(\lambda + |q_b|)} \leqslant \int \limits^{\eta}_\Lambda d\lambda \, \frac{P^{\lambda}_{k}|_{\eta_\tau=0}}{\lambda + |q_b|} \leqslant P^{\Lambda}_{k},\quad k=2l-2.
\end{equation}
\hfill$\blacksquare$
\subsubsection{$\Gamme^{c\bar{c}AA}$ and $\Gamme^{c\bar{c}c\bar{c}}$} The renormalization condition is $\Gamme^{M\Lambda_0;\Vec{\phi}}_l(0)=0$. For $p_2,p_3 \in \mathbb{R}^4$ and $\Lambda^\prime:=\max(\Lambda,\eta(\Vec{p}))$ equation \eqref{eq_anti} gives $\Gamme^{\Lambda^\prime \Lambda_0;\Vec{\phi}}_l(0_{\bar{c}},p_2,p_3)=0$, where the subscript $\bar{c}$ indicates the  momentum of the antighost.
\begin{equation}
|\Gamme^{\Lambda^\prime \Lambda_0;\Vec{\phi}}_l(\Vec{p})|\leqslant \int \limits^1_0 dt \, |p_{\bar{c}}|\,|\partial_{\bar{c}}  \Gamme^{\Lambda^\prime \Lambda_0;\Vec{\phi}}_l(tp_{\bar{c}},p_2,p_3)|\,.
\end{equation}
Substituting $|\partial_{\bar{c}}  \Gamme^{\Lambda^\prime \Lambda_0;\Vec{\phi}}_l|$ with the bound of theorem \ref{thm_1} and using inequality~\eqref{eq_int7} we obtain
\begin{equation}
|\Gamme^{\Lambda^\prime \Lambda_0;\Vec{\phi}}_l(\Vec{p})| \leqslant \Big(1+\log_+\frac{|\Vec{p}|}{\Lambda^\prime + \eta}\Big)P^{\Lambda^\prime\Lambda^\prime}_{2l-1}(\Vec{p})\Big|_{\eta=0} \leqslant P^{\Lambda}_{2l}(\Vec{p})\,.
\end{equation}
If $\Lambda^\prime=\Lambda$ the proof is finished. Otherwise we use \eqref{eq_flw0}.
\subsubsection{$\partial_A \Gamme^{c\bar{c}A}$} The renormalization condition is $\partial_A \Gamme^{M\Lambda_0;c \bar{c} A}_l(0)=0$. For $p_A \in \mathbb{R}^4$ and $\Lambda^\prime=\max(\Lambda,\eta(\Vec{p}))$ equation \eqref{eq_anti} gives $\partial_A \Gamme^{\Lambda^\prime \Lambda_0;c \bar{c} A}_l(0,p_A)=0$.
\begin{equation}
|\partial_A \Gamme^{\Lambda^\prime \Lambda_0;c\bar{c}A}_l(\Vec{p})|\leqslant \int \limits^1_0 dt \, |p_{\bar{c}}|\, |\partial_{\bar{c}}  \partial_A \Gamme^{\Lambda^\prime \Lambda_0;c \bar{c} A}_l(tp_{\bar{c}},p_A)|\,.
\end{equation}
We substitute $|\partial_{\bar{c}}  \partial_A \Gamme^{\Lambda^\prime \Lambda_0;c \bar{c} A}_l(\Vec{p}_t)|$ with the bound from theorem \ref{thm_1} with the choice $w^\prime(2)=(0,1,0)$, see after \eqref{eq_16bnd} for the definition of $w^\prime$. Then
\begin{equation}
Q^{\Lambda^\prime;(0,1,1)}_{c\bar{c}A} \in \left\{\frac{1}{\Lambda^\prime + t|p_{\bar{c}}|},\frac{1}{\Lambda^\prime + |tp_{\bar{c}} + p_A|}\right\}\,,
\end{equation}
and using \eqref{eq_int7} we obtain the inequality
\begin{equation}
|\partial_A \Gamme^{\Lambda^\prime \Lambda_0;c\bar{c}A}_l(\Vec{p})|\leqslant P^{\Lambda^\prime\Lambda^\prime}_{2l}(\Vec{p})\leqslant P^{\Lambda}_{2l}(\Vec{p})\,.
\end{equation}
 If $\Lambda^\prime=\Lambda$ the proof is finished. If not we use \eqref{eq_flw0}.
\subsubsection{Renormalization at $\Lambda=0$} \label{sec_20r}
First we consider the marginal terms $\Gamme^{c c}_{\omega;l}$, $\Gamme^{c A}_{\gamma;l}$, $\Gamme^{AAAA}_l$, $\partial_{\bar{c}} \Gamme^{c\bar{c}A}_l$,  $\partial \Gamme^{AAA}_l$, $\partial\partial \Gamme^{AA}_l$,  $\partial \partial \Gamme^{c \bar{c}}_l$, here denoted by $\Gamme^{\Vec{\phi};w}_{\Vec{\varkappa};l}$. The marginal terms $\Gamme^{\Vec{\phi}}_{1\Vec{\varkappa};l}$ and $\Gamme^{\Vec{\phi}}_{\beta;l}$ will be discussed later.

Let $\{\delta^s\ubar{e}\}_r$ be a basis at the renormalization point $\Vec{q}$. We define $\Gamme^{\Vec{\phi};w}_{\Vec{\varkappa};l}(\Vec{q})$  in the following way:
\begin{equation}
\Gamme^{0 \Lambda_0;\Vec{\phi};w}_{\Vec{\varkappa};l}(\Vec{q}):=\sum \limits_{t \in \{\delta^s\}_r} r_t\, t + \sum \limits_{t \alpha \in \{\delta^s \ubar{e}^{k>0}\}_r} \zeta_t \,t\,,
\end{equation}
where the coefficients $r_t$ are fixed by the renormalization conditions, see appendix~\ref{sec_ren} and hypothesis~\ref{rc3}, and the remaining coefficients $\zeta_t$ are defined using lemma \ref{lem_pr}.  Then, from the bounds on the irrelevant terms and lemma \ref{lem_rn}, see \eqref{eq_73}, it follows that $\Gamme^{\Vec{\phi};w}_{\Vec{\varkappa};l}(\Vec{q})$ complies with theorem \ref{thm_1} at loop order $l$. Let $\Lambda^\prime:=\max(\Lambda,\eta(\Vec{p}))$. It is easy to verify the following inequalities
\begin{align}
\int \limits^{\Lambda^\prime}_0 d\lambda \frac{\log^k_+ \frac{M}{\lambda}}{\lambda + M} &< k! + \log_+ \frac{\Lambda^\prime}{M},& \int \limits^{\Lambda^\prime}_0 d\lambda \frac{\log^k_+ \frac{\lambda}{M}}{\lambda +M}&<2\Big(1+ \log^{k+1}_+ \frac{\Lambda^\prime}{M}\Big)\,.\label{eq_int01}
\end{align}
Recalling \eqref{eq_8l}, \eqref{eq_8r} we obtain the following bound
\begin{equation}
|\dot{\Gamme}^{\lambda\Lambda_0;\Vec{\phi};w}_{\Vec{\varkappa};l}(\Vec{q})|\leqslant  \lambda\frac{\mathcal{P}^{(0)}_{2l-2}(\log_+ \frac{M}{\lambda})+\mathcal{P}^{(1)}_{2l-2}(\log_+(\frac{\lambda}{M}))}{(\lambda + M)^2}\,.\label{eq_int02}
\end{equation}
This implies that
\begin{equation}
|\Gamme^{\Lambda^\prime \Lambda_0;\Vec{\phi};w}_{\Vec{\varkappa};l}(\Vec{q})-\Gamme^{0\Lambda_0;\Vec{\phi};w}_{\Vec{\varkappa};l}(\Vec{q})|\leqslant \int \limits^{\Lambda^\prime}_{0} d\lambda \; |\dot{\Gamme}^{\lambda\Lambda_0;\Vec{\phi};w}_{\Vec{\varkappa};l}(\Vec{q})| \leqslant \mathcal{P}^{(1)}_{2l-1}\Big(\log_+ \frac{\Lambda}{M}\Big).
\end{equation}
Using inequality \eqref{eq_int7} we get
\begin{align}
|\Gamme^{\Lambda^\prime\Lambda_0;\Vec{\phi};w}_{\Vec{\varkappa};l}(0)-\Gamme^{\Lambda^\prime\Lambda_0;\Vec{\phi};w}_{\Vec{\varkappa};l}(\Vec{q})|&\leqslant \sum^{n-1}_{j=1} I_j(\Vec{q})\leqslant P^{\Lambda^\prime}_{2l}(0),\\
|\Gamme^{\Lambda^\prime\Lambda_0;\Vec{\phi};w}_{\Vec{\varkappa};l}(\Vec{p})-\Gamme^{\Lambda^\prime\Lambda_0;\Vec{\phi};w}_{\Vec{\varkappa};l}(0)|&\leqslant \sum^{n-1}_{j=1} I_j(\Vec{p})\leqslant P^{\Lambda}_{2l}(\Vec{p}),
\end{align}
where $I_j(\Vec{q})$ is the interpolation along the vector $q_j$,
\begin{equation}
I_j(\Vec{q})=\int \limits^1_0 dt \, |q_j|\, |\partial_j \Gamme^{\Lambda^\prime\Lambda_0;\Vec{\phi};w}_{\Vec{\varkappa};l}(\sum^{j-1}_{i=1}\Vec{q}_i+t\Vec{q}_j)|. \label{eq_path}
\end{equation}
Here to each vector $q_i\in \mathbb{R}^4$ is associated $\Vec{q}_i\in \mathbb{P}_n$ whose components are $(\Vec{q}_i)_{k}=-q_i \delta_{k,0} + q_i \delta_{k,i}$. Once again we have substituted $|\partial_j \Gamme^{M\Lambda_0;\Vec{\phi};w}_{\Vec{\varkappa};l}|$ with the bound from theorem \ref{thm_1}. If $\Lambda=\Lambda^\prime$ we stop here. If not we use \eqref{eq_flw0}.

The remaining marginal terms $\Gamme^{\Lambda \Lambda_0;\Vec{\phi};w}_{X \Vec{\varkappa};l}$ with $X \in \{\beta,1\}$ can be treated similarly.  Note that the bound from theorem \ref{thm_1} for terms of the type $\partial_{\bar{c}} \Gamme^{\bar{c} \Vec{\phi};w}_{1;l}$ is the same as the one for the corresponding terms $\Gamme^{\Vec{\phi};w}_{1\varkappa;l}$. Consequently the proof of the bounds for the marginal terms $\Gamme^{ccA}_{1\gamma;l}$, $\Gamme^{ccc}_{1\omega;l}$, $\partial \Gamme^{cc}_{1\gamma;l}$  is the same as the proof for respectively  $\partial_{\bar{c}} \Gamme^{\bar{c}ccA}_{1;l}$, $\partial \partial_{\bar{c}} \Gamme^{\bar{c}cc}_{1;l}$ which we shall consider now.

Let us denote by $\Gamme^{0\Lambda_0;\Vec{\phi};w}_{X;l}$
any marginal term without insertions $\gamma$ or $\omega$,
and by $\Vec{q}$ the  corresponding renormalization point as
given in appendix~\ref{sec_ren}. We now anticipate the important fact that the relevant renormalization constants comply with the bounds, which will be proven in section \ref{sec_sti}. Then using the bounds on irrelevant terms from theorems~\ref{thm_3}, \ref{thm_4} and lemma \ref{lem_rn}, see~\eqref{eq_73},  we obtain
\begin{equation}
|\Gamme^{0\Lambda_0;\Vec{\phi};w}_{X;l}(\Vec{q})|\leqslant \frac{M}{\Lambda_0} \mathcal{P}^{(1)}_{r_X}\Big(\log_+ \frac{\Lambda_0}{M}\Big)\,,
\end{equation}
in agreement with theorems \ref{thm_3}, \ref{thm_4}. Set $\Lambda^\prime:=\max(\Lambda,M)$. To integrate the FE from $0$ to $\Lambda^\prime$ we substitute the chain of vertex functions with the trees from theorems~\ref{thm_3},~\ref{thm_4}. Then we have
\begin{equation}
|\Gamme^{\Lambda^\prime \Lambda_0;\Vec{\phi};w}_{X,l}(\Vec{q})-\Gamme^{0\Lambda_0;\Vec{\phi};w}_{X,l}(\Vec{q})| \leqslant \frac{\Lambda + M}{\Lambda_0}\mathcal{P}^{(1)}_{r_X}\Big(\log_+ \frac{\Lambda_0}{M}\Big).\label{eq_413a}
\end{equation}
Using inequality \eqref{eq_int7} we obtain
\begin{align}
|\Gamme^{\Lambda^\prime\Lambda_0;\Vec{\phi};w}_{X;l}(0)-\Gamme^{\Lambda^\prime\Lambda_0;\Vec{\phi};w}_{X;l}(\Vec{q})|&\leqslant \frac{\Lambda + M}{\Lambda_0}P^{\Lambda^\prime \Lambda_0}_{r_X}(0),\label{eq_413b}\\
|\Gamme^{\Lambda^\prime\Lambda_0;\Vec{\phi};w}_{X;l}(0)-\Gamme^{\Lambda^\prime\Lambda_0;\Vec{\phi};w}_{X;l}(\Vec{p})|&\leqslant F^{\Lambda \Lambda_0}_{s_X}(\Vec{p})P^{\Lambda \Lambda_0}_{r_X}(\Vec{p}).\label{eq_413c}
\end{align}
Note here that as compared to the case $X=\beta$, in the case $X=1$ there appears an additional factor
\begin{equation}
\frac{\Lambda^\prime + |p_*|}{\Lambda^\prime + |p_a|}\leqslant 1+2\frac{|\Vec{p}|}{\Lambda + M}\,,
\end{equation}
in the tree bounds from theorem \ref{thm_4}. This factor leads to the polynominal $\mathcal{P}^{(2)}_{s_1}$ in the bounds of theorem \ref{thm_4}.

If $\Lambda=\Lambda^\prime$ the proof is finished. If not we integrate downwards using the FE, substitute the chain with the tree bound from theorems \ref{thm_3}, \ref{thm_4} and use~\eqref{eq_int4b} to get
\begin{align}
|\Gamme^{\Lambda \Lambda_0;\Vec{\phi};w}_{\beta;l}(\Vec{p})-\Gamme^{M \Lambda_0;\Vec{\phi};w}_{\beta;l}(\Vec{p})|&\leqslant F^{\Lambda \Lambda_0}_{0}(\Vec{p}) P^{\Lambda \Lambda_0}_{2l-1}(\Vec{p})\,,\\
|\Gamme^{\Lambda \Lambda_0;\Vec{\phi};w}_{1;l}(0)-\Gamme^{M\Lambda_0;\Vec{\phi};w}_{1;l}(\Vec{p})|&\leqslant \sum \limits_{\tau \in \mathcal{T}_{1\Vec{\phi}}} Q^{\Lambda;w}_\tau\, F^{\Lambda \Lambda_0}_{l-1} P^{\Lambda \Lambda_0}_{3l-2}(\Vec{p})\,.
\end{align}
\subsection{Strictly relevant terms}
If $n_\varkappa=0$ the notation $\Gamme^{\Vec{\phi};w}_{\Vec{\varkappa};l}$ stands for $\Gamme^{c\bar{c}A}_l$,  $\Gamme^{AAA}_l$, $\partial \Gamme^{AA}_l$,  $\partial \Gamme^{c \bar{c}}_l$. In the case $n_\varkappa=1$, it stands for $\Gamme^c_{\gamma;l}$. Moreover we impose $\Gamme^{0\Lambda_0\Vec{\phi};w}_{\Vec{\varkappa};l}(0)=0$ and denote by $\Vec{p}$ arbitrary momenta with corresponding $\eta(\Vec{p})$. We integrate the FE upwards from 0 to $\Lambda$ and substitute $|\dot{\Gamme}^{\lambda\Lambda_0\Vec{\phi};w}_{\Vec{\varkappa};l}|$ with the tree bound of theorem~\ref{thm_1}. Then
\begin{equation*}
|\Gamme^{\Lambda\Lambda_0\Vec{\phi};w}_{\Vec{\varkappa};l}(0)|\leqslant \int \limits^\Lambda_0 d\lambda \; |\dot{\Gamme}^{\lambda\Lambda_0\Vec{\phi};w}_{\Vec{\varkappa};l}(0)|\leqslant \int \limits^{\Lambda}_0 d\lambda \, \lambda^{d-1}P^{\lambda \Lambda}_{2l-2}(0)\leqslant \int \limits^{\Lambda + \eta}_0 d\lambda \, \lambda^{d-1}P^{\lambda \Lambda}_{2l-2}(0)\,,
\end{equation*}
where $d>0$. Inequality \eqref{eq_int5} then gives
\begin{equation}
|\Gamme^{\Lambda\Lambda_0\Vec{\phi};w}_{\Vec{\varkappa};l}(0)|\leqslant (\Lambda + \eta)^d P^{\Lambda+\eta\;\Lambda}_{2l-2}(0). \label{eq_111}
\end{equation}
Substituting $\partial \Gamme^{\Lambda\Lambda_0\Vec{\phi};w}_{\Vec{\varkappa};l}$ with the bound from theorem \ref{thm_1} we obtain
\begin{align}
|\Gamme^{\Lambda\Lambda_0\Vec{\phi};w}_{\Vec{\varkappa};l}(\Vec{p})-\Gamme^{\Lambda\Lambda_0\Vec{\phi};w}_{\Vec{\varkappa};l}(0)|\leqslant& \int \limits^1_0 dt \, |p_i| |\partial_i \Gamme^{\Lambda\Lambda_0\Vec{\phi};w}_{\Vec{\varkappa};l}(t\Vec{p})|\nonumber\\
\leqslant& |\Vec{p}| \int \limits^1_0 dt \, (\Lambda + t|\Vec{p}|)^{d-1} P^{\Lambda}_{2l}(t\Vec{p}) \leqslant I+I^\prime,\label{eq_31228}
\end{align}
where
\begin{align}
I:=& |\Vec{p}| (\Lambda + |\Vec{p}|)^{d-1}\int \limits^1_0 dt \,  \mathcal{P}^{(0)}_{2l}(\log_+ \frac{\max(M, |\Vec{p}|)}{\Lambda + t\eta(\Vec{p})}),\\
I^\prime:=&(\Lambda + |\Vec{p}|)^{d} \, \mathcal{P}^{(1)}_{2l}(\log_+ \frac{\Lambda}{M}).
\end{align}
The calculations given in \eqref{eq_int5} yield
\begin{equation}
I\leqslant |\Vec{p}|(\Lambda + |\Vec{p}|)^{d-1} \, \mathcal{P}^{(0)}_{2l}(\log_+ \frac{\max(M, |\Vec{p}|)}{\Lambda + \eta(\Vec{p})}).
\end{equation}
This implies
\begin{equation}
|\Gamme^{\Lambda\Lambda_0\Vec{\phi};w}_{\Vec{\varkappa};l}(\Vec{p})-\Gamme^{\Lambda\Lambda_0\Vec{\phi};w}_{\Vec{\varkappa};l}(0)| \leqslant  (\Lambda + |\Vec{p}|)^{d}P^{\Lambda}_{2l}(\Vec{p}). \label{eq_112}
\end{equation}
Combining  \eqref{eq_111} and \eqref{eq_112} proves the bounds of theorem \ref{thm_1} to loop order $l$.
\subsubsection{$ \partial \Gamme^{\phi^3}_1$, $\Gamme^{\phi^4}_1$ and $\Gamme^{cc}_{1\gamma}$}
The goal of this section is to explain the expression for the polynomial degree~$r_1$. Subsequently we denote by $\Gamme^{\Vec{\phi};w}_1$ the following terms: $\partial \Gamme^{c AA}_1$, $\partial \Gamme^{c c \bar{c}}_1$, $\Gamme^{c AAA}_1$, $\Gamme^{c Ac \bar{c}}_1$ and we impose vanishing renormalization conditions at the origin, $\Gamme^{0 \Lambda_0;\Vec{\phi};w;}_1(0)=0$. From the bounds of theorem~\ref{thm_4} one realises that the analysis for $\Gamma^{cc}_{1\gamma}$ is similar to $\partial_{\bar{c}}\Gamme^{cc \bar{c}}_1$. For an arbitrary $\Vec{p}\in \mathbb{P}_n$ let $\Lambda^\prime:=\max(\Lambda,\eta(\Vec{p}))$. Using \eqref{eq_8l}, \eqref{ric_11} 
\begin{align}
|\Gamme^{\Lambda^\prime \Lambda_0\Vec{\phi};w}_{1;l}(0)|\leqslant&(\Lambda + \eta) \, \frac{\Lambda + M}{\Lambda_0}\, P^{\Lambda+\eta\;\Lambda_0}_{3(l-1)}(0).
\end{align}
We extend $\Gamme^{\Lambda^\prime \Lambda_0;\Vec{\phi};w}_{1;l}$ from 0 to $\Vec{p}$ using the usual path given in equation \eqref{eq_path} (with $\Vec{q} \mapsto \Vec{p}$), where we substitute $\partial \Gamme^{\Lambda^\prime \Lambda_0\Vec{\phi};w}_{1;l}$ with the tree bound of theorem \ref{thm_4}. The rational factor in this bound makes these terms different from other strictly relevant terms, for example $\partial \partial \Gamme^{c A}_{1}$, $\Gamme^{cA}_{\beta}$ or $\partial \Gamme^{c}_{\beta}$. Then using inequality \eqref{eq_int7} we have
\begin{equation*}
|\Gamme^{\Lambda^\prime \Lambda_0;\Vec{\phi};w}_{1;l}(\Vec{p})-\Gamme^{\Lambda^\prime \Lambda_0;\Vec{\phi};w}_{1;l}(0)|\leqslant (\Lambda + |\Vec{p}|)\Big(1 + \log_+ \frac{|\Vec{p}|}{\Lambda + \eta}\Big)F^{\Lambda \Lambda_0}_l(\Vec{p})\, P^{\Lambda \Lambda_0}_{3l-1}(\Vec{p}).
\end{equation*}
If $\Lambda=\Lambda^\prime$ the proof is finished. If not we repeat the steps used to show \eqref{eq_flw0} integrating the FE from $\eta$ to $\Lambda$ and using \eqref{eq_int4}
\begin{equation*}
|\Gamme^{\Lambda \Lambda_0;\Vec{\phi};w}_{1;l}(\Vec{p})-\Gamme^{\eta \Lambda_0;\Vec{\phi};w}_{1;l}(\Vec{p})|\leqslant |\Vec{p}|\, F^{\Lambda \Lambda_0}_{l-1}(\Vec{p})\, P^{\Lambda \Lambda_0}_{3l-2}(\Vec{p}).
\end{equation*}
\subsubsection{$ \Gamme^{\Vec{\phi};w}_\beta$, $\Gamme^{\phi^2;w}_1$  and $ \Gamme^{\phi^3}_1$}
In this section we briefly discuss the remaining strictly relevant terms $\Gamme^{\Vec{\phi};w}_\beta$, $\Gamme^{\phi^3}_1$ and $\Gamme^{\phi^2;w}_1$ denoting all of them by $\Gamme^{\Vec{\phi};w}_X$ with $X \in \{\beta,1\}$. We impose renormalization conditions $\Gamme^{0 \Lambda_0;\Vec{\phi};w}_X(0)=0$ and integrate the FE from 0 to~$\Lambda$. We use the bounds of theorems \ref{thm_3},\ref{thm_4} and then for an arbitrary $\Vec{p}$ extend the integration up to $\Lambda + \eta(\Vec{p})$
\begin{equation}
|\Gamme^{\Lambda  \Lambda_0;\Vec{\phi};w}_X(0)|\leqslant (\Lambda+\eta)^{d_X}\frac{\Lambda+M}{\Lambda_0}P^{\Lambda+\eta \;\Lambda_0}_{r_X}(0)\,.
\end{equation}
Integration along the path given in equation \eqref{eq_path} and using \eqref{eq_int5} yields
\begin{equation}
|\Gamme^{\Lambda  \Lambda_0;\Vec{\phi};w}_X(\Vec{p})-\Gamme^{\Lambda  \Lambda_0;\Vec{\phi};w}_X(0)|\leqslant (\Lambda + |\Vec{p}|)^{d_X} F_{s_X}(\Vec{p}) P^{\Lambda \Lambda_0}_{s_X}(\Vec{p}).
\end{equation}
\subsection{Convergence}\label{sec_conv}
We first prove the bounds for $\partial_{\Lambda_0}\Gamme^{\Vec{\phi}}$. Then we proceed with the other functions $\partial_{\Lambda_0} \Gamme^{\Vec{\phi}}_{\Vec{\varkappa}}$ ascending in the number of insertions $n_\varkappa$. We use the same inductive scheme as before, based on the FE.

We start with the irrelevant terms integrating the FE from $\Lambda_0$ to $\Lambda$, using the boundary conditions $\Gamme^{\Lambda_0 \Lambda_0;\Vec{\phi};w}_l=0$ and applying the derivative wrt~$\Lambda_0$.
\begin{equation}
\partial_{\Lambda_0} \Gamme^{\Lambda \Lambda_0;\Vec{\phi};w}_{\Vec{\varkappa};l}= - \dot{\Gamme}^{\Lambda_0 \Lambda_0;\Vec{\phi};w}_{\Vec{\varkappa};l} + \int \limits^\Lambda_{\Lambda_0} d \lambda \,\partial_{\Lambda_0} \dot{\Gamme}^{\lambda \Lambda_0;\Vec{\phi};w}_{\Vec{\varkappa};l}.
\end{equation}
To bound the first term of the expression we substitute into the FE the irrelevant tree bound
\begin{equation}
|\dot{\Gamme}^{\Lambda_0 \Lambda_0;\Vec{\phi};w}_{\Vec{\varkappa};l}|\leqslant \frac{1}{\Lambda_0}\sum_{\tau \in \mathcal{T}_{\Vec{\varkappa}\Vec{\phi}}} Q^{\Lambda_0;w}_\tau P^{\Lambda_0 \Lambda_0}_{2(l-1)}.
\end{equation}
If $n_\varkappa=0$ then $\tau \in \mathcal{T}_{\Vec{\phi}}$ and $\theta(\tau) >0$. Consequently, recalling \eqref{eq_pi}
\begin{equation}
\Pi^{\Lambda_0}_{\tau,\theta}(\Vec{p}) \leqslant \frac{\Lambda + |\Vec{p}|}{\Lambda_0}\Pi^{\Lambda}_{\tau,\theta}(\Vec{p}), \quad\mbox{and thus}\quad Q^{\Lambda_0;w}_\tau \leqslant \frac{\Lambda + |\Vec{p}|}{\Lambda_0} Q^{\Lambda;w}_\tau.
\end{equation}
Otherwise, the denominator $\Lambda_0 + |p_e|$ with $e_\varkappa \in E_\varkappa$ gives the inequality
\begin{equation}
\frac{1}{\Lambda_0 + |p_\varkappa|}\leqslant \frac{\Lambda + |\Vec{p}|}{\Lambda_0} \frac{1}{\Lambda + |p_\varkappa|}.
\end{equation}
In both cases this yields
\begin{equation}
|\dot{\Gamme}^{\Lambda_0 \Lambda_0;\Vec{\phi};w}_{\Vec{\varkappa};l}|\leqslant \frac{\Lambda + |\Vec{p}|}{\Lambda^2_0} \sum_{\tau \in \mathcal{T}_{\Vec{\varkappa}\Vec{\phi}}} Q^{\Lambda;w}_\tau P^{\Lambda \Lambda_0}_{2(l-1)}.
\end{equation}
To analyse the second term we apply $\partial^w_p\partial_{\Lambda_0}$ to the chain of vertex functions given in definition \ref{def_chain}. This gives a chain with the element $\partial_{\Lambda_0}((\partial^{w_1}_p \mathbf{C})\Gamme^{;w_2}_{l^\prime})$ $l^\prime < l$ which we bound using \eqref{eq_dS3}, \eqref{eq_16C} and theorem \ref{thm_2}.
\begin{equation*}
|\partial_{\Lambda_0}((\partial^{w_1}_p \mathbf{C})\Gamme^{\lambda \Lambda_0;\zeta\Vec{\phi}\bar{\zeta};w_2}_{\Vec{\varkappa};l^\prime})|\leqslant \,|\partial_{\Lambda_0}\partial^{w_1}_p\mathbf{C}|\,|\Gamme^{\lambda \Lambda_0;\zeta\Vec{\phi}\bar{\zeta};w_2}_{\Vec{\varkappa};l^\prime}|+|\partial^{w_1}_p\mathbf{C}|\,|\partial_{\Lambda_0}\Gamme^{\lambda \Lambda_0;\zeta\Vec{\phi}\bar{\zeta};w_2}_{\Vec{\varkappa};l^\prime}|
\end{equation*}
\begin{equation*}
\leqslant \frac{c}{(\lambda+|p_{\zeta}|)^{2 + \|w_1\|}} \Big(\frac{\lambda + |p_{\zeta}| }{\Lambda^2_0}|\Gamme^{\lambda \Lambda_0;\zeta\Vec{\phi}\bar{\zeta};w_2}_{\Vec{\varkappa};l^\prime}|+|\partial_{\Lambda_0}\Gamme^{\lambda \Lambda_0;\zeta\Vec{\phi}\bar{\zeta};w_2}_{\Vec{\varkappa};l^\prime}|\Big)
\end{equation*}
\begin{equation}
\leqslant \frac{1}{(\lambda+|p_\zeta|)^{2}} \frac{\lambda +M+ |\Vec{p}| }{\Lambda^2_0}\sum \limits_{\tau \in \mathcal{T}_{\Vec{\varkappa}\zeta \Vec{\phi} \bar{\zeta}}} Q^{\lambda; w_2 + w_1}_\tau P^{\lambda \Lambda_0}_{r(l^\prime)}(\Vec{p}) \,.
\end{equation}
We proceed as for the proof of inequality \eqref{eq_16irr} substituting expression \eqref{eq_irr} with the integral
\begin{equation}
\int \limits^{\Lambda_0}_\Lambda d\lambda \frac{(\lambda + M+|\Vec{p}|) P^{\lambda\Lambda_0}_{2l-2}|_{\eta_\tau=0}}{\Lambda^2_0(\lambda+|p_a|)(\lambda + |p_b|)}\leqslant \frac{\Lambda + M+ |\Vec{p}|}{\Lambda^2_0}\frac{1}{\Lambda + |\hat{p}|}P^{\Lambda\Lambda_0}_{2l-1}(\Vec{p}),
\end{equation}
see the explanation after \eqref{eq_irr}. If $\tau \in \mathcal{T}_{\varkappa\phi \phi}$ we  always choose $\hat{p}=p_\varkappa$ in order to preserve the factor $\Lambda + |p_\varkappa|$ in the denominator of $Q^{\Lambda}_{\tau}$.

For the marginal terms we shall integrate the FE upwards from 0 to $\Lambda$. For the terms with the antighost we use renormalization conditions \eqref{eq_16cdt1}
\begin{align}
\Gamme^{M\Lambda_0;c\bar{c}AA}_l(0)&=0,&\Gamme^{M\Lambda_0;c\bar{c} c \bar{c}}_l(0)&=0,&\partial_A \Gamme^{M\Lambda_0;c \bar{c}A}_l(0)&=0.
\end{align}
Using equation \eqref{eq_anti} we obtain that for these terms at $\Lambda^\prime=\max(\Lambda,\eta(\Vec{p}))$
\begin{align}
&|\partial_{\Lambda_0}\Gamme^{\Lambda^\prime\Lambda_0;\Vec{\phi};w}_l(\Vec{p})|\leqslant\int \limits^1_0 dt \, |p_{\bar{c}}|\, |\partial_{\Lambda_0} \partial_{\bar{c}}\Gamme^{\Lambda^\prime\Lambda_0;\Vec{\phi};w}_l(tp_{\bar{c}},...)|\nonumber \\
&\quad \leqslant \frac{\Lambda^\prime + M+|\Vec{p}|}{\Lambda^2_0} \sum_{\tau \in \mathcal{T}_{\Vec{\phi}}} \int \limits^1_0 dt \, |p_{\bar{c}}|\, Q^{\Lambda^\prime;w+1_{\bar{c}}}_{\tau}P^{\Lambda^\prime\Lambda_0}_{2l-1}\leqslant \frac{\Lambda + M+|\Vec{p}|}{\Lambda^2_0} P^{\Lambda\Lambda_0}_{2l}\,,
\end{align}
where we have substituted $\partial_{\Lambda_0} \partial_{\bar{c}} \Gamme^{\Vec{\phi};w}_l$ with the bound of theorem \ref{thm_2} and applied inequality \eqref{eq_int7}.

The remaining marginal terms $\Gamme^{AAAA}$, $\partial_{\bar{c}}\Gamme^{c\bar{c}A}$, $\Gamme^{\phi \phi}_{\varkappa}$, $\partial \partial \Gamme^{\phi \phi}$, $\partial \Gamme^{c}_\gamma$,     are renormalized at
$\Lambda=0$ and nonvanishing momentum $\Vec{q}$,
chosen in $\mathbb{M}^{s}_n$ in all cases but
$\Gamme^{AAAA}$ for which $\Vec{q}\in\mathbb{M}^{cp}_4$.
See appendix \ref{sec_ren} for the list of all
relevant terms and their renormalization points.
Since the renormalization constants are independent of $\Lambda_0$, their derivative wrt $\Lambda_0$ vanishes: it follows that the coefficients of $\delta$-tensors in the decomposition of $\partial_{\Lambda_0}\Gamme^{0\Lambda_0;\Vec{\phi};w}_{\Vec{\varkappa};l}(\Vec{q})$ vanish. Hence using lemma \ref{lem_rn} and the bounds on irrelevant terms we have
\begin{equation}
|\partial_{\Lambda_0}\Gamme^{0\Lambda_0;\Vec{\phi};w}_{\Vec{\varkappa};l}(\Vec{q})|\leqslant \frac{M}{\Lambda^2_0}\mathcal{P}^{(1)}_{2l-1}(\log_+ \frac{\Lambda_0}{M}).
\end{equation}
We integrate the FE from 0 to $\Lambda^\prime$ and substitute the chain with the tree bound. Using inequalities \eqref{eq_int01}, \eqref{eq_int02} it is easy to get the following bound
\begin{equation}
|\partial_{\Lambda_0}\Gamme^{\Lambda^\prime\Lambda_0;\Vec{\phi};w}_{\Vec{\varkappa};l}(\Vec{q}) - \partial_{\Lambda_0}\Gamme^{0\Lambda_0;\Vec{\phi};w}_{\Vec{\varkappa};l}(\Vec{q})|\leqslant \frac{\Lambda + M}{\Lambda^2_0}\mathcal{P}^{(1)}_{2l-1}(\log_+ \frac{\Lambda_0}{M}).
\end{equation}
Integrating back and forth along the path given in equation \eqref{eq_path}, substituting the irrelevant term $\partial_{\Lambda_0}\partial \Gamme^{\Lambda^\prime \Lambda_0 \Vec{\phi};w}_{\Vec{\varkappa};l}$ with its bounds and using inequality~\eqref{eq_int7} we obtain
\begin{align}
|\partial_{\Lambda_0}\Gamme^{\Lambda^\prime\Lambda_0;\Vec{\phi};w}_{\Vec{\varkappa};l}(0)-\partial_{\Lambda_0}\Gamme^{\Lambda^\prime\Lambda_0;\Vec{\phi};w}_{\Vec{\varkappa};l}(\Vec{q})|&\leqslant \frac{\Lambda + M}{\Lambda^2_0}P^{\Lambda^\prime \Lambda_0}_{2l}(0),\\
|\partial_{\Lambda_0} \Gamme^{\Lambda^\prime\Lambda_0;\Vec{\phi};w}_{\Vec{\varkappa};l}(\Vec{p})-\partial_{\Lambda_0}\Gamme^{\Lambda^\prime\Lambda_0;\Vec{\phi};w}_{\Vec{\varkappa};l}(0)|&\leqslant \frac{\Lambda + M + |\Vec{p}|}{\Lambda^2_0}P^{\Lambda \Lambda_0}_{2l}(\Vec{p}).
\end{align}
 If $\Lambda=\Lambda^\prime$ the proof of the bounds on the marginal terms is complete. Otherwise we integrate the FE downwards from $\eta$ to $\Lambda$ and repeat the arguments given to prove inequality \eqref{eq_flw0} with a minor change in the integrand
\begin{equation}
\int \limits^{\eta}_\Lambda d\lambda \,  \frac{\lambda + M +|\Vec{p}|}{\Lambda^2_0}\frac{\lambda \, P^{\lambda \Lambda_0}_{k}|_{\eta_\tau=0}}{(\lambda + |p_a|)(\lambda + |p_b|)} \leqslant \frac{M + |\Vec{p}|}{\Lambda^2_0}\int \limits^{\eta}_\Lambda d\lambda \, \frac{2P^{\lambda \Lambda_0}_{k}|_{\eta_\tau=0}}{\lambda + |p_b|}.
\end{equation}

For the strictly relevant terms we integrate the FE from 0 to $\Lambda$ substituting the vertex functions and propagators with their bounds and extending the upper limit of integration to $\Lambda + \eta$
\begin{equation}
|\partial_{\Lambda_0}\Gamme^{\Lambda\Lambda_0\Vec{\phi};w}_{\Vec{\varkappa};l}(0)|\leqslant \frac{\Lambda+M}{\Lambda^2_0}\int \limits^{\Lambda + \eta}_0 d\lambda \, \lambda^{d-1}P^{\lambda\Lambda_0}_{2l-2}(0),
\end{equation}
where $d>0$. Using inequality \eqref{eq_int5} we obtain
\begin{equation}
|\partial_{\Lambda_0}\Gamme^{\Lambda\Lambda_0\Vec{\phi};w}_{\Vec{\varkappa};l} (0)|\leqslant \frac{\Lambda+M}{\Lambda^2_0} (\Lambda + \eta)^{d} P^{\Lambda+\eta \;\Lambda_0}_{2l-2}(0).
\end{equation}
To extend to momentum $\Vec{p}$ we proceed as in \eqref{eq_31228}, the only change being an additional factor of $\frac{\Lambda+M +| \Vec{p}|}{\Lambda^2_0}$.
\section{Restoration of the STI}\label{sec_sti}
As mentioned before theorems \ref{thm_3}, \ref{thm_4} we now consider all nontrivial marginal terms $\Gamma^{\Lambda \Lambda_0;\Vec{\phi};w}_{X\Vec{\varkappa};l}$ with $X \in \{\beta, 1\}$ at $\Lambda=0$. We want to show that these terms verify the bounds of theorems~\ref{thm_3},~\ref{thm_4}. In this section, since $\Lambda=0$, we will omit the parameters~$\Lambda$,~$\Lambda_0$ in the notations wherever this is not ambiguous, i.e. we write $\Gamme^{\Vec{\phi}}$ for $\Gamme^{0\Lambda_0;\Vec{\phi}}$.

The subsequent relations are obtained by projecting the AGE \eqref{eq_21gh} and the STI \eqref{eq_21st} on the respective monomial in the fields (for example, $c$ in section~\ref{gh_c}) to read off the lhs from the rhs, taken at the renormalization point. We will establish appropriate relations in order to make the coefficients of the $\delta$-tensors wrt the monomial basis at the renormalization point satisfy the bounds of these two theorems. In this analysis we make particular use of the consistency conditions, see \ref{407s}. In section~\ref{sec_329r} we prove the existence of a solution for the above mentioned system of relations that does not depend on the UV cutoff. In the remaining sections we treat the different marginal terms one by one.
\subsubsection{Smallness relations}
It is helful to introduce the notion of \textit{small} terms, which vanish in the limit~$\Lambda_0 \to \infty$. For fixed loop order $l$ and $X \in \{\beta, 1\}$, a homogeneous function $f(\Vec{p},M,\Lambda_0)$ of mass dimension $[f]$ is said \textit{small} on a subset $Y \subset \mathbb{M}_n$, and denoted by~$f \stackrel{X,Y,l}{\sim} 0$, if for all $w \in \mathbb{W}_n$ with $\|w\|\leqslant [f]$, there exists a polynomial $\mathcal{P}^{(1)}_{r_X}$ of degree $r_X([f]-\|w\|,l)$, see theorems \ref{thm_3} and \ref{thm_4}, such that the following bound holds for all $\Lambda_0 \geqslant M$ and all $\Vec{p} \in Y$, see~\eqref{eq_2ren}:
\begin{equation}
|\partial^w_p f (\Vec{p},M,\Lambda_0)| \leqslant \frac{M^{1+ [f]  - \|w\|}}{\Lambda_0}\mathcal{P}^{(1)}_{r_X}(\log_+ \frac{\Lambda_0}{M}).\label{eq_1s}
\end{equation}

Furthermore, $f \stackrel{X,Y,l}{\sim} g$ iff $[f]=[g]$ and $f-g \stackrel{X,Y,l}{\sim} 0$. Because both relations $\stackrel{1,Y,l}{\sim}$ and $\stackrel{\beta,Y,l}{\sim}$ only differ by the degree of polynomials we have $f \stackrel{\beta,Y,l}{\sim} g \implies f \stackrel{1,Y,l}{\sim} g$. Since the loop order $l$ and the renormalization point~$\Vec{q}$ are evident from the context, we write:
\begin{align}
f \stackrel{Y}\sim g&\quad\mbox{for}\quad f \stackrel{1,Y,l}{\sim} g,& f \stackrel{\beta,Y}{\sim} g&\quad\mbox{for}\quad  f \stackrel{\beta,Y,l}{\sim} g,\label{eq_88sim}\\ 
f \sim g&\quad\mbox{for}\quad  f \stackrel{1,\{\Vec{q}\}}{\sim} g,& f \bim g &\quad\mbox{for}\quad  f \stackrel{\beta,\{\Vec{q}\}}{\sim} g. 
\end{align}

Theorem \ref{thm_1} implies that for every vertex function $\Gamme^{\Vec{\phi}}_{\Vec{\varkappa}}(\Vec{p})$  there exists a constant~$c$ such that $\forall w\in\mathbb{W}_n$, $\forall \Vec{p} \in \mathbb{M}_n$, $\forall \Lambda_0 \geqslant M$
\begin{equation}
|\partial^w \Gamme^{\Vec{\phi}}_{\Vec{\varkappa}}(\Vec{p})| \leqslant c  M^{4 - 2n_\varkappa - \N  - \|w\|}\,.
\end{equation}
Using also that
\begin{equation}
|\partial^w(\sigma_{0\Lambda_0}-1)|< c_w \, \frac{1}{\Lambda^w_0}\,,\quad |\sigma_{0\Lambda_0}-1|< c_0\,\frac{M}{\Lambda_0}\,,
\end{equation}
 the terms on rhs of the STI and the AGE  satisfy the relations:
\begin{align}
\Gamme^{\Vec{\phi}_1;w_1} \partial^{w_2} (\sigma_{0\Lambda_0} \Gamme^{\Vec{\phi}_2}_{\varkappa})&\stackrel{\beta,Y}{\sim}\Gamme^{\Vec{\phi}_1;w_1} \Gamme^{\Vec{\phi}_2;w_2}_{\varkappa}\,,\\
\partial^{w} (\sigma_{0\Lambda_0} \Gamme^{\Vec{\phi}})&\stackrel{\beta,Y}{\sim}\Gamme^{\Vec{\phi};w}\,,\\
\partial^{w} (p \sigma_{0\Lambda_0} \Gamme^{\Vec{\phi}}_\gamma)&\stackrel{\beta,Y}{\sim}p \Gamme^{\Vec{\phi};w}_{\gamma}\,.
\end{align}
This fact will be useful in the calculations underlying the following sections.
\subsubsection{The functional $\F_{1, rel}$} \label{sec_327w}
In this section we introduce the notation for the renormalization constants for the functional $\F_1$~\eqref{eq_407a}. For this purpose we define the auxiliary functional
\begin{align}
\F^{\Lambda_0}_{1,rel}:=&\epsilon^{dab} \langle  U^{\gamma cc}_\sigma| \check{\gamma}^d_\sigma \check{c}^a \check{c}^b\rangle + \epsilon^{sab} \epsilon^{sde} u^{\gamma Acc} \langle  \check{\gamma}^a_\sigma \check{A}^b_\sigma \check{c}^d \check{c}^e\rangle\nonumber\\
&+ \langle U^{cA}_\sigma |\check{c}^d\check{A}^d_\sigma\rangle + \epsilon^{dab} \langle  U^{\bar{c}cc} |\check{\bar{c}}^d\check{c}^a \check{c}^b\rangle + \epsilon^{dab} \langle  U^{cAA}_{\mu \nu} |\check{c}^d\check{A}^a_\mu\check{A}^b_\nu\rangle\nonumber\\
&+\langle U^{cAAA}_{\rho \mu \nu} |\check{c}^d\check{A}^d_\rho\check{A}^a_\mu\check{A}^a_\nu \rangle + \langle U^{\bar{c}ccA}_{1;\mu} |\check{\bar{c}}^b \check{c}^b\check{c}^a \check{A}^a_\mu \rangle  \nonumber\\
&+ \langle U^{\bar{c}ccA}_{2;\mu} |\check{\bar{c}}^a\check{c}^b\check{c}^b \check{A}^a_\mu \rangle + \epsilon^{bed} u^{c\bar{c}cAA}_{1} \langle  \check{c}^b\check{\bar{c}}^e\check{c}^d \check{A}^a_\mu \check{A}^a_\mu \rangle \nonumber\\
&+ \epsilon^{bed} u^{c\bar{c}cAA}_{2} \langle  \check{c}^b\check{\bar{c}}^a\check{c}^d \check{A}^a_\mu \check{A}^e_\mu  \rangle + \epsilon^{bed} u^{c\bar{c}cAA}_{3}  \langle \check{c}^a\check{\bar{c}}^e\check{c}^d \check{A}^a_\mu \check{A}^b_\mu \rangle,
\end{align}
where 
\begin{align}
U^{\gamma cc}_\sigma(l,p,q):=&i(p+q)_\sigma u^{\gamma cc},\\
U^{cA}_\sigma(l,p):=&ip_\sigma p^2 u^{cA},\\ 
U^{\bar{c}cc}(l,p,q):=&(p^2 + q^2)u^{\bar{c}cc}_1 + 2pq u^{\bar{c}cc}_2,\\
U^{cAA}_{\mu \nu}(l,p,q):=&(p_\mu p_\nu - q_\mu q_\nu)u^{cAA}_1 + \delta_{\mu \nu} (p^2-q^2)u^{cAA}_2,\\
U^{cAAA}_{\rho \mu \nu}(l,k,p,q):=&i(u^{cAAA}_1 k_\rho  + u^{cAAA}_2 p_\rho+u^{cAAA}_2q_\rho) \delta_{\mu \nu}\nonumber\\
&+i(u^{cAAA}_3 k_\nu + u^{cAAA}_4 q_\nu + u^{cAAA}_5 p_\nu)\delta_{\rho \mu} \nonumber\\
&+i(u^{cAAA}_3 k_\mu + u^{cAAA}_4 p_\mu + u^{cAAA}_5 q_\mu)\delta_{\rho \nu},\label{eq_ucAAA}\\
U^{\bar{c}ccA}_{1;\mu}(l,k,q,p):=& ip_\mu u^{\bar{c}ccA}_1 +  ik_\mu u^{\bar{c}ccA}_2 + iq_\mu u^{\bar{c}ccA}_3,\\
U^{\bar{c}ccA}_{2;\mu}(l,k,q,p):=&i(k_\mu - q_\mu)u^{\bar{c}ccA}_4\,,
\end{align}
and the $u$'s are functions of $\Lambda_0$ defined by the marginal renormalization conditions
\begin{equation}
\F^{0\Lambda_0;\Vec{\phi};w}_{1;\Vec{\varkappa}}(\Vec{q})=\F^{\Lambda_0;\Vec{\phi};w}_{1,rel;\Vec{\varkappa}}(\Vec{q})+\sum_{t \in \{\delta^s \ubar{e}^{k>0}\}_r} \zeta^{\Lambda_0}_t\, t\,.
\end{equation}
Here $2n_\varkappa+ \N+\|w\|=5$, $\Vec{q}$ is the renormalization point defined in section~\ref{sec_ren}, $\ubar{e}=(e_i)_{i \in [m]}$ is an orthogonal basis for the linear span of $\Vec{q}$, $r$ is the tensor rank of $\F^{0\Lambda_0;\Vec{\phi};w}_{1}(\Vec{q})$. The $\zeta^{\Lambda_0}_t$ are the uniquely defined coefficients of tensors $t$. Note that we implicitly set to zero all
constants associated to strictly relevant renormalization
conditions for $\F_1$.
These constants are not needed because,
thanks to hypothesis~\ref{rc1},
the RHS of the STI and AGE at the current loop order
vanish at zero momenta.

The renormalization points $\vec{q}$ are chosen in agreement with the hypotheses of lemma \ref{lem_rn}. From lemma \ref{lem_rn}, theorem \ref{thm_3}, and the irrelevant bounds of theorem \ref{thm_4}, for the marginal terms one has
\begin{equation}
\F^{0\Lambda_0;\Vec{\phi};w}_{1;\Vec{\varkappa}}(\Vec{q})\sim \F^{\Lambda_0;\Vec{\phi};w}_{1,rel;\Vec{\varkappa}}(\Vec{q}).
\end{equation}
\subsubsection{Consistency conditions}\label{407s}
Here we establish the consistency conditions
implied by the nilpotency,
see \eqref{eq_cohom} and \eqref{eq_404a}.
Below we will rely on the validity of theorem~\ref{thm_4}
at loop orders $l’<l$ for all terms
and at the current order $l$ only for irrelevant terms:
these properties are true in our inductive scheme.
Recall definitions  \eqref{eq_409s}, \eqref{eq_2ren} and \eqref{eq_88sim}.
Using the AGE~\eqref{eq_21gh},
the bounds of theorems~\ref{thm_1}-\ref{thm_3},
and \eqref{eq_404a} we get
\begin{equation}
\Big(S \Gamme_{\beta}\Big)^{\Vec{\phi}}\,\stackrel{\mathbb{M}_n}{\sim}\, 0, \quad\mbox{and thus}\quad  \Big(\frac{\delta}{\delta \tilde{c}} \F_{1}\Big)^{\Vec{\phi}}\, \stackrel{\mathbb{M}_n}{\sim}\, 0\,,\label{eq_88gh}
\end{equation}
where $\Vec{\phi}=(\phi_1,...,\phi_{n-1})$ and $\phi_i \in \{A,c,\bar{c}\}$. Equation \eqref{eq_88gh}, theorems~\ref{thm_1},~\ref{thm_2},  and the bounds of theorem \ref{thm_4} for irrelevant terms yield
\begin{align}
\Big(\frac{\delta}{\delta \tilde{c}} \F_{1,rel}\Big)^{\Vec{\phi}}& \stackrel{\mathbb{M}_n}{\sim}0,&\Vec{\phi} &\in \{(c,c), (c,c,A)\}\,.\label{eq_26gh}
\end{align}
See section~\ref{sec_327w} for the definition of $\F_{1,rel}$ and of the constants $u^\Phi$. In section~\ref{st_cccAA} it will be shown that $u^{c\bar{c}cAA}_{1,2,3} \sim 0$. Equation~\eqref{eq_26gh} then gives
\begin{align}
u^{\gamma c c}&\sim u^{\bar{c}cc}_1 \sim u^{\bar{c}cc}_2,&-2u^{\gamma A cc}& \sim u^{\bar{c}ccA}_1 \sim u^{\bar{c}ccA}_2 \sim u^{\bar{c}ccA}_3,&u^{\bar{c}ccA}_4& \sim 0. \label{eq_cccA}
\end{align}
Let us exploit \eqref{eq_cohom} to obtain more constraints
on the renormalization constants $u^\Phi$. At loop order $l$
\begin{equation}
S \F_{1;l}=S_0 \F_{1;l} + \sum_{l^{\prime}<l} S_{l-l^{\prime}} \F_{1;l^{\prime}}.
\end{equation}
By induction $(S_{l-l^{\prime}} \F_{1;l^{\prime}})^{\Vec{\phi}}\stackrel{\mathbb{M}_n}{\sim} 0$ for all loop orders $l^\prime < l$. Then equation \eqref{eq_cohom} implies that 
\begin{equation}
(S_0 \F_{1;l})^{\Vec{\phi}} \, \stackrel{\mathbb{M}_{n}}{\sim} \, 0,\label{eq_404b}
\end{equation}
where
\begin{align}
S_0=&\langle \tilde{\delta}_{A^d_\sigma}\Tamma_0,\sigma_{0 \Lambda_0}\tilde{\delta}_{\gamma^d_\sigma} \rangle + \langle \tilde{\delta}_{\gamma^d_\sigma}\Tamma_0,\sigma_{0 \Lambda_0}\tilde{\delta}_{A^d_\sigma} \rangle\nonumber\\
&- \langle \tilde{\delta}_{c^d}\Tamma_0,\sigma_{0 \Lambda_0}\tilde{\delta}_{ \omega^d} \rangle - \langle \tilde{\delta}_{\omega^d}\Tamma_0,\sigma_{0 \Lambda_0}\tilde{\delta}_{c^d} \rangle,\label{eq_404s}
\end{align}
and recalling notation for $\tilde{\delta}_\phi$ from \eqref{eq_409t}
\begin{align*}
\tilde{\delta}_{A^d_\sigma(q)}\Tamma_0=&g\epsilon^{adb} (\langle \check{\gamma}^a_\sigma \check{c}^b;q\rangle + \langle i p_\sigma \check{\bar{c}}^a(p) \check{c}^b;q\rangle) + A^d_{\sigma^\prime}(q)(\delta_{\sigma^\prime \sigma} q^2-q_{\sigma^\prime} q_\sigma)\sigma^{-1}_{0 \Lambda_0}(q^2) \nonumber\\
&+3 \epsilon^{dab} \langle   F^{AAA}_{\sigma \mu \nu}(q,\cdot,\cdot)|\check{A}^a_\mu \check{A}^b_\nu;q\rangle + 4R^{AAAA}_{\sigma  \rho \mu \nu} \langle \check{A}^d_\rho \check{A}^a_\mu \check{A}^a_\nu;q\rangle,\\
\tilde{\delta}_{\gamma^d_\sigma(q)}\Tamma_0=& iq_\sigma c^d(q)+  g\epsilon^{dab} \langle \check{A}^a_\sigma \check{c}^b;q \rangle,\\
\tilde{\delta}_{c^d(q)}\Tamma_0=&g\epsilon^{dab}\langle i p_\mu \check{\bar{c}}^a(p)\check{A}^b_\mu  + \check{\gamma}^a_\mu \check{A}^b_\mu + \check{\omega}^b \check{c}^a;q\rangle - iq_\mu \gamma^d_\mu(q)-\bar{c}^d(q) q^2 \sigma^{-1}_{0 \Lambda_0}(q^2),\\
\tilde{\delta}_{\omega^d(q)}\Tamma_0=& \frac{1}{2} g\epsilon^{dab} \langle \check{c}^a\check{c}^b;q \rangle.
\end{align*}
Here the notation $\langle i p_\sigma \check{\bar{c}}^a(p) \check{c}^b;q\rangle$ corresponds to $\langle \phi_1 \phi_2 ;q\rangle$ with  $\phi_1(p)= i p_\sigma\check{\bar{c}}^a(p)$, $\phi_2=\check{c}^b$.\\
For all $\Vec{\phi}$, $\Vec{\varkappa}$ and $w$ such that $\N + 2n_\varkappa +\|w\|=6$  we have
\begin{equation}
(S_0 \F_{1;l})^{\Vec{\phi};w}_{\Vec{\varkappa}}=(S_0 \F_{1,rel;l})^{\Vec{\phi};w}_{\Vec{\varkappa}}  + \sum_{\pi} (-)^{\pi_a} S^{\Vec{\phi}_1;w_1}_{0,\Vec{\varkappa}_1} \Delta^{\Vec{\phi_2};w_2}_{\Vec{\varkappa}_2;l},\label{S0F1}
\end{equation}
where $\Delta^{\Lambda \Lambda_0}_l:=\F^{\Lambda\Lambda_0}_{1;l}-\F^{\Lambda_0}_{1,rel;l}\,$, the  sum runs over the permutations $\pi=(\pi_\phi,\pi_\varkappa,\pi_w)$ such that $\Vec{\phi}_{\pi_\phi}=\Vec{\phi}_1\oplus \Vec{\phi}_2$, $\Vec{\varkappa}_{\pi_\varkappa}=\Vec{\varkappa}_1\oplus \Vec{\varkappa}_2$, $w_{\pi_w}=w_1 + w_2$, and $\pi_a$ is the number of transpositions $mod \, 2$ of anticommuting variables in the permutation $\pi$. 
Using~\eqref{eq_404s}, for the terms in the sum on the rhs of~\eqref{S0F1} we have
\begin{align}
|S^{\Vec{\phi}_1;w_1}_{0,\Vec{\varkappa}_1}\Delta^{\Vec{\phi_2};w_2}_{\Vec{\varkappa}_2;l}|&\leqslant |\Tamma^{A\Vec{\phi}_1;w^{\prime}_1}_{\Vec{\varkappa}_1;0} ,\sigma^{w^{\prime \prime}_1}_{0 \Lambda_0}\Delta^{\Vec{\phi_2};w_2}_{\Vec{\varkappa}_2\gamma;l}| + |\Tamma^{c\Vec{\phi}_1;w^{\prime}_1}_{\Vec{\varkappa}_1;0} ,\sigma^{w^{\prime \prime}_1}_{0 \Lambda_0}\Delta^{\Vec{\phi_2};w_2}_{\Vec{\varkappa}_2\omega;l}|\nonumber\\
&+|\Tamma^{\Vec{\phi}_1;w^{\prime}_1}_{\Vec{\varkappa}_1\gamma;0} ,\sigma^{w^{\prime \prime}_1}_{0 \Lambda_0}\Delta^{A\Vec{\phi_2};w_2}_{\Vec{\varkappa}_2;l}| + |\Tamma^{\Vec{\phi}_1;w^{\prime}_1}_{\Vec{\varkappa}_1\omega;0} ,\sigma^{w^{\prime \prime}_1}\Delta^{c\Vec{\phi_2};w_2}_{\Vec{\varkappa}_2;l}|.\label{eq_405a}
\end{align}
Let us show that the lhs of \eqref{eq_405a} is \textit{small} on $\mathbb{M}_n$. Using section~\ref{sec_327w} and the bounds on irrelevant terms of theorems \ref{thm_3}, \ref{thm_4} we see that for all marginal terms 
\begin{equation}
\Delta^{\Vec{\phi};w}_{\Vec{\varkappa};l}\sim 0,\quad\mbox{and thus}\quad\Delta^{\Vec{\phi};w}_{\Vec{\varkappa};l}\, \stackrel{\mathbb{M}_n}{\sim} \,0.\label{eq_405b}
\end{equation}
The relation on the rhs can be obtained by adapting the interpolation in equations \eqref{eq_413a}-\eqref{eq_413c}.
Define $\N_i:=|\Vec{\phi}_i|$ and $n_{\varkappa_i}:=|\Vec{\varkappa}_i|$. Consider the sum of the first and second term on the rhs. If $2n_{\varkappa_2}+\N_2+\|w_2\|\geqslant 3$ then the bounds of theorems~\ref{thm_1},~\ref{thm_3},~\ref{thm_4} and \eqref{eq_405b} imply that the sum is \textit{small}. On the other hand if $2n_{\varkappa_1}+\N_1+\|w_1\|>3$ then $\|w^{\prime \prime}_1\|>0$ and the bounds of theorems~\ref{thm_1},~\ref{thm_3} also give that the sum is \textit{small}. The analysis of the sum of the third and fourth term on the rhs is similar. If $2n_{\varkappa_2}+\N_2+\|w_2\|\geqslant 4$ then the bounds of theorems~\ref{thm_1},~\ref{thm_3},~\ref{thm_4} and \eqref{eq_405b} imply that the sum is \textit{small}. If $2n_{\varkappa_1}+\N_1+\|w_1\|> 2$ then $\|w^{\prime \prime}_1\|>0$ and using the bounds of theorems~\ref{thm_1},~\ref{thm_3} we obtain again that the sum is \textit{small}. It follows that the lhs of \eqref{eq_405a} is \textit{small}. This fact and  \eqref{eq_404b} imply that $(S_0 \F_{1,rel;l})^{\Vec{\phi};w} \stackrel{\mathbb{M}_{n}}{\sim} 0$ for all marginal terms, which leads to the following equations 
\begin{align}
g u^{\gamma cc} & \sim - u^{\gamma A cc},&u^{cAA}_1 + u^{cAA}_2&\sim \frac{gu^{cA}}{2}, \label{eq_cAA}\\
u^{cAAA}_1&\sim u^{cAAA}_2,&u^{cAAA}_3\sim u^{cAAA}_4&\sim u^{cAAA}_5. \label{eq_cAAA}
\end{align}
\subsubsection{Existence of a constant solution}\label{sec_329r}
By our convention (which is the standard one) the renormalization constants that are solutions of the relations listed in \ref{rc3} are supposed not to depend on $\Lambda_0$.
We give here a proof of this property, which is not evident because these relations contain nontrivial functions of $\Lambda_0$, here denoted by $\zeta^{\Lambda_0}_{\Phi}$. The relations corresponding to the marginal terms $\Gamme^{\Vec{\phi};w}_{1 \Vec{\varkappa}}$, $\Gamme^{\Vec{\phi};w}_{\beta \Vec{\varkappa}}$ have respectively the general form
\begin{align}
c^\Phi + \varrho \, C^\Phi_1 \varrho +  \zeta^{\Lambda_0}_{\Phi} &\sim 0,& \zeta^{\Lambda_0}_{\Phi}&:= \varrho \, C^\Phi_2 \zeta^{\Lambda_0} + \zeta^{\Lambda_0} C^\Phi_3 \zeta^{\Lambda_0}\,,\\
c^\Phi +  V^\Phi_1 \varrho +  \zeta^{\Lambda_0}_{\Phi} &\bim 0,& \zeta^{\Lambda_0}_{\Phi}&:= V^\Phi_2 \zeta^{\Lambda_0}\,.
\end{align}
Here $\varrho=(r^{\phi...}_i, R_i, \Sigma^{AA}_{L}, \Sigma^{AA}_{T}, \Sigma^{\bar{c}c}  )$ denotes the relevant terms for vertex functions, see \eqref{eq_11r} for the list of $r^{\phi...}_i$ and appendixes~\ref{sec_gamma},~\ref{sec_gamma2} for the remaining terms. The sequence $\zeta^{\Lambda_0}$ stands for the irrelevant terms listed in appendixes~\ref{sec_gamma},~\ref{sec_gamma2} and for the  derivative  of $r^{\phi...}_i$, $R_i$ wrt scalar products of momenta. Finally, $c^\Phi$ is a constant, $V^\Phi_{1,2}$ are constant vectors,  and $C^\Phi_{1,2,3}$ are constant matrices.

At loop order $l$,
the terms $\zeta^{\Lambda_0}_{\Phi,l}$ depend only on
$\varrho_{l'}$ of loop order $l'<l$:
this property holds because
each $\zeta^{\Lambda_0}_{\Phi,l}$ is at least linear in the $\zeta^{\Lambda_0}$ and because all the $\zeta^{\Lambda_0}_{l=0}$ vanish.
Moreover, at order $l$ for each relation we have a distinct renormalization constant. Consequently, the aforementioned relations have a solution. The existence of a solution $\varrho_l$ independent of $\Lambda_0$ follows immediately if the limit $\lim_{\Lambda_0\to\infty}\zeta^{\Lambda_0}_{\Phi,l}$ exists: in this case it is enough to choose a solution of the following equations
\begin{align}
c^\Phi + \varrho \, C^\Phi_1 \varrho +  \zeta^{\infty}_{\Phi} &= 0,&c^\Phi +  V^\Phi_1 \varrho +  \zeta^{\infty}_{\Phi} &= 0.
\end{align}
The convergence of $\zeta^{\Lambda_0}_{\Phi,l}$
relies on the validity of the bounds of theorem \ref{thm_2}
up to order $l$ for all irrelevant terms $\Gamme^{\Vec{\phi};w}_{\Vec{\varkappa}}$ and up to order $l-1$ for 
all the relevant ones. This property holds because in our inductive scheme at fixed loop order the irrelevant terms are treated before the relevant ones.
\subsection{$\Gamme^{c}_\beta$}\label{gh_c}The renormalization point is $\Vec{q}=(-\bar{q},\bar{q})\in \mathbb{M}^s_2$, see \eqref{eq_9sym}.
\begin{align}
\Gamme^{c^a}_{\beta^b}(p)&\bim \sigma_{0 \Lambda_0}(p^2)\,\Gamma^{c^a \bar{c}^b}(p)+i p_\mu \Gamma^{c^a}_{\gamma^b_\mu}(p) \bim -\delta^{ab} f(p^2)\,,\\
f(x)&:= x \left(1+\Sigma^{\bar{c}c}(x) - R_1(x)\right). \label{eq_331d}
\end{align}
For the marginal term we obtain
\begin{equation}
-\frac{\delta^{ab}}{3} \, \Gamme^{c^a;p_\mu p_\nu}_{\beta^b}(p) \bim 2\delta_{\mu \nu} f^\prime(p^2) + 4p_\mu p_\nu f^{\prime \prime}(p^2) \,.
\end{equation}
The coefficient of $\delta_{\mu \nu}$ is \textit{small} at the renormalization point iff
\begin{align}
f^\prime=1-r^{\bar{c}c} - R_1 - \zeta^{\Lambda_0}_{\beta c} &\bim 0,& \zeta^{\Lambda_0}_{\beta c}(p^2)&:=p^2\frac{\partial R_1(p^2)}{\partial p^2}\,. \label{eq_gh_c}
\end{align}
This gives the renormalization condition for $R_1$.
\subsection{$\Gamme^{cA}_\beta$}The renormalization point is $\Vec{q}=(\bar{k},\bar{p},\bar{q})\in \mathbb{M}^s_3$, see \eqref{eq_9sym}. With $k=-p-q$,
\begin{align}
\Gamme^{c^a A^d_\mu}_{\beta^b}(p,q)&\bim \Gamma^{A^d_\mu c^a \bar{c}^b}(p,k)-ik_\rho \Gamma^{c^a A^d_\mu}_{\gamma^b_\rho}(p,q)\\
&=i\epsilon^{dab} \mathcal{I}^{0\Lambda_0}_\mu(p,q)\,,\\
\mathcal{I}^{0\Lambda_0}_\mu(p,q)&=k_\mu R^{A\bar{c}c}_1(k,p)+p_\mu r^{A\bar{c}c}_2(k,p) -gk_\rho F^{\gamma A c}_{\rho  \mu}(q,p)\,.
\end{align}
Let $\Delta^{\Lambda\Lambda_0}_{\mu\nu}:=\mathcal{I}^{\Lambda\Lambda_0;p_\nu}_\mu-\mathcal{I}^{\Lambda\Lambda_0;q_\nu}_\mu$. At zero external momenta and $\Lambda=M$ we have $\Delta^{M\Lambda_0}_{\mu \nu}(0)=0$. Then using the bounds of theorem \ref{thm_3} we get
\begin{equation}
|\Delta^{M\Lambda_0}_{\mu \nu}(\Vec{q})|\leqslant \int \limits^1_0 dt\,M |\partial \partial \Gamma^{M\Lambda_0;c A}_{\beta}| \bim 0 \,.
\end{equation}
The term $\Delta^{\Lambda\Lambda_0}_{\mu \nu}$ obeys the FE, see \eqref{eq_20ge}. It remains to integrate the FE from 0 to $\Lambda$ and use inequality \eqref{eq_int5} to obtain
\begin{equation}
\Delta^{0\Lambda_0}_{\mu \nu}(\Vec{q})-\Delta^{M\Lambda_0}_{\mu \nu}(\Vec{q})\bim 0,\quad\mbox{and thus}\quad\Delta^{0\Lambda_0}_{\mu\nu}(\Vec{q}) \bim 0\,.
\end{equation}
Hence in the monomial basis $\{\delta^s Q^k\}_2$ with $Q=(\bar{p},\bar{q})$ the $\delta$-component of $\mathcal{I}^{A\bar{c}c;q_\nu}_\mu$ is \textit{small} at the renormalization point if the following condition holds
\begin{equation}
R^{A\bar{c}c}_1 - gR_2-\zeta^{\Lambda_0}_{\beta cA} \bim 0\,.\label{eq_gh_cA}
\end{equation}
This gives the renormalization condition for $R^{A\bar{c}c}_1$.
\subsection{$\Gamme^{cAA}_\beta$ and $\Gamme^{cc\bar{c}}_\beta$}The renormalization point is $\Vec{q}=(\bar{k},\bar{l},\bar{p},\bar{q})\in \mathbb{M}^s_4$, see \eqref{eq_9sym}.
\begin{align}
\Gamme^{c^a A^t_\mu A^s_\nu}_{\beta^b}(l,p,q)&\bim \Gamma^{c^a \bar{c}^b A^t_\mu A^s_\nu}(k,p,q)-ik_\rho \Gamma^{c^a A^t_\mu A^s_\nu}_{\gamma^b_\rho}(l,p,q)\\
\Gamme^{c^a c^t \bar{c}^s}_{\beta^b}(l,p,q)&\bim \Gamma^{c^a \bar{c}^b c^t \bar{c}^s}(k,p,q)-ik_\rho \Gamma^{c^a c^t \bar{c}^s}_{\gamma^b_\rho}(l,p,q)
\end{align}
 At $\Lambda=M$ it follows from property~\eqref{eq_21anti} that these terms vanish at zero momenta. Denoting the renormalization point by $\Vec{q}$, using the bounds of theorem \ref{thm_3} and integrating the FE from $M$ to 0 we obtain
\begin{align}
|\Gamme^{M\Lambda_0;\Vec{\phi}}_{\beta}(\Vec{q})|\leqslant& \int \limits^1_0 dt\,|\Vec{q}| |\partial \Gamma^{M\Lambda_0;\Vec{\phi}}_{\beta}(t\Vec{q})|\bim 0\,,\\
|\Gamme^{0\Lambda_0;\Vec{\phi}}_{\beta}(\Vec{q})-\Gamme^{M\Lambda_0;\Vec{\phi}}_{\beta}(\Vec{q})|\leqslant& \frac{M}{\Lambda_0}\mathcal{P}^{(1)}_{2(l-1)}(\log_+\frac{\Lambda_0}{M})\bim 0\,.
\end{align}
\subsection{$\Gamme^{c  \bar{c} c \bar{c}  c}_1$ and $\Gamme^{c A A A A}_1$}\label{st_cAAAA}
These functions do not have nonvanishing marginal terms:
\begin{align}
\epsilon^{dab}\langle \check{c}^d\check{A}^a_\mu\check{A}^b_\nu\check{A}^s_\mu\check{A}^s_\nu\rangle&=0,&\epsilon^{dab}\langle \check{c}^d\check{\bar{c}}^s\check{c}^a\check{\bar{c}}^s\check{c}^b\rangle&=0\,.
\end{align}
\subsection{$\Gamme_1^{\bar{c} c c A A}$} \label{st_cccAA}
From equation \eqref{eq_21anti} it follows that for $\Lambda=M$ the function vanishes if the antighost momentum is zero. Using the bounds of theorem \ref{thm_4} first we obtain at the renormalization point $|\Gamme_1^{M \Lambda_0;\bar{c} c c A A}(\Vec{q})| \sim 0$ where $\Vec{q} \in \mathbb{M}^s_5$, and then integrating the FE from $M$ to 0 we show that the term is \textit{small} at $\Lambda=0$.
\subsection{$\Gamme^{c A}_1$}The renormalization point is $\Vec{q}=(-\bar{q},\bar{q})\in \mathbb{M}^s_2$, see \eqref{eq_9sym}. 
\begin{align}
\Gamme^{c^a A^b_\mu}_1(p)& \sim i \delta^{ab}  F^{AA}_{T;\mu\nu}(p)  R_1 p_\nu=i \delta^{ab} p_{\mu} f(p^2)\,,\\
f(x)&:=\frac{1}{\xi} xR_1(x)\Sigma^{AA}_L(x)\,.
\end{align}
The marginal term satisfies
\begin{equation*}
\Gamme^{c^a A^b;ppp}_1(p) \sim  i\delta^{ab}\Big(2 f^\prime(p^2)(\Sigma_{t \in  \{\delta^2\}_4} t)  + 4f^{\prime \prime}(p^2) (\Sigma_{t \in \{\delta pp\}_4} t)  + 8 pppp f^{\prime \prime \prime}(p^2)\Big)\,.
\end{equation*}
For the coefficient of  $\delta$-tensors we have
\begin{equation}
\xi f^\prime(p^2)=R_1(p^2) \Big(\Sigma^{AA}_L(p^2) + p^2 \frac{\partial \Sigma^{AA}_L(p^2)}{\partial p^2}\Big) + p^2 \Sigma^{AA}_L(p^2) \frac{\partial R_1(p^2)}{\partial p^2}\,.
\end{equation}
Recalling the definition of $r^{AA}_{1,2}$ in appendix \ref{sec_gamma},
\begin{equation}
r^{AA}_2(p^2) + r^{AA}_1(p^2) + p^2 \frac{\partial r^{AA}_2(p^2)}{\partial p^2}= \frac{1}{\xi}\Big(\Sigma^{AA}_L(p^2) + p^2 \frac{\partial \Sigma^{AA}_L(p^2)}{\partial p^2}\Big)\,.
\end{equation}
We  then obtain the following sufficient condition
\begin{equation}
u^{cA} \sim 0 \Longleftrightarrow R_1(r^{AA}_2 + r^{AA}_1) + \zeta^{\Lambda_0}_{cA} \sim 0\,,\label{eq_st_cA}
\end{equation}
where
\begin{equation}
\zeta^{\Lambda_0}_{cA}(p^2):=p^2 \Big(R_1(p^2)\frac{\partial r^{AA}_2(p^2)}{\partial p^2} + \frac{1}{\xi}\Sigma^{AA}_L(p^2)\frac{\partial R_1(p^2)}{\partial p^2}\Big)\,.
\end{equation}
See section~\ref{sec_327w} for the definition of $u^{cA}$. Relation \eqref{eq_st_cA} gives us the renormalization condition for $r^{AA}_2$.
\subsection{$\Gamme^{c A A}_1$}The renormalization point is $\Vec{q}=(\bar{k},\bar{p},\bar{q})\in \mathbb{M}^s_3$, see \eqref{eq_9sym}. With $k=-p-q$,
\begin{align}
\Gamme^{c^s A^a_\mu A^b_\nu}_1(p,q) \sim &\Gamma^{A^t_\rho A^a_\mu A^b_\nu}(p,q)\Gamma^{c^s}_{\gamma^t_\rho}(k)+\sum \limits_{Z_2} F^{AA}_{T;\mu \rho}(p)\Gamma^{c^s A^b_\nu}_{\gamma^a_\rho}(k,q)\,,\label{eq_29cAA}
\end{align}
where the sum $\sum_{Z_2}$ runs over all cyclic permutations of $\{(\mu,p,a),(\nu,q,b)\}$. The marginal terms are: $\Gamme^{c A A;pp}_1$, $\Gamme^{c A A;pq}_1$, $\Gamme^{c A A;qq}_1$. Using equation \eqref{eq_cAA} we see that $u^{cAA}_2 \sim 0 \implies u^{cAA}_1 \sim 0$. Acting with $\partial_p\partial_p$ on both sides of~\eqref{eq_29cAA} we obtain
\begin{align}
u^{cAA}_2 &\sim 0&& \Longleftrightarrow &gR_2\left(1+r^{AA}_1\right) - 2R_1R^{AAA} + \zeta^{\Lambda_0}_{cAA} &\sim 0\,.\label{eq_st_cAA}
\end{align}
This gives the renormalization condition for $R_2$.
\subsection{$\Gamme_1^{\bar{c} c c A}$  and $\Gamme^{c c A}_{1 \gamma}$}The renormalization point is $\Vec{q}=(\bar{l},\bar{k},\bar{q},\bar{p}) \in \mathbb{M}^s_4$, see \eqref{eq_9sym}. With $l=-k-q-p$,
\begin{align}
\Gamme_1^{\bar{c}^ac^bc^dA^s_\mu}(k,q,p)\sim&\, \Gamma^{c^d \bar{c}^a A^s_\mu A^t_\rho}(l,p,k)\Gamma^{c^b}_{\gamma^t_\rho}(k)-\Gamma^{c^b \bar{c}^a A^s_\mu A^t_\rho}(l,p,q)\Gamma^{c^d}_{\gamma^t_\rho}(q) \nonumber\\
&+\Gamma^{A^t_\rho c^d \bar{c}^a}(q,l)\Gamma^{A^s_\mu c^b}_{\gamma^t_\rho}(p,k) -\Gamma^{A^t_\rho c^b \bar{c}^a}(k,l)\Gamma^{A^s_\mu c^d}_{\gamma^t_\rho}(p,q) \nonumber\\
&+F^{AA}_{T;\mu \rho}(p)\Gamma^{ c^b c^d \bar{c}^a}_{\gamma^s_\rho}(k,q,l)+\Gamma^{c^t \bar{c}^a A^s_\mu}(l,p)\Gamma^{c^b c^d}_{\omega^t}(k,q) \nonumber\\
&+\Gamma^{c^t \bar{c}^a}(l)\sigma_{0 \Lambda_0}(l^2)\Gamma^{c^b c^d  A^s_\mu}_{\omega^t}(k,q,p)\,.
\end{align}
From equation \eqref{eq_cccA} it follows that $u^{\bar{c}ccA}_4\sim 0$ and 
\begin{equation}
u^{\bar{c}ccA}_1 \sim 0 \implies u^{\gamma A cc}\sim 0, u^{\bar{c}ccA}_2\sim 0, u^{\bar{c}ccA}_3 \sim 0\,.
\end{equation}
Consequently, we need only one condition
\begin{align}
u^{\bar{c}ccA}_1& \sim 0&& \Longleftrightarrow &g(R_2-R_3) R^{A\bar{c}c}_1 + \zeta^{\Lambda_0}_{\bar{c}ccA} &\sim 0\,.\label{eq_st_cccA}
\end{align}
This gives the renormalization condition for $R_3$.
\subsection{$\Gamme_1^{\bar{c} c c}$ and $\Gamme^{c c}_{1\gamma}$} \label{st_ccc}
From equations \eqref{eq_cccA}, \eqref{eq_cAA} we have
\begin{equation}
u^{\bar{c}ccA}_1 \sim 0 \implies u^{\gamma Acc} \sim 0 \implies u^{\gamma cc} \sim 0 \implies u^{\bar{c}cc}_{i \in \{1,2\}}\sim 0.
\end{equation}
Consequently, the marginal contribution to the functions is \textit{small}.
\subsection{$\Gamme_1^{c A A A}$}\label{st_cAAA}The renormalization point is $\Vec{q}=(\bar{l},\bar{k},\bar{q},\bar{p}) \in \mathbb{M}^{cp}_4$, see \eqref{eq_9cp}. With $l=-k-q-p$,
\begin{align}
\Gamme_1^{c^s A^t_\mu A^b_\nu A^d_\rho}(p,q,k)\sim&\Gamma^{A^a_\sigma A^t_\mu A^b_\nu A^d_\rho}(p,q,k)\Gamma^{c^s}_{\gamma^a_\sigma}(l) \nonumber\\
&+\sum \limits_{Z_3}\Gamma^{A^a_\sigma A^t_\mu A^b_\nu}(p,q)\Gamma^{A^d_\rho c^s}_{\gamma^a_\sigma}(k,l) \nonumber\\
&+\sum \limits_{Z_3} F^{AA}_{T;\rho \alpha}(k) \Gamma^{c^s A^t_\mu A^b_\nu}_{\gamma^d_\alpha}(l,p,q),
\end{align}
where $F^{AA}_T$ is defined in \eqref{def_aa_t}, and the sum $\sum_{Z_3}$ runs over all cyclic permutations of $\{(d,\rho,k),(b,\nu,q),(t,\mu,p)\}$. From \eqref{eq_cAAA} it follows that we need two equations
\begin{align}
u^{cAAA}_1&\sim 0&& \Longleftrightarrow& 8R_1R^{AAAA}_2-4gR_2R^{AAA} + \zeta^{\Lambda_0}_{cAAA,1}& \sim 0\,,\\
u^{cAAA}_3&\sim 0&& \Longleftrightarrow& 4R_1R^{AAAA}_1+2gR_2R^{AAA} + \zeta^{\Lambda_0}_{cAAA,3}& \sim 0\,.\label{eq_st_cAAA}
\end{align}
These equations give the renormalization conditions for $R^{AAAA}_{1,2}$.
\section{Acknowledgements}
We want to thank the reviewer for careful reading of the paper, useful remarks and criticism to which we paid serious attention. 

A.N. Efremov is indepted to the Institute for Theoretical Physics of the University of Leipzig whose facilities and financial support were used during the work on the final version of the manuscript.

We are especially greatful to S. Hollands for the time he devoted to study this work, numerous discussions and warm hospitality.

We also thank H. Gies from the University of Jena for his interest in our paper, valuable remarks and questions.
\begin{appendices}
\section{Properties of Gaussian measures} \label{sec_gauss}
In the following $d\nu=d\nu_{C}(A)d\nu_{S}(c,\bar{c})$ is the measure given in \eqref{def_cf}, \eqref{def_gauss}
\begin{gather}
d\nu_{C_1+C_2}(A)f(A)=d\nu_{C_1}(A_1)d\nu_{C_2}(A_2)f(A_1+A_2),\\
d\nu_{t C}(A)f(A) = d\nu_C(A)f(t^{\frac{1}{2}} A),\\
d\nu_C(A-\delta A)=d\nu_C(A)e^{-\frac{1}{2\hbar} \langle \delta A, C^{-1}_{\Lambda \Lambda_0} \delta A \rangle}e^{\frac{1}{\hbar}\langle A, C^{-1}_{\Lambda \Lambda_0} \delta A \rangle},\\
\frac{d}{d\Lambda}d\nu_{C}f(A)=\frac{1}{2}d\nu_{C}\langle \frac{\delta}{\delta A}, \hbar\dot C^{\Lambda \Lambda_0} \frac{\delta}{\delta A} \rangle f(A),\\
d \nu_C\, G(A) \left(AC^{-1}_{\Lambda \Lambda_0}-\hbar\frac{\delta }{\delta A} \right) F(A)  =d \nu_C \left( \hbar \frac{\delta}{\delta A} G(A) \right)  F(A). \label{eq_parts}
\end{gather}
When integrating over Grassmann variables one obtains
\begin{gather}
d\nu_{S_1+S_2}(c,\bar{c})f(\bar{c},c)=d\nu_{S_1}(c_1,\bar{c}_1)d\nu_{S_2}(c_2,\bar{c}_2)f(\bar{c}_1+\bar{c}_2,c_1+c_2),\\
d\nu_{t S}(c,\bar{c})f(\bar{c},c) = d\nu_S(c,\bar{c})f(t^{\frac{1}{2}} \bar{c},t^{\frac{1}{2}} c),\\
d\nu_S(c-\delta c,\bar{c}-\delta \bar{c})=d\nu(c,\bar{c})e^{\frac{1}{\hbar}\langle \delta \bar{c},S^{-1}_{\Lambda \Lambda_0} \delta c \rangle}e^{-\frac{1}{\hbar}(\langle \bar{c},S^{-1}_{\Lambda \Lambda_0}\delta c \rangle + \langle \delta \bar{c},S^{-1}_{\Lambda \Lambda_0}c \rangle)},\\
\frac{d}{d\Lambda}d\nu_{S}f(\bar{c},c)=d\nu_{S}\langle \frac{\delta}{\delta c}, \hbar \dot{S}^{\Lambda \Lambda_0} \frac{\delta}{\delta \bar{c}} \rangle f(\bar{c},c),\\
d\nu_{S}\,\tilde{G}(\bar{c},c) \left(-\bar{c}S^{-1}_{\Lambda \Lambda_0} + \hbar \frac{\delta}{\delta c}\right) \tilde{F}(\bar{c},c)=d\nu_{S} \left(\hbar \frac{\delta_R}{\delta c} \tilde{G}(\bar{c},c) \right) \tilde{F}(\bar{c},c),\label{eq_parts2a}\\
d\nu_{S}\, \tilde{G}(\bar{c},c) \left(S^{-1}_{\Lambda \Lambda_0}c + \hbar \frac{\delta}{\delta \bar{c}}\right) \tilde{F}(\bar{c},c)  = d\nu_{S} \left( \hbar\frac{\delta_R}{\delta \bar{c}} \tilde{G}(\bar{c},c) \right) \tilde{F}(\bar{c},c),\label{eq_parts2b}
\end{gather}
where right functional derivatives are distinguished from left ones by the label $R$. Properties \eqref{eq_parts}, \eqref{eq_parts2a}, \eqref{eq_parts2b}
are proved for
\begin{align}
G=&e^{\frac{i}{\hbar}\langle j,A \rangle},&F&=e^{\frac{i}{\hbar}\langle j^\prime,A \rangle},&\tilde{G}&=e^{\frac{i}{\hbar}(\langle \bar{c},\eta \rangle + \langle \bar{\eta}, c \rangle)},&\tilde{F}&=e^{\frac{i}{\hbar}(\langle \bar{c}, \eta^\prime \rangle + \langle \bar{\eta}^\prime, c \rangle)}\,,
\end{align}
and extended to polynomials in the fields by functional differentiation.
\section{Chains of vertex functions} \label{ex_AAcc}
For the purpose of example we give the complete list of reduced chains which appear in the loop integrals for $\dot{\Gamma}^{AAc\bar{c};w}$, together with the corresponding "dotted” propagators. The external fields are underlined. Moreover, $\sum_{i=0}^{k-1} w_i=w$, $k$ being the number of vertex functions in each chain.
\begin{equation}\begin{array}{ll}
\dot{C}\Gamma^{A \ubar{A} A;w_0}C\Gamma^{A \ubar{A} A;w_1} C\Gamma^{A \ubar{c} \bar{c};w_2}S\Gamma^{c \wbar{c} A;w_3},&\dot{C}\Gamma^{AA \ubar{A} \ubar{A} \ubar{c} \wbar{c};w},\\
\dot{S}\Gamma^{c\ubar{A}\bar{c};w_0}S\Gamma^{c \ubar{A} \bar{c};w_1} S\Gamma^{Ac\wbar{c};w_2}C\Gamma^{\ubar{c} \bar{c}A;w_3},&\dot{S}\Gamma^{c \bar{c} \ubar{A} \ubar{A} \ubar{c} \wbar{c} ;w},\\
\dot{C}\Gamma^{A \ubar{A} \ubar{A} A;w_0} C\Gamma^{A \ubar{c} \bar{c};w_1}S\Gamma^{c \wbar{c} A;w_2},&\dot{C}\Gamma^{A A \ubar{A} \ubar{c} \wbar{c};w_0} C \Gamma^{A \ubar{A} A;w_1},\\
\dot{S}\Gamma^{c\ubar{A} \ubar{A} \bar{c};w_0}S\Gamma^{Ac\wbar{c};w_1}C\Gamma^{\ubar{c} \bar{c}A;w_2},&\dot{S}\Gamma^{\ubar{A} \ubar{c} \wbar{c} c \bar{c};w_0} S \Gamma^{\ubar{A} c \bar{c};w_1},\\
\dot{C}\Gamma^{A \ubar{A} \ubar{c} \bar{c};w_0} S\Gamma^{\ubar{A} c \bar{c};w_1}S\Gamma^{A c \wbar{c};w_2},&\dot{C}\Gamma^{A \ubar{A} \ubar{A} c \wbar{c};w_0} S\Gamma^{A \ubar{c} \bar{c};w_1},\\
\dot{S}\Gamma^{\bar{c} \ubar{A} \ubar{c} A;w_0}C\Gamma^{A \ubar{A} A;w_1}C\Gamma^{A c \wbar{c} ;w_2},&\dot{S}\Gamma^{A \ubar{A} \ubar{A} c \wbar{c} ;w_0} C\Gamma^{A \ubar{c} \bar{c};w_1},\\
\dot{C}\Gamma^{A \ubar{A} \wbar{c} c;w_0} S\Gamma^{A \ubar{c} \bar{c};w_1}C\Gamma^{A \ubar{A} A;w_2},&\dot{C}\Gamma^{A \ubar{A} \ubar{A} \ubar{c} \bar{c};w_0} S\Gamma^{A c \wbar{c};w_1},\\
\dot{S}\Gamma^{c \ubar{A} \wbar{c} A;w_0}C\Gamma^{A \ubar{c} \bar{c};w_1}S\Gamma^{\ubar{A} c \bar{c} ;w_2},&\dot{S}\Gamma^{A \ubar{A} \ubar{A} \ubar{c} \bar{c};w_0} C\Gamma^{A c \wbar{c};w_1},\\
\dot{C}\Gamma^{A \ubar{c} \wbar{c} A;w_0} C\Gamma^{A \ubar{A} A;w_1}C\Gamma^{A \ubar{A} A;w_2},&\dot{C}\Gamma^{A \ubar{A} c \wbar{c};w_0} S\Gamma^{A \ubar{A} \ubar{c} \bar{c};w_1},\\
\dot{S}\Gamma^{c \ubar{c} \wbar{c} \bar{c};w_0}S\Gamma^{\ubar{A} c \bar{c};w_1}S\Gamma^{\ubar{A} c \bar{c} ;w_2},&\dot{S}\Gamma^{A \ubar{A} c \wbar{c};w_0}C\Gamma^{A \ubar{A} \ubar{c} \bar{c} ;w_1},\\
\dot{C}\Gamma^{A \ubar{A} \ubar{A} A;w_0} C\Gamma^{A \ubar{c} \wbar{c} A;w_1},&\dot{S}\Gamma^{\ubar{A} \ubar{A} c \bar{c};w_0}S\Gamma^{c \bar{c} \ubar{c} \wbar{c} ;w_1}.\\
\end{array}\end{equation}
\section{Tensors}\label{sec_t}
For the definition of the tensor monomial sets $\{\delta^s q^n\}$, $\{\delta^s q^n\}_r$ see beginning of page \pageref{sec_21bnd}.
\begin{lemma}\label{lem_monom}Let $q=(q_1,...,q_m)$ where $q_i \in \mathbb{R}^D$ are $m \in \mathbb{N}$ linearly independent vectors. Then the tensor monomials $\{\delta^s q^n\}_r$ of positive rank $r=2s+n\leqslant 2(D-m)+1$ are linearly independent,
\begin{equation}
\sum \limits_{t \in \{\delta^s q^n\}_r} c_t \, t=0\implies c_t=0,\;\forall t. \label{eq_monom}
\end{equation}
\end{lemma}
\paragraph{Proof} 
Observe that, for $m,r,D\in\mathbb{N}$,
the inequality $r\leqslant 2(D-m)+1$
is equivalent to $m+s\leqslant D$ for all $s,n\in\mathbb{N}$
such that $r=n+2s$.
Let $I:=\{1,...,r\}$.
Let $\mathcal{P}_{s}$ be the set of all divisions of the set $I$ in $m+s$ pairwise-disjoint, possibly-empty sets,
\begin{equation}
I=\big(\cup_{j=1}^m V_j \big)\;\bigcup\;\big(\cup_{k=1}^{s}S_k\big),
\end{equation}
such that $S_k=\{s_k^1,s_k^2\}$, $s_k^1<s_k^2$,
and $\min{S_1}<...<\min{S_s}$.
There is a bijection that maps a division
$(V_j,S_k)\in\mathcal{P}_{s}$ to a tensor monomial $t\in\{\delta^s q^n\}$, constructed by the relation
\begin{equation}
t_{\mu_1...\mu_r}=
\prod_{j=1}^m \prod_{v_j\in V_j} q_{j;\mu_{v_j}}
\prod_{k=1}^s \delta_{\mu_{s_k^1}\mu_{s_k^2}}.
\end{equation}
Let us first prove the statement of the lemma for orthonormal $q_j$. In an appropriate basis of $\mathbb{R}^D$, their components are
\begin{align}\label{qkcomp}
q_{j;\mu}&=\delta_{j\mu},& j\in \{1,...,m\},\quad \mu \in \{1,...,D\}.
\end{align}
Let us assume that $\sum_{t} c_t\, t=0$, with $t\in \{\delta^s q^n\}_r$. We will proceed by proving inductively that $c_{t}=0$ for all $t\in \{\delta^s q^n\}$, from $s=D-m$ down to $s=0$. Fix $\bar{s}\le D-m$ and assume that
$c_t=0$ for all $t$ involving more than $\bar{s}$ Kronecker's tensors (which is vacuously true for $\bar{s}=D-m$, due to the rank constraint).
Let us prove that $c_{\bar{t}}=0$ for an arbitrary
$\bar{t}\in\{\delta^{\bar{s}} q^{\bar{n}}\}$,
which is associated to a division
$(\bar{V}_j,\bar{S}_k)\in\mathcal{P}_{\bar{s}}$.
Fix the values of the indices $\bar{\mu}_i$, with $i\in I$, by
\begin{align}
\bar{\mu}_i&=\left \{ \begin{matrix} j,&\mbox{if }\exists j \mbox{ such that }i\in \bar{V}_j,\\ m+k,&\mbox{if } \exists k \mbox{ such that } i \in \bar{S}_k,\end{matrix} \right. \label{eq_17idx}
\end{align}
Note that this choice is possible because $m+\bar{s}\leqslant D$.
It is enough to show
that whenever $t_{\bar{\mu}_1...\bar{\mu}_r}\neq0$ for $t\in\{\delta^s q^n\}$ and $s\leqslant\bar{s}$ (i.e.~$n\geqslant \bar{n}$) then $s=\bar{s}$ and $t=\bar{t}$:
in fact this property,
the inductive hypothesis, and the vanishing of the sum
$\sum_t c_t\, t$ imply that $c_{\bar{t}}=0$.
To prove the aforementioned property,
introduce the division $(V_j, S_k)\in\mathcal{P}_{s}$ defining the tensor $t$ and, using~\eqref{qkcomp}, correspondingly write
\begin{equation}\label{tNotVan}
0\neq t_{{\bar{\mu}}_1...{\bar{\mu}}_r}=
\prod_{j=1}^m \prod_{v_j\in V_j} \delta_{j{\bar{\mu}}_{v_j}}
\prod_{k=1}^s \delta_{{\bar{\mu}}_{s_k^1}{\bar{\mu}}_{s_k^2}}.
\end{equation}
Relations \eqref{eq_17idx} and \eqref{tNotVan} imply that
$V_j\subseteq \bar{V}_j$ for all $j$, which, together with the inductive condition $n\geqslant\bar{n}$, leads to $n=\bar{n}$ and, because the rank $r$ is fixed, to $s=\bar{s}$.
Relations \eqref{eq_17idx}, \eqref{tNotVan}, and $s=\bar{s}$ imply that there is an injective map $f:I\to I$ such that
$S_j=\bar{S}_{f(j)}$. By definition of the $S_j$, it then follows
that $\min\bar{S}_{f(1)}<...<\min\bar{S}_{f(1)}$: this is only possible if $f$ is the identity, which concludes the first part of the proof.

Let us now prove the statement for $m$ linearly independent vectors $p_1,...,p_m$. The sum $\sum_{t} c_t\, t=0$, with $t\in\{\delta^s p^n\}_r$, may be rewritten as
\begin{equation}
\sum_{2s+n=r}\,\sum_{1\leqslant k_1,...,k_n\leqslant m}\,\sum_{\pi}\, c_{k_1,...,k_n;\pi} \, \prod^n_{j=1} (p_{k_j})_{\mu^\pi_{j}}\,\prod^{s}_{j'=1}\delta_{\mu^\pi_{m+2j'-1}\mu^\pi_{m+2j'}}
=0,
\end{equation}
where $\mu^\pi_j:=\mu_{\pi(j)}$ and the sum over $\pi$ runs over the right coset of permutation groups $S_r\backslash(S_n {\times} S_s {\times} S^s_2 )$. Expressing the $p_k$ in terms of $m$ orthonormal vectors $q_{k^\prime}$, $p_k=A_{k k^\prime}q_{k^\prime}$, gives a tensor transformation leading to the coefficients in the $\{\delta^s q^n\}_r$ basis:
\begin{equation}
c^{\prime}_{k^\prime_1,...,k^\prime_n;\pi}=c_{k_1,...,k_n;\pi}\prod^n_{j=1} A_{k_j k^\prime_j}.
\end{equation}
The validity of equation \eqref{eq_monom} for the $q_k$ implies that $c^{\prime}_{k^\prime_1,...,k^\prime_n;\pi}=0$, which, by invertibility of the matrix $A \in GL(m,\mathbb{R})$, gives $c_{k_1,...,k_n;\pi}=0$.
\hfill$\blacksquare$\\
Note that for $m=n_\phi-1$ and $D=4$, the condition for linear independence of the monomials in $\{\delta^s q^k\}_r$ reads
\begin{equation}
r \leqslant 2(4-(n_\phi-1))+1=11-2n_\phi.
\end{equation}
\begin{lemma}\label{lem_monom2}
Let $q=(q_1,...,q_m)$
with $q_i\in\mathbb{R}^D$ and $m$ nonnegative integer.
The tensor monomials $\{\delta^s q^n\}_r$ of positive rank
$r\geqslant 2(D-m+1)$  are linearly dependent.
\end{lemma}
\paragraph{Proof}
The signature of a permutation $\pi$ is denoted by $(-)^\pi$.
\begin{itemize}
\item
The Gram matrix $g:= g(t_1,\dots,t_k)$
of $k$ tensors $t_i$ of equal rank
is defined by its components:
$g_{ij}:=(t_i,t_j)$ for all $i,j\in[k]$.
The scalar product \eqref{scalar_product} is $O(D)$-invariant:
$(\underline{R}\,t,\underline{R}\,u)=(t,u)$
for all tensors $t,u$ of rank $r$
and all
$\underline{R}\,:=R\otimes\cdots\otimes R$ ($r$ times)
with $R\in O(D)$. As a consequence,
$g(\underline{R}\,t_1,\dots,\underline{R}\,t_k)
=g(t_1,\dots,t_k)$
for $k$ tensors of rank $r$.
It is a well-known fact that tensors of equal rank
$t_1,\dots,t_k$
are linearly independent iff their Gram matrix 
$g(t_1,\dots,t_k)$ is invertible.
It follows that
$t_1,\dots,t_k$ are linearly independent iff
$\underline{R}\,t_1,\dots,\underline{R}\,t_k$
are linearly independent for some $\underline{R}$ as above
stated.
\item
We assume that $m>0$ because
whenever $m=0$ for each $s\geqslant D+1$ one has
\begin{flalign}\label{ldep-relation2ter}
\sum_{\pi\in S_s}
(-)^\pi 
\delta_{\mu_1\nu_{\pi(1)}}
\cdots
\delta_{\mu_s\nu_{\pi(s)}}=0\,.
\end{flalign}
\item
We assume that $q_1,\dots,q_m$ are linearly independent
because otherwise
there exist $c_i$ not all vanishing such that
$\sum_{i\in[m]}c_i\, q_i=0$,
which, for every $r>0$, yields
\begin{flalign}\label{ldep-relation3}
\sum_{i_1,\dots,i_r\in[m]} c_{i_1}\cdots c_{i_r}\,q_{i_1}\otimes\cdots\otimes q_{i_r}=0\,.
\end{flalign}
\item
If the tensors $\{\delta^s q^n\}_r$
are linearly dependent
then
the tensors $\{\delta^s q^n\}_{r'}$
are linearly dependent,
for every $r'> r$.
Proof:
linear dependence of the tensors $\{\delta^s q^n\}_r$ yields
\[
\sum_{t\in \{\delta^s q^n\}_r} c_t\, t=0\,,
\]
where not all $c_t$ vanish.
Linear independence of $q_1,\dots,q_m$
implies that $q_1\neq0$.
The proof is concluded by
applying $r'-r$ times the tensorial product $\otimes q_1$,
which gives
\[
\sum_{t\in \{\delta^s q^n\}_r}
c_t\,t\otimes (q_1)^{\otimes r'-r}=0\,.
\]
\item
For $q_1,\dots, q_D$ linearly independent
the statement of the lemma follows from the previous fact
and a well-known relation involving the Gram matrix
$g:=g(q_1,\dots,q_D)$:
\begin{flalign}\label{ldep-Kron=}
\delta_{\mu\nu}=
\sum_{i,j\in[D]}q_{i,\mu}\, (g^{-1})_{ij}\,q_{j,\nu}\,.
\end{flalign}
\item
It is then enough to prove the statement of the lemma for
${r=2(D-m+1)}$, ${0<m<D}$,
and $q_1,\dots,q_m$ linearly independent and
such that $q_i^\mu=0$ for all ${\mu\in[m+1: D]}$ and
${i\in[m]}$.
There exist
$q_{m+1},\dots,q_D$ orthonormal vectors such that
$q_i^\mu=\delta_i^\mu$ for all
${\mu\in[D]}$ and $i\in[m+1 : D]$.
Relation \eqref{ldep-Kron=} then gives
\begin{flalign}\label{ldep-deltaperp}
\delta_{\mu\nu}=
\sum_{i,j\in[m]}q_{i,\mu}\, (g_{\sslash}^{-1})_{ij}\,q_{j,\nu}
+\delta^\perp_{\mu\nu}\,,
\end{flalign}
where $g_{\sslash}:=g(q_1,\dots,q_m)$ and
$\delta^\perp_{\mu\nu}
:=\sum_{i\in [m+1 : D]} q_{i,\mu}q_{i,\nu}$.
For $s=D-m+1$, one has
\begin{flalign}\label{ldep-relation}
\sum_{\pi\in S_{s}}
(-)^\pi \,
\delta^\perp_{\mu_{m+1}\nu_{m+\pi(1)}}
\cdots
\delta^\perp_{\mu_{m+s}\nu_{m+\pi(s)}}=0\,.
\end{flalign}
Combining \eqref{ldep-deltaperp} and \eqref{ldep-relation}
concludes the proof of the lemma.
\end{itemize}
\hfill$\blacksquare$\\

The following lemma states a necessary condition for a regular, $O(4)$-invariant tensor field.
\begin{lemma}\label{lem_pr}Let $f(\ubar{y})$ be a regular, $O(4)$-invariant tensor field of rank $r$ where $\ubar{y}:=(y_1,...,y_m)$ with $y_j \in \mathbb{R}^4$. Assume that the tensor monomials $\{\delta^s\ubar{y}^k\}_{r}$ as well as $\{\delta^s\ubar{y}^k\}_{r+1}$ are linearly independent pointwise for all $\ubar{y} \in O$ where $O$ is some open set. Then on $O$ we have
\begin{align}
f&=\sum \limits_{t \in \{\delta^s\ubar{y}^k\}_{r}} f_{t}\, t\,,&\partial_j f&=\sum \limits_{t^\prime \in \{\delta^s\ubar{y}^k\}_{r+1}} f_{j,t^\prime}\, t^\prime\,. 
\end{align}
Furthermore for every $t \in \{\delta^s \ubar{y}^{k>0}\}_r$ there exist $j$ and $t^\prime \in \{\delta^s \ubar{y}^k\}_{r+1}$ such that $f_t=f_{j,t^\prime}$ on $O$.
\end{lemma}
\paragraph{Proof} For shortness we consider only the case $m=2$. We have
\begin{equation}
f=\sum \limits_{t \in \{\delta^s\}_{r}} u_{t}\, t   + \sum_{t \in \{\delta^s\ubar{y}^{k>0}\}_{r}} \zeta_t \, t,\label{eq_415a}
\end{equation}
where  $u_t$, $\zeta_t$ are regular functions of the scalar parameters $\mathbb{X}=\{\tfrac{1}{2}y^2_1,\tfrac{1}{2}y^2_2,y_1y_2\}$. Apply the operator $\partial_j$ to both sides of \eqref{eq_415a}. The Leibniz rule  gives
\begin{align}
\partial_{j} (u_{t} \delta^s) &= \sum \limits_{x \in \mathbb{X}}  (\partial_x u_{t})\, \delta^s \, \partial_j x,\label{eq_415b}\\
\partial_{j} (\zeta_{t} \delta^s\ubar{y}^k) &= \sum \limits_{x \in \mathbb{X}}  (\partial_x \zeta_{t})\, \delta^s \, \ubar{y}^k  \partial_j x \;+ \zeta_{t} \delta^s \, \partial_j \ubar{y}^k,\label{eq_415c}
\end{align}
where $\partial_j x\in\{0,y_1,y_2\}$, $\ubar{y}^k:=y_{i_1}... y_{i_k}$ with $i_l \in \{1,2\}$, and
\begin{equation}
\partial_j\ubar{y}^k=\partial_j \prod^k_{l=1} y_{i_l} = \sum^k_{q=1}  \delta_{j,i_q}\,  \prod^k_{\substack{l=1\\l\neq q}} y_{i_l}.
\end{equation}
The nonvanishing tensor monomials arising at a given $j\in\{1,2\}$ from the rhs of \eqref{eq_415b} and \eqref{eq_415c} have rank $r+1$, are linearly independent by assumption, and are pairwise different. Each coefficient $\zeta_{t}$ of the tensor $\delta^s y_{i_1}...y_j...y_{i_k}$ in the decomposition \eqref{eq_415a} appears also as the coefficient of $\delta^s y_{i_1}...\delta ...y_{i_k}$  in the decomposition \eqref{eq_415c} for $\partial_j f$. \hfill$\blacksquare$

Lemma \ref{lem_rn} below relies on lemma~\ref{lem_monom},
which shows that, for $m$ linearly independent vectors
$\underline{e}=(e_1,\cdots,e_{m})$,
the relation $r+1\leqslant 9-2m$ is a
sufficient condition for the linear independence
of the tensor monomials $\{\delta^s\underline{e}^k\}_{r}$
and $\{\delta^s\underline{e}^k\}_{r+1}$, see lemma~\ref{lem_monom2} for a necessary condition. The renormalization points in appendix \ref{sec_ren} are chosen to comply with the aforementioned relation. The proof of lemma \ref{lem_rn} is in the same spirit as the one of the preceding lemma~\ref{lem_pr}.
\begin{lemma}\label{lem_rn}Let $F$ be a regular, $O(4)$-invariant tensor field of rank $r\in \{2,4\}$ on~$\mathbb{P}_n$.
Let be given $\Vec{q} \in \mathbb{M}_n$ and $m\geqslant 2$ linearly independent vectors $\underline{e}=(e_1,\cdots,e_{m})$, such that $\mathrm{span}(\Vec{q})=\mathrm{span}(\ubar{e})$.
Assume that $r+1\leqslant 9-2m$. By lemma~\ref{lem_monom} there exist unique coefficients $F_t$ such that $F(\Vec{q})=\Sigma_{t \in \{\delta^s \underline{e}^k\}_r}\, F_t\, t$.
Furthermore,
\begin{align}
|F(\Vec{q})|&\leqslant c \max\Big( \big(|F_t|\big)_{t \in \{\delta^s \}_r},(M|\partial_k F (\Vec{q})|)_{k\in[n-1]}\Big),\label{eq_73}
\\
\big|\sum_{t \in \{\delta^s \underline{e}^{k>0}\}_r}\,F_t\,t\big|&
\leqslant c\,  \max\Big( (M|\partial_k F (\Vec{q})|)_{k\in[n-1]}\Big)\,.
\label{eq_73bis}
\end{align}
The bounds hold with the same constant $c$ for all $F$ of equal rank.
\end{lemma}
\paragraph{Proof} The coefficients $\big(|F_t|\big)_{t \in \{\delta^s \}_r}$ in the basis $\{\delta^s \ubar{e}^k\}_r$ do not depend on the choice of the vectors $\ubar{e}$. Hence it is enough to prove \eqref{eq_73} in the case when $e_ie_j=M^2\delta_{ij}$. For simplicity we assume that $m=2$, the extension to other $m$ being clear.\\By hypothesis, there exists a $(n-1)\times 2$ matrix $L$ such that  $q_k=L_{ki} e_i$ and
\begin{equation}
|L|:= \sqrt{L_{ki} L_{ki}}=\frac{1}{M}\sqrt{L_{ki} L_{kj}e_ie_j}=\frac{1}{M}\Big(\sum^{n-1}_{k=1} q^2_k\Big)^{\frac{1}{2}}\leqslant (n-1)^{\frac{1}{2}}.\label{eq_26a}
\end{equation}
Denote by $\mathbb{E}\subset{\mathbb{R}}^4$ the linear span of the vectors $q_0,..,q_{n-1}$.
The matrix $L$ induces a linear map ${L:\mathbb{E}^2 \to \mathbb{R}^{4(n-1)}}$, $\ubar{y}\mapsto L\ubar{y}$, where $\ubar{y}=(y_1,y_2)$ and $(L\ubar{y})_k=L_{ki}y_i$. We also define an auxiliary function on $\mathbb{E}^2$: 
$f(\ubar{y}):=F(L\ubar{y})$.
Setting $\partial_{y_i}:=\partial/\partial y_i$ and $\partial_k:=\partial/\partial p_k$, the Cauchy--Schwarz inequality and \eqref{eq_26a} imply that,
\begin{equation}\label{eqdyGa}
\textstyle{\sum}_{i=1}^2|\partial_{y_i} f(\ubar{e})|^2\leqslant |L|^2 \; \sum_{k=1}^{n-1}|\partial_k F(\Vec{q})|^2 \leqslant (n-1)^2 \max \limits_k (|\partial_k F(\Vec{q})|^2).
\end{equation}
\noindent
For all $y_1,y_2 \in \mathbb{E}$, denote as usual by  $\{\delta^s \ubar{y}^k\}$ the set  of all monomials being a tensor product of $k$ vectors in
$\ubar{y}=(y_1,y_2)$ and of $s$ Kronecker tensors, and by  $\{\delta^s \ubar{y}^k\}_r$ the union of 
all $\{\delta^s \ubar{y}^k\}$ such that $2s+k=r$.
By \eqref{eq_monom}, whenever $r\leqslant 5$ and $y_1,y_2$ are linearly independent, the elements of $\{\delta^s \ubar{y}^k\}_r$ are
linearly independent. In this case, we label the tensor monomials by fixing  a family of disjoint sets $A_{k,s}$ and a family of bijections 
$\alpha\mapsto t_\alpha$ from each $A_{k,s}$ to $\{\delta^s \ubar{y}^k\}$. Furthermore, we define the auxiliary sets
\begin{align}
A^r&:=\bigcup \limits_{2s+k=r}A_{k,s}\,,
&A^r_+&:=\bigcup \limits_{2s+k=r,\,k\geqslant 1}A_{k,s}\,,
&A^r_0&:=A^r\setminus A^r_+\,.\label{eq_415d}
\end{align}
For $r \in \{2,4\}$ and $\ubar{y}=(y_1,y_2)$ in an open neighborhood of $\ubar{e}$ (in $\mathbb{E}^2$) for which $y_1,y_2$ are linearly independent, we write the following tensor decomposition:
\begin{equation}
f(\ubar{y})=\sum \limits_{\alpha \in A^r_0} u_{\alpha} t_\alpha   + \sum_{\alpha\in A^r_+} \zeta_{\alpha} t_\alpha,\label{eq_24g}
\end{equation}
where  $u_\alpha$, $\zeta_\alpha$ are regular functions of the scalar parameters $\mathbb{X}=\{\tfrac{1}{2}y^2_1,\tfrac{1}{2}y^2_2,y_1y_2\}$.
 Evaluating \eqref{eq_24g} at $\ubar{y}=\ubar{e}$ and using the general fact that $|t|=2^s M^k$ for ${t\in\{\delta^s \ubar{e}^k\}}$ we obtain
\begin{equation}
|f(\ubar{e})|\leqslant \sum_{\alpha \in A^r_0}  2^s\, |u_{\alpha}| +\sum_{\alpha \in A^r_+} |\zeta_{\alpha}|\,2^s M^k.\label{eq_26gao}
\end{equation}
We now want to prove the existence of a constant $c_1>0$ such that
\begin{equation}
\sum_{\alpha \in A^r_+} |\zeta_{\alpha}|\,2^s M^k
\;\leqslant\;
c_1 M \Big(\sum_{i\in\{1,2\}}|\partial_{y_i}f(\ubar{e})|^2\Big)^{1/2}.
\label{eq_24bnd}
\end{equation}
Apply the operator $\partial_{y_i}$ to both sides of \eqref{eq_24g}. The Leibniz rule  gives
\begin{align}
\partial_{y_i} (u_{\alpha} \delta^s) &= \sum \limits_{x \in \mathbb{X}}  (\partial_x u_{\alpha})\, \delta^s \, \partial_{y_i} x,\label{eq_24t0}\\
\partial_{y_i} (\zeta_{\alpha} \delta^sy^k) &= \sum \limits_{x \in \mathbb{X}}  (\partial_x \zeta_{\alpha})\, \delta^s \, y^k  \partial_{y_i} x \;+ \zeta_{\alpha} \delta^s \, \partial_{y_i}y^k,\label{eq_24t1}
\end{align}
where $\partial_{y_i} x\in\{0,y_1,y_2\}$, $y^k=y_{i_1}... y_{i_k}$ with $i_l \in \{1,2\}$, and
\begin{equation}\label{eq_24dyn}
\partial_{y_i}y^k=\partial_{y_i} \prod^k_{l=1} y_{i_l} = \sum^k_{j=1}  \delta_{i,i_j}\,  \prod^k_{\substack{l=1\\l\neq j}} y_{i_l}.
\end{equation}
Fix $\ubar{y}=\ubar{e}$. The nonvanishing tensor monomials arising at a given $i\in\{1,2\}$ from the rhs of \eqref{eq_24t0} and \eqref{eq_24t1} have rank $r+1\in\{3,5\}$,  are pairwise different, and  are a subset of the tensor monomials in $A^{r+1}$,  themselves linearly independent by \eqref{eq_monom}.
Denote by $B^{r+1}_i\subset A^{r+1}_+$ the subset labeling the monomials of type
$\delta^s\delta_{i,i_j}\prod_{l\neq j}y_{i_l}$
arising from \eqref{eq_24t1} and \eqref{eq_24dyn} at the given $i$.
By construction, we can define the maps
${\pi_i:B^{r+1}_i\to A^{r}_+}$ with $i\in\{1,2\}$
such that:\\
i)
if $t_{\pi_i(\beta)}\in\{\delta^s \ubar{e}^k\}$
then $t_{\beta}\in\{\delta^{s+1} \ubar{e}^{k-1}\}$
(in this case, $|t_{\pi_i(\beta)}|=2^sM^k$ and $|t_{\beta}|=2^{s+1}M^{k-1}$);\\
ii) for each $\beta\in B^{r+1}_i$
the coefficient of $t_\beta$ in \eqref{eq_24t1} and that of 
$t_{\pi_i(\beta)}$ in \eqref{eq_24g} are the same, namely
$\zeta_{\pi_i(\beta)}$;\\
iii)
$\pi_1(B^{r+1}_1)\cup\pi_2(B^{r+1}_2)=A^{r}_+$.\\
The following bound holds at $\ubar{y}=\ubar{e}$ for every tensor $\Psi=\sum_\beta \Psi_\beta t_\beta/|t_\beta|$ with $\beta\in A^{r'}$ and $r'\leqslant5$, and for every nonempty $B\subset A^{r'}$:
\begin{equation}
|\Psi|^2=(\Psi,\Psi)=\Psi_\beta^*\, G_{\beta \beta^\prime} \,\Psi_{\beta^\prime}\geqslant \lambda_\mathrm{1}\sum_{\beta \in A^{r'}} |\Psi_\beta|^2 \geqslant \lambda_\mathrm{1} \sum_{\beta \in B} |\Psi_\beta|^2,\label{eq_26gram}
\end{equation}
where $\lambda_\mathrm{1}>0$ is the smallest eigenvalue of the Gram matrix of components $G_{\beta\beta^\prime}:=(t_\beta,t_{\beta'})/(|t_\beta||t_{\beta'}|)$, which is positive definite by \eqref{eq_monom}.
Application of \eqref{eq_26gram} to $\Psi_i:=\partial_{y_i} f(\ubar{e})$ for each $i\in\{1,2\}$,  with $r'=r+1$ and $B=B^{r+1}_i$, gives
\begin{equation}
\sum_{i\in\{1,2\}}|\partial_{y_i} f(\ubar{e})|^2
\geqslant
\lambda_\mathrm{1}
\sum\limits_{\substack{i\in\{1,2\}\\\beta \in B^{r+1}_i}}
|\zeta_{\pi_i(\beta)}|^2 |t_\beta|^2
\geqslant
\frac{4\lambda_\mathrm{1}}{M^2}
\sum_{\alpha \in A^{r}_+} |\zeta_\alpha|^2 \,(2^{s} M^{k})^2
,\label{sumpaf}
\end{equation}
from which follows the bound \eqref{eq_24bnd}.
\\
Inequalities \eqref{eq_26gao}, \eqref{sumpaf} and \eqref{eqdyGa} lead to the bounds~\eqref{eq_73}, \eqref{eq_73bis}, with a constant
\begin{equation}
c=\max\Big(2^{\frac{r}{2}}|A_0|, (n-1)\sqrt{\frac{|A^r_+|}{4\lambda_1}}\Big)\,.
\end{equation}
\hfill$\blacksquare$
\section{Basic estimates}\label{sec_b}
\begin{lemma}Let $0\leqslant q \leqslant p \leqslant P$, $k \in \mathbb{N}$. Then $\exists C_k>0$ such that
\begin{equation}
\int \limits^{+\infty}_\Lambda  \frac{d\lambda \; \log^k_+ \frac{P}{\lambda}}{(\lambda + p) (\lambda + q)}\leqslant C_k\frac{1+\log^{k+1}_+ \frac{P}{\Lambda + q}}{\Lambda + p + q}.\label{eq_twin}
\end{equation}
\end{lemma}
\paragraph{Proof} Let $I^k$ be the left hand side of the inequality and
\begin{equation}
I^k_{[a,b]}:=\int \limits^{b}_a  \frac{d\lambda \; }{(\lambda + p) (\lambda + q)}\log^k_+ \frac{P}{\lambda}.
\end{equation}
We begin with the case $k \geqslant 1$.
\begin{itemize}
\item $\Lambda > P$, $I^k=0$.
\item $q \leqslant \Lambda \leqslant P$, $I^k=I^k_{[\Lambda,P]}$.
\begin{align}
I^k\leqslant& \frac{1}{\Lambda + p}  \int \limits^{P}_\Lambda \frac{d \lambda}{\lambda + q}\log^k \frac{2P}{\lambda + q}\leqslant \frac{2}{k+1}\frac{1}{\Lambda + p+q}\log^{k+1} \frac{2P}{\Lambda + q}\nonumber\\
&\leqslant \frac{A_k}{\Lambda + p+q} (\log^{k+1}_+ \frac{P}{\Lambda + q} + 1),
\end{align}
where
\begin{equation}
A_k:=\frac{2 (\log 2 + 1)^{k+1}}{k+1},
\end{equation}
and we have used the inequality
\begin{equation}
\frac{\log_+ x + \log 2}{(1+\log^n_+ x)^{\frac{1}{n}}}<1 + \log 2.
\end{equation}
\item $\Lambda<q \leqslant P$, $I^k=I^k_{[\Lambda,q]} + I^k_{[q,P]}$.
\begin{align}
I^k_{[q,P]}\leqslant& \frac{1}{\Lambda + p} \int \limits^{P}_q \frac{d \lambda}{\lambda + q} \log^k \frac{2P}{\lambda+ q}\leqslant \frac{1}{\Lambda + p} \int \limits^{P}_\Lambda \frac{d \lambda}{\lambda + q} \log^k \frac{2P}{\lambda+ q}\nonumber\\
&\leqslant \frac{A_k}{\Lambda + p+q} (\log^{k+1}_+ \frac{P}{\Lambda + q} + 1),\\
I^k_{[\Lambda,q]}\leqslant& \frac{1}{(\Lambda + p)(\Lambda+q)} \int \limits^q_0 d \lambda \;\log^k \frac{P}{\lambda}< \frac{q  k!}{(\Lambda + p)(\Lambda+q)} \sum \limits^k_{j=0} \frac{1}{j!}\log^j \frac{P}{q}\nonumber\\
&< \frac{e k!}{\Lambda + p}( \log^k \frac{2P}{\Lambda +q} + 1)< \frac{2 e k!}{\Lambda + p + q}( \log^k_+ \frac{P}{\Lambda +q} + 2).
\end{align}
\end{itemize}
It remains to consider the case $k=0$.
\begin{itemize}
\item $\Lambda > p$. This implies $2(\lambda + q)>\lambda+p +2q> \lambda+p$, $\forall \lambda>\Lambda$.
\begin{equation}
I^0= 2 \int \limits^{+\infty}_\Lambda \frac{d \lambda}{(\lambda+p)^2}<\frac{4}{\Lambda + p + q}.
\end{equation}
\item $\Lambda \leqslant p$, $I^0=I^0_{[\Lambda,p]} + I^0_{[p,+\infty]}$.
\begin{align}
I^0_{[p,+\infty]}<&2 \int \limits^{+\infty}_p \frac{d \lambda}{(\lambda+p)^2}<\frac{1}{p}<\frac{3}{\Lambda +p+q},\\
I^0_{[\Lambda,p]}<&\frac{1}{\Lambda+p}\int \limits^{p}_\Lambda \frac{d \lambda}{\lambda+q}<\frac{2}{\Lambda + p + q} \log\frac{2P}{\Lambda + q}\nonumber\\
&<\frac{2}{\Lambda + p + q}(\log 2+ \log_+ \frac{P}{\Lambda + q}).
\end{align}
\end{itemize}
\hfill$\blacksquare$
\begin{lemma}Let $0 \leqslant q \leqslant p$, $\eta:=\min(M,q)$, $k \in \mathbb{N}$. Then $\exists C_k>0$ such that
\begin{equation}
\int \limits^{+\infty}_\Lambda  \frac{d\lambda \; \log^k_+ \frac{\lambda}{M}}{(\lambda + p) (\lambda + q)} \leqslant C_k \frac{1+ \log^{k+1}_+ \frac{p}{\Lambda + \eta}+ \log^{k+1}_+ \frac{\Lambda}{M}}{\Lambda + p + q}. \label{eq_twin2}
\end{equation}
\end{lemma}
\paragraph{Proof} Denote by $I^k$  the lhs of equation \eqref{eq_twin2}. If $k=0$ then the inequality follows from \eqref{eq_twin} with $P=p$. Let $k \geqslant 1$ and $\mu=\max(\Lambda,M)$.
\begin{itemize}
\item $p \leqslant \Lambda$.
\begin{equation}
I^k \leqslant \int \limits^{+\infty}_{\mu} d \lambda  \frac{2\log^k \frac{\lambda}{M}}{(\lambda + p)^2} \leqslant \int \limits^{+\infty}_{\mu} d \lambda f^\prime(\lambda)\leqslant 2^{k+2}k!\sqrt{e} \frac{1 + \log^k_+ \frac{\Lambda}{M}}{\Lambda + p + q},
\end{equation}
\begin{equation}
f(\lambda):=-\frac{k!2^{k+1}}{\lambda + p}\sum^k_{j=0} \frac{\log^k \frac{\lambda}{M}}{j!2^j}.
\end{equation}
\item $\Lambda<p$.
\begin{equation}
J^k:=\int \limits^p_\Lambda d\lambda \frac{\log^k_+ \frac{\lambda}{M}}{\lambda + q},\quad I^k \leqslant \frac{2J^k}{\Lambda + p + q} + \int \limits^{+\infty}_{\mu} d\lambda  \frac{2 \log^k_+ \frac{\lambda}{M}}{(\lambda + p)^2}.
\end{equation}
The integral on the rhs of $I^k$ is exactly the same as in the case $p\leqslant \Lambda$. For the integral $J^k$ using the inequality
\begin{equation}
\log_+\frac{\lambda}{M}\leqslant \log 2 + \log_+ \frac{\lambda}{\Lambda+M} + \log_+ \frac{\Lambda}{M},
\end{equation}
we have
\begin{equation}
J^k<3^{k+1} \left(1 + \log^{k+1}_+ \frac{p}{\Lambda+\eta} + \log^{k+1}_+ \frac{\Lambda}{M}\right).
\end{equation}
\end{itemize}
\hfill$\blacksquare$
\begin{lemma}For $0<\Lambda<\eta\leqslant M$, $k \in \mathbb{N}$
\begin{equation}
\int \limits^{\eta}_\Lambda d\lambda  \frac{1}{\lambda + \eta} \log^k_+ \frac{\lambda}{M} < 1 . \label{eq_int3}
\end{equation}
\end{lemma}
\paragraph{Proof} Denote by $I^k$ the lhs of equation \eqref{eq_int3}. If $k>0$ then $I^k=0$. It remains to consider the integral
\begin{equation}
\int \limits^{\eta}_\Lambda \frac{d \lambda}{\lambda + \eta} = \log\frac{2\eta}{\Lambda +\eta} < \log 2.
\end{equation}
\hfill$\blacksquare$
\begin{lemma}Let $0<\Lambda<\eta\leqslant P$, $k \in \mathbb{N}$. Then $\exists C_k>0$ such that
\begin{equation}
\int \limits^{\eta}_\Lambda  \frac{d\lambda }{\lambda + \eta} \log^k_+ \frac{P}{\lambda} <  C_k\Big(1 + \log^k_+ \frac{P}{\Lambda + \eta}\Big). \label{eq_int4}
\end{equation}
\end{lemma}
\paragraph{Proof} Denote by $I^k_{[\Lambda,\eta]}$ the lhs of equation \eqref{eq_int4}.
\begin{equation}
I^k_{[\Lambda,\eta]} < I^k_{[0,\eta]}<\frac{1}{\eta}\int \limits^\eta_0 d\lambda\, \log^k_+ \frac{P}{\lambda} < 2k! e^2 \Big( 1 + \log^k_+ \frac{P}{\eta +\Lambda} \Big).
\end{equation}
\hfill$\blacksquare$
\begin{lemma}Let $0<\Lambda< M \leqslant P$, $0 \leqslant \eta \leqslant M$, $k \in \mathbb{N}$. Then $\exists C_k>0$ such that
\begin{equation}
\int \limits^{M}_\Lambda  \frac{d\lambda }{\lambda + \eta} \log^k_+ \frac{P}{\lambda} <  C_k\Big(1 + \log^{k+1}_+ \frac{P}{\Lambda + \eta}\Big). \label{eq_int4b}
\end{equation}
\end{lemma}
\paragraph{Proof} Denote by $I^k_{[\Lambda,M]}$ the lhs of equation \eqref{eq_int4b}. If $\Lambda \geqslant \eta$ then
\begin{equation}
I^k_{[\Lambda,M]} \leqslant \int \limits^M_\Lambda \frac{d\lambda }{\lambda + \eta} \log^k_+ \frac{2P}{\lambda + \eta} \leqslant \frac{2^{k+1}}{k+1} \Big( 1 + \log^{k+1}_+ \frac{P}{\Lambda + \eta}\Big).
\end{equation}
If $\eta \geqslant \Lambda$ then $I^k < I^k_{[0,\eta]} + I^k_{[\eta,M]}$ where
\begin{align}
I^k_{[0,\eta]}&=\int \limits^\eta_0 \frac{d\lambda}{\lambda + \eta} \log^k_+ \frac{P}{\lambda} < 2k! e^2 \Big(1 + \log^k_+ \frac{P}{\eta + \Lambda}\Big),\\
I^k_{[\eta,M]}&=\int \limits^M_\eta \frac{d\lambda}{\lambda + \eta} \log^k_+ \frac{2P}{\lambda + \eta}<\frac{2^{k+1}}{k+1} \Big( 1 + \log^{k+1}_+ \frac{P}{\Lambda + \eta}\Big).
\end{align}
\hfill$\blacksquare$
\begin{lemma}For $q \geqslant 0$, $k \in \mathbb{N}$, $P>0$ there exists a constant $C_k>0$ such that
\begin{equation}
\int \limits^{\Lambda_0}_\Lambda  \frac{d\lambda }{\lambda + q} \log^k_+ \frac{P}{\lambda} <  C_k\Big(1 + \log^{k+1}_+ \frac{P}{\Lambda + q} + \log_+ \frac{\Lambda_0}{\Lambda + q}\Big). \label{eq_int4a}
\end{equation}
\end{lemma}
\paragraph{Proof} Denote by $I^k_{[\Lambda,\Lambda_0]}$ the lhs of equation \eqref{eq_int4a}. If $k=0$ then
\begin{equation}
I^0_{[\Lambda,\Lambda_0]}< 1 + \log_+ \frac{\Lambda_0}{\Lambda + q}.
\end{equation}
For $k>0$ and $\Lambda \geqslant q$
\begin{equation}
I^k_{[\Lambda,\Lambda_0]}< \frac{2^{k+1}}{k+1}\Big(1 +\log^{k+1}_+ \frac{P}{\Lambda + q}\Big).
\end{equation}
If $k>0$ and $\Lambda<q$ then $I^k_{[\Lambda,\Lambda_0]}< I^k_{[0,q]} + I^k_{[q,\Lambda_0]}$ where
\begin{align}
I^k_{[0,q]}&<\frac{1}{q}\int \limits^q_0 d\lambda \,\log^k_+\frac{P}{\lambda}< k!e^2\Big( 1 + \log^{k+1}_+ \frac{P}{\Lambda +q}\Big),\\
I^k_{[q,\Lambda_0]}&<\frac{2^{k+1}}{k+1}\Big(1 +\log^{k+1}_+ \frac{P}{\Lambda + q}\Big).
\end{align}
\hfill$\blacksquare$
\begin{lemma}Let $a,d>0$, $b\geqslant 0$ and $m,k \in \mathbb{N}$. Then $\exists C_{k,m}>0$ such that
\begin{equation}
\int \limits^{a}_0  dx \; x^m \log^k_+ \frac{d}{b + x} \leqslant  C_{k,m}a^{m+1}\Big(1 + \log^k_+ \frac{d}{a+b}\Big). \label{eq_int5}
\end{equation}
\end{lemma}
\paragraph{Proof} By direct calculations it is easy to show that
\begin{align}
x^m \log^k_+ \frac{d}{b + x}&\leqslant f^\prime,&f&:=\frac{k!x^{m+1}}{(m+1)^{k+1}}\sum^k_{j=0} \frac{(m+1)^j}{j!} \log^j_+ \frac{d}{b+x}.
\end{align}
Consequently, the lhs of equation \eqref{eq_int5} is bounded above by $f(a)$,
\begin{align}
f(a)=&\frac{k!}{(m+1)^{k+1}}a^{m+1} \sum^k_{j=0} \frac{(m+1)^j}{j!}\log^j_+ \frac{d}{a+b}\nonumber\\
&\leqslant \frac{e^{m+1} k!}{(m+1)^{k+1}} a^{m+1} \Big(1+\log^k_+ \frac{d}{a+b}\Big).
\end{align}
\hfill$\blacksquare$
\begin{lemma}Let $p,q \in \mathbb{R}^4$, $\Lambda^\prime>0$, $\Lambda^\prime \geqslant \eta \geqslant 0$. Then
\begin{equation}
\int \limits^{1}_0  dt \; \frac{|p|}{\Lambda^\prime + |tp + q|} \leqslant  2\Big(\log 4 + \log_+ \frac{|p|}{\Lambda^\prime + \eta}\Big).\label{eq_int7}
\end{equation}
\end{lemma}
\paragraph{Proof} Let $I_{[0,1]}$ denote the lhs of equation \eqref{eq_int7}.  There exists $t_1 \in [0,1]$ such that $|tp + q|\geqslant |p||t-t_1|$ for all $t \in [0,1]$.
(To prove this fact: write $q=-t_q\, p+ q^\perp$, with $p.q^\perp=0$, in such a way that
$|t\,p-q|\geqslant|t-t_q|\,|p|$; then set
$t_1:=0$ for $t_q<0$,
$t_1:=1$ for $t_q>1$,
and $t_1=t_q$ otherwise.)
\begin{align}
I_{[0,1]}=&I_{[0,t_1]}+I_{[t_1,1]}\leqslant 2 \log\frac{\Lambda^\prime+|p|}{\Lambda^\prime}\leqslant 2 \log 4\max(1,\frac{|p|}{\Lambda^\prime+\eta})\nonumber\\
&\leqslant 2 \Big(\log 4 +  \log_+\frac{|p|}{\Lambda^\prime + \eta}\Big).
\end{align}
\hfill$\blacksquare$
\begin{lemma}Let $r>0$, $w \in \mathbb{N}$ and $x,y \in \mathbb{R}^4$ 
\begin{equation}
\frac{e^{-r x^2}}{(1+|x-y|)^w}\leqslant \frac{w!\max(2,1+\frac{1}{2r})^w}{(1+|y|)^w}.\label{eq_frac}
\end{equation}
\end{lemma}
\paragraph{Proof} Choosing the Cartesian coordinate system such that one of the basis vectors $e_L$ is along $y$ we have 
\begin{align}
\frac{e^{-r x^2}}{(1+|x-y|)^w}&\leqslant f(t),&f(t):=\frac{e^{-r t^2}}{(1+|t-t_0|)^w}.
\end{align}
where $t$, $t_0$ are the longitudinal coordinates, $x=x_T + t e_L$ and $y=t_0 e_L$, $t_0=|y|$. If $t\geqslant t_0$ then $f$ is strictly decreasing. If $t \leqslant t_0$ then $f(t)=g(t)$ where
\begin{align}
g(t)&:=\frac{e^{-r t^2}}{(t_1-t)^w},&t_1&:=1+t_0.
\end{align}
If $t < t_1$ then g is either increasing or has a local maximum at $t_-$.
\begin{align}
g^\prime(t) &\leqslant 0 \mbox{ if } (\Delta \geqslant 0) \wedge (t_- \leqslant t \leqslant t_+),&t_{\mp}&:=\frac{t_1 \mp \sqrt{\Delta}}{2},\\g^\prime(t) &\geqslant 0  \mbox{ otherwise },& \Delta&:=t^2_1-2\frac{w}{r}.
\end{align}
Consequently, $f(t) \leqslant \max(g(t_-),g(t_0))$ where
\begin{align}
g(t_-)&=\frac{e^{-rt^2_-}}{t^w_+}\leqslant  \frac{2^w}{t^w_1}=\frac{2^w}{(1+t_0)^w},&g(t_0)&=e^{-rt^2_0}\leqslant\frac{w!(1+\frac{1}{2r})^w}{(1+t_0)^w}.
\end{align}
\hfill$\blacksquare$
\begin{lemma}Let $r>0$. There is a constant $C$ such that
\begin{equation}
e^{-r\frac{s^2}{\Lambda^2}}\log_+ \frac{\max(M,\sqrt{p^2 + s^2})}{\Lambda}< C+\frac{1}{2}\log_+\frac{1}{r}+\log_+ \frac{\max(M,p)}{\Lambda}. \label{eq_plog}
\end{equation}
\end{lemma}
\paragraph{Proof} Using the following inequality
\begin{equation}
\frac{\max(M,\sqrt{p^2 + s^2})}{\Lambda}\leqslant 2\max(1,\frac{s}{\Lambda})\max(1,\frac{\max(M,p)}{\Lambda}),
\end{equation}
we bound the lhs of the statement by
\begin{equation}
\log_+\frac{\max(M,p)}{\Lambda} + \log 2 + \frac{1}{2}\log_+\frac{1}{r} + \frac{1}{2}e^{-z}\log_+ z,\quad z:=r\frac{s^2}{\Lambda^2}.
\end{equation}
The inequality $e^{-z}\log_+z < e^{-1}$ finishes the proof.\hfill$\blacksquare$
\begin{lemma} Let $x,y,m \geqslant 0$, $\mathcal{P}^{(0)}(x)=\sum \limits^n_{k=0} c_k x^k$ a polynomial of the degree $n$, $\log_m x:=\log_+ \max(x,m)$. Then there exist polynomials $\mathcal{P}^{(1)}$, $\mathcal{P}^{(2)}$ of the degree $n$ such that 
\begin{equation}
\mathcal{P}^{(0)}(\log_m \sqrt{y^2 + x^2} )\leqslant \mathcal{P}^{(1)}(\log_m y) + \mathcal{P}^{(2)}(\log_m x).
\end{equation}
\end{lemma}
\paragraph{Proof} Substitution of the inequalities
\begin{equation}
\max(\sqrt{y^2+x^2},m) \leqslant \max(y + x,m) \leqslant \max(y,m) + \max(x,m),
\end{equation}
\begin{equation}
\max(a+b,1) \leqslant \max(a,1) + \max(b,1) \leqslant 2\max(a,1)\max(b,1),
\end{equation}
into the definition $\log_+ a := \log \max(a,1)$ yields
\begin{equation}
\log_m \sqrt{y^2 + x^2} \leqslant \log_m y + \log_m x+\log2.
\end{equation}
This gives
\begin{equation}
\mathcal{P}^{(0)}(\log_m \sqrt{x^2 + y^2}) \leqslant \sum \limits^n_{k=0} c_k 3^k \left( \log^k_m y + \log^k_m x + 1 \right).
\end{equation}
\hfill$\blacksquare$
\begin{lemma}\label{lem_402a}For a fixed $s \in \mathbb{N}$ there exists a constant $c$ such that $\forall u \leqslant w_{max}$ and $\forall x \in\mathbb{R}^4$
\begin{equation}
|\partial^u   \left( x^{\otimes s}(1-e^{-x^4})\right)| \leqslant c \, \left\{\begin{matrix} |x|^{s + 1 -u},&u \leqslant s ,\\ 1, & otherwise. \end{matrix}\right.\label{eq_402t}
\end{equation}
\end{lemma}
\paragraph{Proof}First we consider the case $u\leqslant s$
\begin{equation}
|\partial^u (x^{\otimes s}(1-e^{-x^4}))|\leqslant|\partial^{u} x^{\otimes s}|(1-e^{-x^4}) +  \sum_{u_2>0} \frac{u! 4^{u_2}}{u_1! u_2!}|\partial^{u_1} x^{\otimes s+ 3u_2}|\, e^{-x^4}.\label{eq_402a}
\end{equation}
For the derivatives on the right we have
\begin{equation}
|\partial^{u_1} x^{\otimes s+ 3u_2}|\leqslant (s+3u_2)! \,\frac{|x|^{s-u+1}}{(s-u+1)!}\frac{|x|^{4u_2-1}}{(4u_2-1)!}\,.\label{eq_402b}
\end{equation}
Furthermore, $\forall k \in \mathbb{N}$ 
\begin{equation}
\frac{|x|^k}{k!}e^{-x^4} < \sum^{\infty}_{n=0} \frac{|x|^n}{n!}e^{-x^4}=e^{-x^4+|x|}<e.\label{eq_402c}
\end{equation}
Noting that $1-e^{-x^4} \leqslant |x|$, we obtain the bound in the case $u\leqslant s$
\begin{equation}
\frac{s!|x|^{s+1-u}}{(s-u)!} +  \frac{e |x|^{s+1-u}}{(s-u+1)!} \sum_{u_2>0} \frac{u! 4^{u_2} (s+3u_2)!}{u_1! u_2!}\leqslant c |x|^{s+1-u}\,.
\end{equation}
For $u>s$ using \eqref{eq_402c} we have
\begin{equation}
\sum \frac{u! 4^{u_2}(s+3u_2)!}{u_1! u_2! }\frac{|x|^{s+3u_2-u_1} }{(s+3u_2-u_1)!}\,e^{-x^4} \leqslant (s+3u)! 5^{u} e \leqslant c\,.
\end{equation}
\hfill$\blacksquare$
\begin{lemma}For a fixed $s \in \mathbb{N}\backslash\{0\}$ there exists a constant $c$ such that for all $u,v \leqslant w_{max}$ and all $x=(x_1,x_2)$ with $x_i \in\mathbb{R}^4$
\begin{equation}
|\partial^v_2 \partial^u_1 \left(x_2 \otimes x^{\otimes s-1}_1 (1-e^{-x^4_1})\right)| \leqslant c \, \left\{\begin{matrix} |x|^{s + 1 -u-v},&u+v \leqslant s ,\\ |x|+1, & otherwise. \end{matrix}\right. \label{eq_331b}
\end{equation}
\end{lemma}
\paragraph{Proof}First let $v \in \{0,1\}$. Using lemma \ref{lem_402a} we obtain for $0\leqslant u \leqslant s$
\begin{align}
|\partial^u_1 (x_2 \otimes x^{\otimes s-1}_1(1-e^{-x^4_1}))|&\leqslant c |x|^{s+1-u}\,,\\
|\partial_2 \partial^u_1 (x_2 \otimes x^{\otimes s-1}_1(1-e^{-x^4_1}))|&\leqslant c |x|^{s-u}\,,
\end{align}
and for $u> s$ we also have
\begin{align}
|\partial^u_1 (x_2 \otimes x^{\otimes s-1}_1(1-e^{-x^4_1}))|&\leqslant c|x|\,,\\
|\partial_2 \partial^u_1 (x_2 \otimes x^{\otimes s-1}_1(1-e^{-x^4_1}))|&\leqslant c\,.
\end{align}
Finally, for $v>1$ $\partial^v_2 \partial^u_1 (x_2 \otimes x^{\otimes s-1}_1(1-e^{-x^4_1}))=0$.\hfill$\blacksquare$
\begin{lemma}For $s \in \{0,1\}$ there exists a constant $c$ such that for all $u,v \leqslant w_{max}$ and all $x=(x_1,x_2)$ with $x_i \in\mathbb{R}^4$
\begin{equation}
|\partial^v_2 \partial^u_1 \left(x^{\otimes s}_1(e^{-x^4_1}-e^{-(x_1 +x_2)^4})\right)| \leqslant c \, \left\{\begin{matrix} |x|^{s+1-u-v},&u+v\leqslant s,\\ |x|+1, & otherwise. \end{matrix}\right.\label{eq_402s}
\end{equation}
\end{lemma}
\paragraph{Proof}First let $s=0$. For $u=v=0$ put $y=x_1 +x_2$ and assume $|x_1| \leqslant |y|\leqslant 1$
\begin{equation}
e^{-x^4_1}-e^{-y^4} =e^{-x^4_1}(1-e^{-(y^4-x^4_1)})\leqslant y^4-x^4_1\leqslant |y|\leqslant 2|x|\,.\label{eq_402d}
\end{equation}
Inequality \eqref{eq_402c} implies that 
\begin{equation}
\partial^w e^{-x^4}< e (3w)!5^4.\label{eq_402f} 
\end{equation}
Consequently for $u+v>0$
\begin{equation}
|\partial^v_2 \partial^u_1 (e^{-x^4_1}-e^{-(x_1 +x_2)^4})| \leqslant |\partial^u_1 e^{-x^4_1}|+ |\partial^v_2 \partial^u_1 e^{-(x_1 +x_2)^4})| \leqslant c_2\,.
\end{equation}
Finally, we consider the case $s=1$. Using \eqref{eq_402d} and \eqref{eq_402f} we get
\begin{equation}
|x_1(e^{-x^4_1}-e^{-(x_1 + x_2)^4})| \leqslant 2|x|^2\,,
\end{equation}
\begin{equation}
|\partial_2 (x_1(e^{-x^4_1}-e^{-{x_1 + x_2}^4}))| \leqslant |x||\partial_2 (e^{-x^4_1}-e^{-(x_1 + x_2)^4})| \leqslant c_3|x|\,,
\end{equation}
\begin{equation}
|\partial_1  (x_1(e^{-x^4_1}-e^{-(x_1 + x_2)^4}))| \leqslant c_4|x|\,,
\end{equation}
and for $u+v>1$
\begin{equation}
|\partial^u_1 \partial^v_2 (x_1(e^{-x^4_1}-e^{-(x_1 + x_2)^4}))| \leqslant  c_5 + |x|c_6\,.
\end{equation}
\hfill$\blacksquare$
\begin{lemma}  Let $0 \leqslant \beta \leqslant 1$, $ x \geqslant 0$ and
\begin{equation}
f(x):=\frac{1}{x} (e^{-\beta x}-e^{-x})
\end{equation}
Then  $\forall w \in \mathbb{N}$
\begin{equation}
|\partial^w f(x)| <  e \frac{w!}{(1 + x)^{w+1}}. \label{eq_f}
\end{equation}
\end{lemma}
\paragraph{Proof} We have an identity
\begin{align}
\partial^w f(x)&=(-1)^w g_w(x),&g_w(x):= \int \limits^1_\beta d \gamma \, \gamma^w e^{-\gamma x},
\end{align}
It follows
\begin{align}
0\leqslant g_w(x) \leqslant&  \int \limits^1_0 d \gamma \, \gamma^w e^{-\gamma (x+1)}e^{\gamma}<e \int \limits^1_0 d \gamma \, \gamma^w e^{-\gamma (x+1)} \nonumber\\
&<e \frac{1}{(x+1)^{w+1}}\int \limits^\infty_0 d z \, z^w e^{-z}=e \frac{\Gamma(w+1)}{(x+1)^{w+1}}.
\end{align}
\hfill$\blacksquare$
\begin{lemma}  Let $0 \leqslant \beta \leqslant 1$, $x \geqslant 0$ and
\begin{equation}
h(x):=\frac{1}{x} (e^{-\beta x^2}-e^{-x^2})
\end{equation}
Then  $\forall w \in \mathbb{N}$, $\forall C>1$ 
\begin{equation}
|\partial^w h(x)|<  w!e(C+1) \frac{(2 \sqrt{e} C)^w}{(1+x)^{w+1}}. \label{eq_h}
\end{equation}
\end{lemma}
\paragraph{Proof} 
\begin{align}
|\partial^w h(x)|&\leqslant |x\partial^w f(x^2)| +  |w\partial^{w-1} f(x^2)|,&f(x^2)&:=\frac{1}{x}h(x). 
\end{align}
Using an auxiliary variable $y$
\begin{align}
\frac{\partial^w}{(\partial x)^w} f(x^2)&=\left(\frac{\partial}{\partial x}+2x\frac{\partial}{\partial y}\right)^w f(y)|_{y=x^2}\nonumber \\
&=\sum \limits^{2k\leqslant w}_{k=0} \frac{w! (2k-1)!!}{(w-2k)! (2k)!}\left(2 \frac{\partial}{\partial y}\right)^k \left(2x \frac{\partial}{\partial y}\right)^{w-2k} f(y)|_{y=x^2} \nonumber\\
&=\sum \limits^{2k\leqslant w}_{k=0} \frac{w!}{(w-2k)! k!}(2x)^{w-2k} \left(\frac{\partial}{\partial y}\right)^{w-k} f(y)|_{y=x^2}.
\end{align}
Equation \eqref{eq_f} gives
\begin{equation}
|x \partial^w f(x^2)|< w! 2^w e \sum \limits^{2k\leqslant w}_{k=0} \frac{1}{k!}\frac{(w-k)!}{(w-2k)!}2^{-2k}\frac{x^{w-2k+1}}{(1+x^2)^{w-k+1}}.
\end{equation}
With an arbitrary constant $C>1$ we have
\begin{align}
\frac{x}{1+x^2}& < \frac{C}{1+x},&(1+x^2)&\geqslant \frac{1}{2}(1+x)^2,&\frac{(w-k)!}{(w-2k)!}&\leqslant w^k.
\end{align}
Consequently,
\begin{equation}
|x \partial^w f(x^2)|< \frac{w!Ce (2C)^w}{(1+x)^{w+1}}  \sum \limits^{2k\leqslant w}_{k=0} \frac{1}{k!} \left(\frac{w}{2C^2}\right)^k< \frac{w!Ce (2 \sqrt{e} C)^w}{(1+x)^{w+1}}.
\end{equation}
Similarly, we obtain
\begin{equation}
|w \partial^{w-1} f(x^2)|<  \frac{2 w! e (2 \sqrt{e} C)^{w-1}}{(1+x)^{w+1}}.
\end{equation}
\hfill$\blacksquare$
\begin{lemma} Let $f(p^2)$ be a scalar function. Then
\begin{equation}
|\prod \limits^{w}_{i=1} \frac{\partial}{\partial p_{\mu_i}} f(p^2)| \leqslant 2^w \sum \limits^{2k \leqslant w}_{k=0} \frac{w!}{(w-2k)! k!} |p|^{w-2k}  |\left(\frac{\partial}{\partial p^2}\right)^{w-k} f(p^2)|.\label{eq_df}
\end{equation}
\end{lemma}
\paragraph{Proof} With the aid of an auxiliary variable $y$
\begin{equation}
\prod \limits^w_{i=1} \frac{\partial}{\partial p_{\mu_i}}f(p^2)=\prod \limits^w_{i=1} \left( \frac{\partial}{\partial p_{\mu_i}} + 2p_{\mu_i} \frac{\partial}{\partial y}\right)f(y)|_{y=p^2}.
\end{equation}
A partial derivative wrt $p_{\mu}$ contributes only if it can be paired with $2p_\mu$ term. Consequently, we can compute the right hand side by considering the possible pairs,
\begin{equation}
(\frac{\partial}{\partial p_{\mu_i}} + 2p_{\mu_i} \frac{\partial}{\partial y})(\frac{\partial}{\partial p_{\mu_j}} + 2p_{\mu_j} \frac{\partial}{\partial y}) = 2 \delta_{\mu_i \mu_j}\frac{\partial}{\partial y} + 2p_{\mu_i} 2p_{\mu_j}\left(\frac{\partial}{\partial y}\right)^2.
\end{equation}
It gives
\begin{equation}
\sum \limits^{2k \leqslant w}_{k=0, \pi} \frac{1}{(w-2k)!(2k)!} \sum \limits_{\sigma,\pi^\prime:=\sigma \pi} \frac{1}{2^k k!} \prod \limits^k_{i=1}\left( 2 \delta_{\pi^\prime_i \pi^\prime_{i+1}} \frac{\partial}{\partial y}\right) \prod \limits^w_{i=2k+1}\left(2p_{\pi_i} \frac{\partial}{\partial y}\right),
\end{equation}
where the outer and inner sums run over $w!$, $(2k)!$ permutations, respectively. Using the inequality $|A_{\Vec{\mu}}B_{\Vec{\nu}}| \leqslant |A_{\Vec{\mu}}| |B_{\Vec{\nu}}|$ we obtain the upper bound. \hfill$\blacksquare$
\begin{lemma}  Let $C>1$ 
\begin{equation}
|\prod \limits^{w}_{i=1} \frac{\partial}{\partial p_{\mu_i}} S^{\Lambda\Lambda_0}(p)|<2 w! e(C+1)\frac{(2^2 e^{\frac{3}{2}} C^2)^w}{(\Lambda+|p|)^{w+2}}.\label{eq_dS}
\end{equation}
\end{lemma}
\paragraph{Proof} We  change the variable $x_\mu =p_\mu /\Lambda$
\begin{equation}
\prod \limits^{w}_{i=1} \frac{\partial}{\partial p_{\mu_i}}  S^{\Lambda\Lambda_0}(p) = \frac{1}{\Lambda^{2w+2}}  \prod \limits^{w}_{i=1} \frac{\partial}{\partial x_{\mu_i}} h(x^2).
\end{equation}
Equations \eqref{eq_df}, \eqref{eq_h} yield
\begin{equation}
|\partial^{w} h(x^2)|<2^w e(C+1) \sum \limits^{2k\leqslant w}_{k=0} \frac{w!}{k!}\frac{(w-k)!}{(w-2k)!}\frac{x^{w-2k}(2\sqrt{e}C)^{w-k}}{(1+x^2)^{w-k+1}}.
\end{equation}
\hfill$\blacksquare$
\begin{lemma} Let $C>1$
\begin{align}
|\prod \limits^{w}_{i=1} \frac{\partial}{\partial p_{\mu_i}} C^{\Lambda\Lambda_0}_{\mu \nu}(p)|<2^3 w! e(C+1 + (\xi+1)(C^2 +3))\frac{(2^2 e^{\frac{3}{2}} C^2)^w}{(\Lambda+|p|)^{w+2}}.\label{eq_dC}
\end{align}
\end{lemma}
\paragraph{Proof}
\begin{equation}
|\partial^w C^{\Lambda\Lambda_0}_{\mu \nu}(p)| \leqslant  4 |\partial^w S^{\Lambda\Lambda_0}(p)| + |\xi -1||\partial^w \frac{p_\mu p_\nu}{p^2}S^{\Lambda\Lambda_0}(p)|.
\end{equation}
The first term is bounded in \eqref{eq_dS}.  Using $x_\mu =p_\mu /\Lambda$ we have an upper bound for the last term
\begin{equation}
x^2 |\partial^w f(x^4)|+2^3 w|x||\partial^{w-1}f(x^4)|+4^2 w(w-1)|\partial^{w-2}f(x^4)|,
\end{equation}
where $f(x^4)$ is the same as in \eqref{eq_f}. Equation \ref{eq_f} gives
\begin{align}
|\partial^w f(x^2)|&<  \frac{2 w! e (2 \sqrt{e} C)^w}{(1+x)^{w+2}},&|\partial^w f(x^4)|&<  \frac{2^3 w! e (2^2 e^{\frac{3}{4}} C^2)^w}{(1+x)^{w+4}}.
\end{align}
Consequently,
\begin{equation}
|\partial^w x_\mu x_\nu f(x^4)|<2^3 (C^2 + 3)ew! \frac{(2^2 C^2 e^{\frac{3}{4}})^w}{(1+x)^{w+2}}.
\end{equation}
\hfill$\blacksquare$
\begin{lemma} For all $p \in \mathbb{R}^4$ there exists a constant $C$ such that
\begin{equation}
|\dot{\mathbf{C}}^{\Lambda\Lambda_0}(p)| \leqslant C\frac{1}{\Lambda^3} e^{-\frac{p^2}{\Lambda^2}}.\label{eq_dS2}
\end{equation}
\end{lemma}
\paragraph{Proof} Using the inequality $xe^{-x^2}\leqslant x e^{1-2x}=xe^{1-x}e^{-x}\leqslant e^{-x}$ we obtain
\begin{align}
|\dot{S}^{\Lambda\Lambda_0}(p)|&=4\frac{1}{\Lambda^3}xe^{-x^2}\leqslant \frac{4}{\Lambda^3}e^{-x}  ,&x:=\frac{p^2}{\Lambda^2},\\
|\dot{C}^{\Lambda\Lambda_0}_{\mu \nu}(p)| &\leqslant  4\frac{1  + |\xi-1|}{\Lambda^3}xe^{-x^2}.
\end{align}
\hfill$\blacksquare$
\begin{lemma}  For all $p \in \mathbb{R}^4$ there exists a constant $C$ such that
\begin{equation}
|\partial^w_p \partial_{\Lambda_0} \mathbf{C}^{\Lambda\Lambda_0}(p)| \leqslant C\frac{1}{\Lambda_0(\Lambda_0 + |p|)^{2 + \|w\|}}.\label{eq_dS3}
\end{equation}
\end{lemma}
\paragraph{Proof} Let $f=e^{-x^4} x x$ then using $|\partial^w_x|g||\leqslant |\partial^w_x g|$ we have
\begin{align}
|\partial^w_p \partial_{\Lambda_0} C^{\Lambda\Lambda_0}(p)|\leqslant&\frac{4(2+|\xi-1|)|\partial^w_x f|}{\Lambda^{\|w\|+3}_0},&|\partial^w_p \partial_{\Lambda_0} S^{\Lambda\Lambda_0}(p)|\leqslant&\frac{4|\partial^w_x f|}{\Lambda^{\|w\|+3}_0},
\end{align}
where introducing $C_1=2^{\frac{3}{2}w_{max}}(4w_{max}+2)!$
\begin{equation}
|\partial^w_x f|\leqslant C_1 e^{-x^4}(|x|^{3\|w\| +2}+1)\leqslant eC_1 e^{-2x^2}(|x|^{3\|w\| +2}+1).
\end{equation}
Then it is easy to see that for $0\leqslant m\leqslant 3w_{max}+2$
\begin{equation}
\max(e^{-x^2}(x^m+1))\leqslant \max(e^{-x^2}x^m) + 1\leqslant  e^{\frac{m}{2}(\log \frac{m}{2} -1)} + 1\leqslant C_2.
\end{equation}
To go further we need the following inequality for all $y \geqslant 0$ and $k \in \mathbb{N}$
\begin{align}
g_k(y)&\leqslant k!,&g_k(y)&:=(1+y)^k e^{-y}, \label{eq_21dS3}
\end{align}
which is obtained looking for the maximum $\bar{y}_k=k-1$ and using $g_0(\bar{y}_0)=1$.
\begin{equation}
g_k(\bar{y}_k)=kg_{k-1}(\bar{y}_k)\leqslant k g_{k-1}(\bar{y}_{k-1}) \implies g_k(y) \leqslant g_k(\bar{y}_k)\leqslant k!
\end{equation}
Defining $C_3:=e^2 C_1 C_2$ and using inequality \eqref{eq_21dS3} we have
\begin{equation}
|\partial^w_x f|\leqslant C_3 e^{-2x} \leqslant C_3\frac{2!}{(1 + |x|)^2} \frac{\|w\|!}{(1+x)^{\|w\|}}\leqslant \frac{2\,C_3 \,w_{max}!}{(1 + |x|)^{2+\|w\|}}.
\end{equation}
\hfill$\blacksquare$
\begin{lemma}Let $\mathcal{C}^{\Lambda\Lambda_0}$ be one of the propagators $S^{\Lambda\Lambda_0}$ or $C^{\Lambda\Lambda_0}_{\mu \nu}$, as defined in \eqref{def_c}, \eqref{def_s}. There are positive constants $c_0$, $c_1$, $d$ such that for all $w \in \mathbb{N}$, $p \in \mathbb{R}^4$, $0<\Lambda \leqslant \Lambda_0$, and with $c_\xi:=c_0 + \xi c_1$,
\begin{align}
|\Big(\prod \limits^{w}_{i=1} \frac{\partial}{\partial p_{\mu_i}} \Big)\mathcal{C}^{\Lambda\Lambda_0}(p)|&<\frac{w!\, d^w c_\xi}{(|p|+\Lambda)^{w+2}},&|\dot{\mathcal{C}}^{\Lambda\Lambda_0}(p)|&<\frac{c_\xi}{\Lambda^3}e^{- \frac{p^2}{\Lambda^2} }\,. \label{eq_16C}
\end{align}
\end{lemma}
\paragraph{Proof} The statement follows from  \eqref{eq_dS}, \eqref{eq_dC}, \eqref{eq_dS2}.\hfill$\blacksquare$
\section{The functional $\Gamma_{n\leqslant 4}$} \label{sec_gamma}
We expand the generating functional $\tilde{\Gamma}^{0\Lambda_0}(\ubar{A}, \ubar{c}, \wbar{c})$ of \eqref{eq_legendre_w} as formal series in $\ubar{A}, \ubar{c}, \wbar{c}$. As usual, we adopt the shorthand notation $\Gamma^{0\Lambda_0}(A,\bar{c},c)$ for $\tilde{\Gamma}^{0\Lambda_0}(\ubar{A}, \ubar{c}, \wbar{c})$.
\begin{align}
\Gamma^{0\Lambda_0}&=\Gamma^{0\Lambda_0}_{n\leqslant 4} + \Gamma^{0\Lambda_0}_{n\geqslant 5},&\Gamma^{0\Lambda_0}_{n\leqslant 4}&:=\sum \limits_{n=1}^4 \Gamma^{0\Lambda_0}_n,
\end{align}
where $n$ counts the number of fields.  The functionals $\Gamma^{0\Lambda_0}_n$ with $n\leqslant4$ contain both relevant and irrelevant terms. We assume hypothesis \ref{rc1}. In general the tensors $\zeta^{\Vec{\phi}}_{\mu_1...\mu_r}$ appearing in the form factors $F^{\Vec{\phi}}_{\mu_1...\mu_r}(\Vec{p})$ are elements of $\mathrm{span}(\{\delta^s \ubar{p}^{k>0}\}_{r>0})$ where $\ubar{p}=(p_1,...,p_{n-1})$.
\begin{enumerate}
\item{\textbf{One-point function}}\\
There are no local terms that preserve Euclidean invariance and global SU(2) symmetry. It follows that $\Gamma_1=0$.
\item{\textbf{Two-point functions}}
\begin{align}
\Gamma^{0\Lambda_0}_2&= \frac{1}{2}\langle F^{AA}_{\mu\nu}\check{A}^a_\mu \check{A}^a_\nu \rangle + \langle F^{\bar{c}c} \check{\bar{c}}^a\check{c}^a \rangle ,\\
F^{AA}_{\mu\nu}(p)&:=(\delta_{\mu\nu}p^2-p_\mu p_\nu)(\sigma^{-1}_{0 \Lambda_0}(p^2)+\Sigma^{AA}_T(p^2)) \nonumber \\
&+\frac{1}{\xi}p_\mu p_\nu(\sigma^{-1}_{0 \Lambda_0}(p^2)+\Sigma^{AA}_L(p^2)),\\
F^{\bar{c}c}(p)&:=-p^2(\sigma^{-1}_{0 \Lambda_0}(p^2) + \Sigma^{\bar{c}c}(p^2))\,.
\end{align}
We assume that the form factors $\Sigma^{AA}$ and $\Sigma^{\bar{c}c}$ include all loop corrections. Note that for the functional $\Tamma^{0\Lambda_0}_2$ we have
\begin{align}
\Tamma^{0\Lambda_0}_2&= \frac{1}{2}\langle \bunderline{F}^{AA}_{\mu\nu}\check{A}^a_\mu \check{A}^a_\nu \rangle + \langle F^{\bar{c}c} \check{\bar{c}}^a\check{c}^a \rangle ,\\
\bunderline{F}^{AA}_{\mu\nu}(p)&:=F^{AA}_{\mu\nu}(p) - \frac{1}{2\xi}p_\mu p_\nu\,.
\end{align}
With $p^2=M^2$ substitution of the above definitions into the expressions $\sigma_{0 \Lambda_0} F^{AA}$, $\sigma_{0 \Lambda_0} F^{\bar{c}c}$ appearing in AGE\eqref{eq_21gh} and STI\eqref{eq_21st} gives
\begin{equation}
\sigma_{0 \Lambda_0}(p^2)\,F^{\bar{c}c}(p)\bim -p^2(1 + \Sigma^{\bar{c}c}(p^2)),
\end{equation}
\begin{equation}
\sigma_{0 \Lambda_0}(p^2)\,\bunderline{F}^{AA}_{\mu\nu}(p)\bim F^{A A}_{T;\mu \nu}(p)\,,
\end{equation}
\begin{equation}
F^{A A}_{T;\mu \nu}(p):=(\delta_{\mu \nu}p^2 - p_\mu p_\nu)(1+\Sigma^{AA}_T(p^2))+\frac{1}{\xi} p_\mu p_\nu \Sigma^{AA}_L(p^2). \label{def_aa_t}
\end{equation}
Using \eqref{eq_17g3} for the functional $\Gamme^{0\Lambda_0}_2$ we have
\begin{align}
\Gamme^{0\Lambda_0}_2&= \frac{1}{2}\langle \mathsf{F}^{AA}_{\mu\nu}\check{A}^a_\mu \check{A}^a_\nu \rangle + \langle \mathsf{F}^{\bar{c}c} \check{\bar{c}}^a\check{c}^a \rangle ,\\
\mathsf{F}^{AA}_{\mu\nu}(p)&:=(\delta_{\mu\nu}p^2-p_\mu p_\nu)\Sigma^{AA}_T(p^2)+\frac{1}{\xi}p_\mu p_\nu\Sigma^{AA}_L(p^2)\,,\\
\mathsf{F}^{\bar{c}c}(p)&:=-p^2\Sigma^{\bar{c}c}(p^2)\,.
\end{align}
For marginal terms we obtain
\begin{align}
\mathsf{F}^{AA;p_\rho p_\sigma}_{\mu\nu}(p)=&2\delta_{\mu \nu}\delta_{\rho \sigma} r^{AA}_1 + 2(\delta_{\mu \rho} \delta_{\nu \sigma} + \delta_{\nu \rho} \delta_{\mu \sigma}) r^{AA}_2+\zeta^{AA}_{\mu \nu \rho \sigma} \;,\\
\mathsf{F}^{\bar{c}c;p_\rho p_\sigma}(p)=&2\delta_{\rho \sigma}r^{\bar{c}c}(p^2) +\zeta^{\bar{c}c}_{\rho \sigma}\;,\\
r^{AA}_1(p^2):=&\Sigma^{AA}_T(p^2) + p^2 \frac{\partial \Sigma^{AA}_T(p^2)}{\partial p^2}\,,\\
r^{AA}_2(p^2):=&\frac{1}{\xi}\Sigma^{AA}_L(p^2) - \Sigma^{AA}_T(p^2)\,,\\
r^{\bar{c}c}(p^2):=&-\Sigma^{\bar{c}c}(p^2) - p^2 \frac{\partial \Sigma^{\bar{c}c}(p^2)}{\partial p^2}\,.
\end{align}
\item{\textbf{Three-point functions}}
\begin{align}
\Gamma^{0\Lambda_0}_3=&\langle \epsilon_{abd} F^{AAA}_{\rho\mu\nu}  \check{A}^a_\rho  \check{A}^b_\mu \check{A}^d_\nu \rangle + \langle \epsilon_{adb}  F^{A \bar{c} c}_\mu \check{A}^a_\mu \check{\bar{c}}^b \check{c}^d \rangle,\\
F^{A \bar{c} c}_\mu(k,p,q):=&ip_\mu R^{A\bar{c}c}_1(p,q) + iq_\mu r^{A\bar{c}c}_2(p,q),\\
R^{A\bar{c}c}_1(p,q):=&g+r^{A\bar{c}c}_1(p,q),\\
F^{AAA}_{\rho\mu\nu}(k,p,q):=&i\delta_{\mu\nu}(p_\rho - q_\rho)R^{AAA}(p,q) + i\delta_{\mu\nu}k_\rho \zeta^{AAA}_-(p,q) \nonumber\\
&+i\zeta^{AAA}_{\rho \mu \nu}(p,q),\\
R^{AAA}(p,q):=&\frac{1}{2}g + r^{AAA}(p,q).
\end{align}
Here $R^{AAA}(p,q)$ is a symmetric function whereas $\zeta^{AAA}_-(p,q)$ is antisymmetric.
\item{\textbf{Four-point functions}}
\begin{align}
\Gamma^{0\Lambda_0}_4=&\langle  F^{AAAA}_{\sigma \rho \mu \nu} \check{A}^b_\sigma \check{A}^b_\rho \check{A}^a_\mu \check{A}^a_\nu +F^{\bar{c}cAA}_{1, \mu \nu} \check{\bar{c}}^b \check{c}^b \check{A}^a_\mu \check{A}^a_\nu \nonumber \\
&+F^{\bar{c}cAA}_{2, \mu \nu} \check{\bar{c}}^a \check{c}^b \check{A}^a_\mu \check{A}^b_\nu +r^{\bar{c}c\bar{c}c} \check{\bar{c}}^b \check{c}^b \check{\bar{c}}^a \check{c}^a \rangle,\\
F^{\bar{c}cAA}_{n, \mu \nu}:=&\delta_{\mu \nu} r^{\bar{c}cAA}_n +\zeta^{\bar{c}cAA}_{n, \mu \nu},\\
F^{AAAA}_{\sigma \rho \mu \nu}:=&R^{AAAA}_{\sigma \rho \mu \nu}+\zeta^{AAAA}_{\sigma \rho \mu \nu},\\
R^{AAAA}_{\sigma \rho \mu \nu}:=&\frac{1}{2}(\delta_{\mu \rho} \delta_{\nu \sigma}+ \delta_{\mu \sigma} \delta_{\rho \nu})R^{AAAA}_1  +\delta_{\mu \nu} \delta_{\sigma \rho} R^{AAAA}_2 \nonumber\\
&+\frac{1}{2}(\delta_{\mu \rho} \delta_{\nu \sigma} - \delta_{\mu \sigma} \delta_{\rho \nu})\zeta^{AAAA}_-,\\
R^{AAAA}_1:=&-\frac{g^2}{4}+r^{AAAA}_1,\\
R^{AAAA}_2:=&\frac{g^2}{4}+r^{AAAA}_2
\end{align}
\end{enumerate}
Here the terms
\begin{equation}
\begin{matrix}
r^{AA}_1,&r^{AA}_2,&r^{\bar{c}c},&R^{A\bar{c}c}_1,&r^{A\bar{c}c}_2,&R^{AAA},\\
R^{AAAA}_1,&R^{AAAA}_2,&r^{\bar{c}cAA}_1,&r^{\bar{c}cAA}_2,&r^{\bar{c}c\bar{c}c},&\label{eq_11r}
\end{matrix}
\end{equation}
are scalar functions of momenta, $\Lambda_0$, and $M$. All 11 renormalization constants are fine-tuned by imposing appropriate renormalization conditions.
\section{The functionals $\Gamme_{\gamma; n\leqslant 2}$ and  $\Gamme_{\omega;n\leqslant 2}$}\label{sec_gamma2}
With $\Gamme^{0\Lambda_0}_{\varkappa}(p):=\tilde{\delta}_{\varkappa(p)} \Gamme^{0\Lambda_0}|_{\varkappa=0}$,
\begin{align}
\Gamme^{0\Lambda_0}_{\gamma^a_\mu; n\leqslant 2}(p)=&R_1 ip_\mu c^a(p)+  g\epsilon^{abs} \langle F^{\gamma A c}_{\mu \nu}| \check{A}^b_\nu\check{c}^s;p \rangle,\\
F^{\gamma A c}_{\mu \nu}(k,q):=&\delta_{\mu \nu}R_2 + \zeta^{\gamma A c}_{\mu \nu}(k,q),\\
\zeta^{\gamma A c}_{\mu \nu}(k,q):=&k_\mu q_\nu \zeta^{\gamma A c}_1 + k_\nu q_\mu\zeta^{\gamma A c}_2 +k_\mu k_\nu \zeta^{\gamma A c}_3 + q_\mu q_\nu \zeta^{\gamma A c}_4,\\
\Gamme^{0\Lambda_0}_{\omega^a; n\leqslant 2}(p)=& \frac{1}{2} g\epsilon^{abs} \langle R_3 |\check{c}^b \check{c}^s;p \rangle.
\end{align}
Here $R_1$, $R_2$, $R_3$ are scalar functions of momenta, $\Lambda_0$, and $M$.
\section{Violated STI for $\Lambda>0$}\label{sec_rr}
In this section we present, omitting the details of the calculation, the extension of some results of section \ref{sec_vsti} to arbitrary $0< \Lambda < \Lambda_0$.
\begin{align}
\tilde{\Gamme}^{\Lambda \Lambda_0}_\beta=&\sigma_{0\Lambda_0}*\Big(\frac{\delta \tilde{\Damma}^{\Lambda \Lambda_0}}{\delta \wbar{c}} - \partial \frac{\delta \tilde{\Damma}^{\Lambda \Lambda_0}}{\delta \gamma}\Big)\,,\label{628a}\\
\tilde{\Gamme}^{\Lambda \Lambda_0}_1=&\langle\frac{\delta \tilde{\Damma}^{\Lambda \Lambda_0}}{\delta \ubar{A}},\sigma_{0\Lambda_0}*\frac{\delta \tilde{\Damma}^{\Lambda \Lambda_0}}{\delta \gamma}\rangle-\langle\frac{\delta \tilde{\Damma}^{\Lambda \Lambda_0}}{\delta \ubar{c}},\sigma_{0\Lambda_0}*\frac{\delta \tilde{\Damma}^{\Lambda \Lambda_0}}{\delta \omega}\rangle \nonumber\\
& - \frac{1}{\xi}\langle \partial \ubar{A},\sigma_{0\Lambda_0}*\frac{\delta \tilde{\Damma}^{\Lambda \Lambda_0}}{\delta \wbar{c}}\rangle + \hbar\tilde{\Delta}^{\Lambda \Lambda_0}\,,\label{628b}\\
\tilde{\Damma}^{\Lambda \Lambda_0}:=&\tilde{\Gamme}^{\Lambda \Lambda_0} + \frac{1}{2}\langle \tilde{\ubar{\Phi}}, \tilde{\mathbf{C}}^{-1}_{0 \Lambda_0}\tilde{\ubar{\Phi}}\rangle\,,\\
\tilde{\Delta}^{\Lambda \Lambda_0}:=&\langle (\sigma_{\Lambda},0,0)\Big(1+\frac{\delta^2 \tilde{\Gamme}^{\Lambda \Lambda_0}}{\delta \tilde{\ubar{\Phi}} \delta \tilde{\ubar{\Phi}}} \hat{\mathbf{1}}\tilde{\mathbf{C}}^{\Lambda \Lambda_0}\Big)^{-1}\frac{\delta^2 \tilde{\Gamme}^{\Lambda \Lambda_0}}{\delta \tilde{\ubar{\Phi}} \delta \gamma }\rangle \nonumber\\
&+\langle (0,\sigma_{\Lambda},0)\Big(1+\frac{\delta^2 \tilde{\Gamme}^{\Lambda \Lambda_0}}{\delta \tilde{\ubar{\Phi}} \delta \tilde{\ubar{\Phi}}} \hat{\mathbf{1}}\tilde{\mathbf{C}}^{\Lambda \Lambda_0}\Big)^{-1}\frac{\delta^2 \tilde{\Gamme}^{\Lambda \Lambda_0}}{\delta \tilde{\ubar{\Phi}}\delta \omega }\rangle \nonumber\\
&-\langle (\sigma_{\Lambda}\partial,0,0)\Big(1+\frac{\delta^2 \tilde{\Gamme}^{\Lambda \Lambda_0}}{\delta \tilde{\ubar{\Phi}} \delta \tilde{\ubar{\Phi}}} \hat{\mathbf{1}}\tilde{\mathbf{C}}^{\Lambda \Lambda_0}\Big)^{-1}\frac{\delta \ubar{c}}{\delta \tilde{\ubar{\Phi}}}\rangle\,.
\end{align}
Note that in relations \eqref{628a} and \eqref{628b} there still appears a convolution with $\sigma_{0\Lambda_0}$ since we have chosen to define the regularized BRST transformation to include a convolution with this function, see~\eqref{eq_20brs}.

Using the bounds of theorem \ref{thm_1} one can show that $\lim \limits_{\Lambda \to 0} \Delta^{\Lambda \Lambda_0}=0$ at nonexceptional momenta. It follows that $\lim \limits_{\Lambda \to 0} \tilde{\Gamme}^{\Lambda \Lambda_0}_1=\tilde{\Gamme}^{0 \Lambda_0}_1$ at nonexceptional momenta. More information can be found in \cite{thesis}.
\newpage
\section{List of the renormalization points}\label{sec_ren}
\begin{center}
\newcommand{\hs}{\hspace{2pt}}
\renewcommand{\arraystretch}{1.2}
\begin{tabular}{cc}
\begin{tabular}[t]{|@{\hs}c@{\hs}|@{\hs}c@{\hs}|@{\hs}c@{\hs}|@{\hs}c@{\hs}|@{\hs}c@{\hs}|@{\hs}c@{\hs}|}
\hline
$X$ & $[X]$ & $n_X$   & $r_{\mathrm{m}}$ & $r_X$
& ren.p.
\\
\hline
$\partial\partial\Gamme^{\bar{c}c}$
& 0 &2 & 7& 0 &$\Lambda=0,\;\vec{q}\in \mathbb{M}^s_2$
\\
\hline
$\partial\Gamme_\gamma^{c}$
& 0 &2 & 7& 2 
&$\Lambda=0,\;\vec{q}\in\mathbb{M}^{s}_2$
\\
\hline
$\partial\partial\Gamme^{AA}$
& 0 &2 & 7& 4 
&$\Lambda=0,\;\vec{q}\in\mathbb{M}^{s}_2$
\\
\hline
$\Gamme_\omega^{cc}$
& 0 &3& 5& 0
&$\Lambda=0,\;\vec{q}\in\mathbb{M}^{s}_3$
\\
\hline
$\partial_{A}\Gamme^{c\bar{c}A}$
& 0 & 3& 5& 2 
&$\Lambda=M,\;\vec{q}=0$
\\
\hline
$\partial_{\bar{c}} \Gamme^{c\bar{c}A}$
& 0 & 3& 5& 2 
&$\Lambda=0,\;\vec{q} \in \mathbb{M}^s_3$
\\
\hline
$\Gamme_\gamma^{cA}$
& 0& 3 & 5& 2 
&$\Lambda=0,\;\vec{q}\in\mathbb{M}^{s}_3$
\\
\hline
$\Gamme_\gamma^{AA}$
& 0 &3 & 5& 3 
&$\Lambda=0,\;\vec{q}\in\mathbb{M}^{s}_3$
\\
\hline
$\partial\Gamme^{AAA}$
& 0 & 3& 5& 4 
&$\Lambda=0,\;\vec{q}\in\mathbb{M}^{s}_3$
\\
\hline
$\Gamme^{\bar{c}c\bar{c}c}$
&0 & 4&3& 0 
&$\Lambda=M,\;\vec{q}=0$
\\
\hline
$\Gamme^{\bar{c}cAA}$
& 0 & 4& 3& 2 
&$\Lambda=M,\;\vec{q}=0$
\\
\hline
$\Gamme^{AAAA}$
& 0 & 4 & 3& 4${}^*$
&$\Lambda=0,\;\vec{q}\in\mathbb{M}^{cp}_4$
\\
\hline\hline
$\partial\partial\Gamme_{\beta}^{c}$&0&2&7&2
&$\Lambda=0,\;\vec{q}\in\mathbb{M}^{s}_2$
\\\hline
$\partial\partial\partial\Gamme_1^{cA}$&0&2&7&4
&$\Lambda=0,\;\vec{q}\in\mathbb{M}^{s}_2$
\\\hline
$\partial\partial\Gamme_1^{c\bar{c}c}$&0&3&5&2
&$\Lambda=0,\;\vec{q}\in\mathbb{M}^{s}_3$
\\\hline
$\partial\Gamme_{1,\gamma}^{cc}$&0&3&5&2
&$\Lambda=0,\;\vec{q}\in\mathbb{M}^{s}_3$
\\\hline
$\partial\partial\Gamme_1^{cAA}$&0&3&5&4
&$\Lambda=0,\;\vec{q}\in\mathbb{M}^{s}_3$
\\\hline
$\Gamme_{\beta}^{c\bar{c}c}$&0&4&3&0
&$\Lambda=0,\;\vec{q}\in\mathbb{M}^{s}_4$
\\\hline
$\Gamme_{1,\omega}^{ccc}$&0&4&3&0
&$\Lambda=0,\;\vec{q}\in\mathbb{M}^{s}_4$
\\\hline
$\Gamme_{\beta}^{cAA}$&0&4&3&2
&$\Lambda=0,\;\vec{q}\in\mathbb{M}^{s}_4$
\\\hline
$\partial\Gamme_1^{c\bar{c}cA}$&0&4&3&2
&$\Lambda=0,\;\vec{q}\in\mathbb{M}^{s}_4$
\\\hline
$\Gamme_{1,\gamma}^{ccA}$&0&4&3&2
&$\Lambda=0,\;\vec{q}\in\mathbb{M}^{s}_4$
\\\hline
$\partial\Gamme_1^{cAAA}$&0&4&3&4${}^*$
&$\Lambda=0,\;\vec{q}\in\mathbb{M}^{cp}_4$
\\\hline
$\Gamme_1^{c\bar{c}c\bar{c}c}$&0&5&1&0
&$\Lambda=0,\;\vec{q}\in\mathbb{M}^{s}_5$
\\\hline
$\Gamme_1^{c\bar{c}cAA}$&0&5&1&$2^*$
&$\Lambda=0,\;\vec{q}\in\mathbb{M}^{s}_5$
\\\hline
$\Gamme_1^{cAAAA}$&0&5&1&4${}^*$
&$\Lambda=0,\;\vec{q}\in\mathbb{M}^{cp}_5$
\\\hline
\end{tabular}
&
\begin{tabular}[t]{|@{\hs}c@{\hs}|@{\hs}c@{\hs}|@{\hs}c@{\hs}|@{\hs}c@{\hs}|@{\hs}c@{\hs}|}
\hline
$X$ & $[X]$ & $n_X$  & $r_X$
& ren.p.
\\
\hline
$\partial\Gamme^{\bar{c}c}$
& 1 &2 & 1 
& $\Lambda=0,\;\vec{q}=0$
\\
\hline
$\partial\Gamme^{AA}$
& 1 &2 & 3 
& $\Lambda=0,\;\vec{q}=0$
\\
\hline
$\Gamme^{\bar{c}cA}$
& 1 & 3& 1 
& $\Lambda=0,\;\vec{q}=0$
\\
\hline
$\Gamme^{AAA}$
& 1 & 3& 3 
& $\Lambda=0,\;\vec{q}=0$
\\
\hline
$\Gamme^{\bar{c}c}$
& 2 &2 & 0 
& $\Lambda=0,\;\vec{q}=0$
\\
\hline
$\Gamme^{AA}$
& 2 &2 & 2 
& $\Lambda=0,\;\vec{q}=0$
\\
\hline\hline
$\partial\Gamme_{\beta}^{c}$&1&2&1&
$\Lambda=0,\;\vec{q}=0$
\\\hline
$\partial\partial\Gamme_1^{cA}$&1&2&3&
$\Lambda=0,\;\vec{q}=0$
\\\hline
$\Gamme_{\beta}^{cA}$&1&3&1&
$\Lambda=0,\;\vec{q}=0$
\\\hline
$\partial\Gamme_1^{c\bar{c}c}$&1&3&1&
$\Lambda=0,\;\vec{q}=0$
\\\hline
$\Gamme_{1,\gamma}^{cc}$&1&3&1&
$\Lambda=0,\;\vec{q}=0$
\\\hline
$\partial\Gamme_1^{cAA}$&1&3&3&
$\Lambda=0,\;\vec{q}=0$
\\\hline
$\Gamme_1^{c\bar{c}cA}$&1&4&1&
$\Lambda=0,\;\vec{q}=0$
\\\hline
$\Gamme_1^{cAAA}$&1&4&3&
$\Lambda=0,\;\vec{q}=0$
\\\hline
$\Gamme_{\beta}^{c}$&2&2&0&
$\Lambda=0,\;\vec{q}=0$
\\\hline
$\partial\Gamme_1^{cA}$&2&2&2&
$\Lambda=0,\;\vec{q}=0$
\\\hline
$\Gamme_1^{c\bar{c}c}$&2&3&0&
$\Lambda=0,\;\vec{q}=0$
\\\hline
$\Gamme_1^{cAA}$&2&3&2&
$\Lambda=0,\;\vec{q}=0$
\\\hline
$\Gamme_1^{cA}$&3&2&1&
$\Lambda=0,\;\vec{q}=0$
\\\hline
\end{tabular}
\end{tabular}
\end{center}

\vspace{2em}
\noindent
List of all terms preserving the global symmetries,
with an arbitrary number of $\gamma,\omega$ insertions,
and
with at most one $\beta$ or $1$ insertion (not both).
Notation:
$[X]$ is the mass dimension of $X$
(reduced Fourier transform);
$r_X$ is the tensor rank;
$n_X$ is the total number of fields and sources
(not including $1$);
$\partial$ stands for a momentum derivative;
``ren.p.'' stands for ``renormalization point''.
A $*$ in the rank entry means that the condition
$r_{\mathrm{m}}:= 11-2n_X\ge r+1$
is violated for a term $X$:
as stated in Lemma \ref{lem_monom} the tensor monomials
$\{\delta^s\underline{q}^k\}_{r+1}$
are not linearly independent for
$\vec{q}=(q_0,\underline{q})\in\mathbb{M}^{s}_{n_X}$,
hence  they are not suitable as a basis
for the form-factor decomposition of $\partial X$. See lemma \ref{lem_rn} and sections~\ref{st_cAAAA},~\ref{st_cccAA},~\ref{st_cAAA}.
\section{List of insertions}\label{sec_op}
\begin{center}
\renewcommand{\arraystretch}{1.1}
\begin{tabular}{ |Sc|Sc|Sc|Sc|}
\hline
X & [X]& $\mathrm{gh}(X)$& def
\\
\hline
$\psi_\mu^a$
& $2$&$1$&\eqref{eq_21brs}
\\
$\gamma_\mu^a$
& $2$&$-1$&
\\
\hline
$\Omega^a$ 
&$2$&$2$&\eqref{eq_21brs}
\\
$\omega^a$ 
&$2$&$-2$&
\\
\hline
$Q_\rho$
& $5 $&$1$&\eqref{eq_328a}
\\
$Q_{\rho\gamma}$
& $3 $&$2$&
\eqref{eq_11rg}\\
$Q_{\rho\omega}$
& $3 $&$3$&
\eqref{eq_11rw}\\
$\rho$& $-1$&$-1$&
\\
\hline
$Q_{\beta}$ &$3 $& $1$&\eqref{eq_10a}
\\
${\beta}$&$1$& $-1$&
\\
\hline
\end{tabular}
\end{center}

\vspace{2em}
\noindent
List of operators and sources,
and their quantum numbers.
Notation:
$[X]$ stands for the mass dimension of $X$ in position space;
the ghost charge of the ghost field is $\mathrm{gh}(c):=1$.
\end{appendices}
\newpage
\bibliographystyle{alpha}
\bibliography{ym}

\begin{thebibliography}{BDM95}

\bibitem[BDM93]{bam2}
M.~Bonini, M.~D'Attanasio, and G.~Marchesini.
\newblock Perturbative renormalization and infrared finiteness in the {W}ilson
  renormalization group: the massless scalar case.
\newblock {\em Nucl. Phys. B}, 409:441 -- 464, 1993.

\bibitem[BDM95]{bam}
M.~Bonini, M.~D'Attanasio, and G.~Marchesini.
\newblock {BRS symmetry for Yang-Mills theory with exact renormalization
  group}.
\newblock {\em Nucl. Phys. B}, 437:163--186, 1995.

\bibitem[Bec96]{bec}
C.~Becchi.
\newblock On the construction of renormalized gauge theories using
  renormalization group techniques.
\newblock {\em arXiv}, 9607188v1, 1996.

\bibitem[Ber66]{ber}
F.A. Berezin.
\newblock {\em The method of second quantization. Translated by {N.
  Mugibayashi} and {A. Jeffrey}}.
\newblock Academic Press New York, 1966.

\bibitem[BRS75]{brs}
C.~Becchi, A.~Rouet, and R.~Stora.
\newblock Renormalization of the abelian {Higgs-Kibble} model.
\newblock {\em Comm. Math. Phys}, 42(2):127--162, 1975.

\bibitem[DF91]{dlfm}
Yu.L. Dalecky and S.V. Fomin.
\newblock {\em Measures and differential equations in infinite-dimensional
  space}.
\newblock Springer, 1991.

\bibitem[Efr17]{thesis}
A.N. Efremov.
\newblock {\em Renormalization of SU(2) Yang-Mills theory with flow equations}.
\newblock PhD thesis, Universit\'e Paris-Saclay, 2017.

\bibitem[FHH16]{fhh}
M.B. Fr{\"o}b, J.~Holland, and S.~Hollands.
\newblock All-order bounds for correlation functions of gauge-invariant
  operators in {Yang-Mills} theory.
\newblock {\em Journal Math. Phys.}, 57(12), 2016.

\bibitem[FP67]{FP}
L.D. Faddeev and V.N. Popov.
\newblock Feynman diagrams for the {Yang-Mills} field.
\newblock {\em Physics Letters B}, 25:29--30, 1967.

\bibitem[GG00]{gg}
P.~Gambino and P.~A. Grassi.
\newblock Nielsen identities of the {SM} and the definition of mass.
\newblock {\em Phys. Rev. D}, 62(7):076002, 2000.

\bibitem[GJ87]{gljf}
J.~Glimm and A.~Jaffe.
\newblock {\em Quantum physics}.
\newblock Springer, 1987.

\bibitem[GK]{ric}
R.~Guida and Ch. Kopper.
\newblock All-order uniform momentum bounds for the massless $\phi^4_4$ field
  theory.
\newblock {\em In preparation}.

\bibitem[GK11]{kogui}
R.~Guida and Ch. Kopper.
\newblock All-order uniform momentum bounds for the massless $\phi^4$ theory in
  four dimensional euclidean space.
\newblock {\em arXiv}, 1103.5692, 2011.

\bibitem[KK92]{keko}
G.~Keller and Ch. Kopper.
\newblock Perturbative renormalization of composite operators via flow
  equations.
\newblock {\em Comm. Math. Phys.}, 148:445--467, 1992.

\bibitem[KKS92]{kks}
G.~Keller, Ch. Kopper, and M.~Salmhofer.
\newblock Perturbative renormalization and effective lagrangians in $\phi^4$ in
  four-dimensions.
\newblock {\em Helv. Phys. Acta}, 65:32--52, 1992.

\bibitem[KKS97]{kks2}
G.~Keller, Ch. Kopper, and C.~Schophaus.
\newblock Perturbative renormalization with flow equations in minkowski space.
\newblock {\em Helv.Phys.Acta}, (70):247--274, 1997.

\bibitem[KM02]{kome}
Ch. Kopper and F.~Meunier.
\newblock Large momentum bounds from flow equations.
\newblock {\em Annales Henri Poincar{\'e}}, 3:435--449, 2002.

\bibitem[KM09]{komu}
Ch. Kopper and V.F. M{\"u}ller.
\newblock Renormalization of spontaneously broken {SU(2) Yang-Mills} theory
  with flow equations.
\newblock {\em Rev. in Math. Phys.}, 21(6):781--820, 2009.

\bibitem[KSZ75]{KSZ75}
H.~Kluberg-Stern and J.B. Zuber.
\newblock Ward identities and some clues to the renormalization of
  gauge-invariant operators.
\newblock {\em Phys. Rev. D}, 12:467--481, Jul 1975.

\bibitem[Lai81]{Lai81}
C.H. Lai, editor.
\newblock {\em Gauge theory of weak and electromagnetic interactions}.
\newblock World Scientific, 1981.

\bibitem[LZJ72]{LZJ}
B.W. Lee and J.~Zinn-Justin.
\newblock Spontaneously broken gauge symmetries. {I,II,III.}
\newblock {\em Phys. Rev. D}, 5:3121--3137, 3137--3155,3155--3160., 1972.

\bibitem[MD96]{ma}
T.R. Morris and M.~D'Attanasio.
\newblock {Gauge Invariance, the Quantum Action Principle, and the
  Renormalization Group}.
\newblock {\em Physics Letters B}, 378:213--221, 1996.

\bibitem[Mor94]{mor}
T.R. Morris.
\newblock The exact renormalization group and approximate solutions.
\newblock {\em Int. Journ. of Mod. Phys. A}, 09(14):2411--2449, 1994.

\bibitem[M{\"u}l03]{mul}
V.F. M{\"u}ller.
\newblock Perturbative renormalization by flow equations.
\newblock {\em Rev. in Math. Phys.}, 15(05):491--558, 2003.

\bibitem[Nie75]{niel}
N.K. Nielsen.
\newblock On the gauge dependence of spontaneous symmetry breaking in gauge
  theories.
\newblock {\em Nucl. Phys. B}, 101:173--188, 1975.

\bibitem[PF67]{fad}
V.N. Popov and L.D. Faddeev.
\newblock Perturbation theory for gauge-invariant fields.
\newblock {\em preprint ITF-67-036 Kiev}, 1967.
\newblock In russian. English translation: NAL-THY-57.

\bibitem[Pol84]{pol}
J.~Polchinski.
\newblock Renormalization and effective {L}agrangians.
\newblock {\em Nucl. Phys. B}, 231:269--295, 1984.

\bibitem[PS85]{pig}
O.~Piguet and K.~Sibold.
\newblock Gauge independence in ordinary {Yang-Mills} theories.
\newblock {\em Nucl. Phys. B}, 253:517--540, 1985.

\bibitem[Ree73]{ree}
M.~Reed.
\newblock Functional analysis and probability theory.
\newblock In G.~Velo and A.~Wightman, editors, {\em Constructive Quantum Field
  Theory}, volume~25 of {\em Lecture Notes in Physics}. Springer, 1973.

\bibitem[RW94]{wet}
M.~Reuter and C.~Wetterich.
\newblock Effective average action for gauge theories and exact evolution
  equations.
\newblock {\em Nucl. Phys. B}, 417:181--214, 1994.

\bibitem[SF91]{slfd}
A.A. Slavnov and L.D. Faddeev.
\newblock {\em Gauge fields: an introduction to quantum theory}.
\newblock Addison-Wesley Publishing Company, 1991.

\bibitem[Sla72]{st2}
A.A. Slavnov.
\newblock Ward identities in gauge theories.
\newblock {\em Theo. and Math. Phys.}, 10(2):99--104, 1972.

\bibitem[Tay71]{taylor}
J.C. Taylor.
\newblock Ward identities and charge renormalization of the {Yang-Mills} field.
\newblock {\em Nucl. Phys. B}, 33:436--444, 1971.

\bibitem[tH71]{thooft}
G.~'t~Hooft.
\newblock Renormalization of massless {Yang-Mills} fields.
\newblock {\em Nucl. Phys. B}, 33:173--199, 1971.

\bibitem[tHV72]{tHV72}
G.~'t~Hooft and M.~Veltman.
\newblock Regularization and renormalization of gauge fields.
\newblock {\em Nucl. Phys. B}, 44:189--213, 1972.

\bibitem[Tyu75]{tyutin}
I.V. Tyutin.
\newblock Gauge invariance in field theory and statistical physics in operator
  formalism.
\newblock {\em arXiv}, 0812.0580, 1975.

\bibitem[Wet93]{Wet2}
C.~Wetterich.
\newblock Exanct evolution equation for the effective potential.
\newblock {\em Physics Letters B}, 301:90--94, 1993.

\bibitem[WH73]{wehoug}
F.~Wegner and A.~Houghton.
\newblock Renormalization group equations for critical phenomena.
\newblock {\em Phys. Rev. A}, 8(1):401--412, 1973.

\bibitem[Wil71]{wilson}
K.G. Wilson.
\newblock Renormalization group and critical phenomena.
\newblock {\em Phys. Rev. B}, 4(9):3174--3183, 1971.

\bibitem[YM54]{YM54}
C.N. Yang and R.L. Mills.
\newblock Conservation of isotopic spin and isotopic gauge invariance.
\newblock {\em Phys. Rev.}, 96:191--195, Oct 1954.

\bibitem[ZJ75]{st}
J.~Zinn-Justin.
\newblock {\em Trends in elementary particle theory}, volume~37.
\newblock Springer Berlin Heidelberg, 1975.
\newblock International summer institute on theoretical physics in Bonn 1974.

\end{thebibliography}
\end{document}